\documentclass[aps,prx,onecolumn,amsmath,nofootinbib,amssymb,11pt,floatfix]{revtex4-2}
\usepackage[T1]{fontenc}
\usepackage[latin9]{luainputenc}
\usepackage[T1]{fontenc}
\usepackage[latin9]{luainputenc}
\usepackage{amsmath,bm}
\usepackage{fancyhdr}
\usepackage{amssymb}
\usepackage{graphicx}
\usepackage{ulem}   
\usepackage{xcolor}
\usepackage{bbold}
\usepackage{comment}
\usepackage[colorlinks, linkcolor=blue, citecolor=blue]{hyperref}
\usepackage[compat=1.1.0]{tikz-feynman}
\tikzset{
pattern size/.store in=\mcSize, 
pattern size = 5pt,
pattern thickness/.store in=\mcThickness, 
pattern thickness = 0.3pt,
pattern radius/.store in=\mcRadius, 
pattern radius = 1pt}
\makeatletter
\pgfutil@ifundefined{pgf@pattern@name@_66zr4cbsr}{
\pgfdeclarepatternformonly[\mcThickness,\mcSize]{_66zr4cbsr}
{\pgfqpoint{0pt}{0pt}}
{\pgfpoint{\mcSize+\mcThickness}{\mcSize+\mcThickness}}
{\pgfpoint{\mcSize}{\mcSize}}
{
\pgfsetcolor{\tikz@pattern@color}
\pgfsetlinewidth{\mcThickness}
\pgfpathmoveto{\pgfqpoint{0pt}{0pt}}
\pgfpathlineto{\pgfpoint{\mcSize+\mcThickness}{\mcSize+\mcThickness}}
\pgfusepath{stroke}
}}

\tikzset{
pattern size/.store in=\mcSize, 
pattern size = 5pt,
pattern thickness/.store in=\mcThickness, 
pattern thickness = 0.3pt,
pattern radius/.store in=\mcRadius, 
pattern radius = 1pt}

\tikzset{
pattern size/.store in=\mcSize, 
pattern size = 5pt,
pattern thickness/.store in=\mcThickness, 
pattern thickness = 0.3pt,
pattern radius/.store in=\mcRadius, 
pattern radius = 1pt}
\makeatletter
\pgfutil@ifundefined{pgf@pattern@name@_zkop8g48k}{
\pgfdeclarepatternformonly[\mcThickness,\mcSize]{_zkop8g48k}
{\pgfqpoint{0pt}{0pt}}
{\pgfpoint{\mcSize+\mcThickness}{\mcSize+\mcThickness}}
{\pgfpoint{\mcSize}{\mcSize}}
{
\pgfsetcolor{\tikz@pattern@color}
\pgfsetlinewidth{\mcThickness}
\pgfpathmoveto{\pgfqpoint{0pt}{0pt}}
\pgfpathlineto{\pgfpoint{\mcSize+\mcThickness}{\mcSize+\mcThickness}}
\pgfusepath{stroke}
}}
\makeatother

\tikzset{
pattern size/.store in=\mcSize, 
pattern size = 5pt,
pattern thickness/.store in=\mcThickness, 
pattern thickness = 0.3pt,
pattern radius/.store in=\mcRadius, 
pattern radius = 1pt}
\makeatletter
\pgfutil@ifundefined{pgf@pattern@name@_66zr4cbsr}{
\pgfdeclarepatternformonly[\mcThickness,\mcSize]{_66zr4cbsr}
{\pgfqpoint{0pt}{0pt}}
{\pgfpoint{\mcSize+\mcThickness}{\mcSize+\mcThickness}}
{\pgfpoint{\mcSize}{\mcSize}}
{
\pgfsetcolor{\tikz@pattern@color}
\pgfsetlinewidth{\mcThickness}
\pgfpathmoveto{\pgfqpoint{0pt}{0pt}}
\pgfpathlineto{\pgfpoint{\mcSize+\mcThickness}{\mcSize+\mcThickness}}
\pgfusepath{stroke}
}}
\makeatother

\tikzset{
pattern size/.store in=\mcSize, 
pattern size = 5pt,
pattern thickness/.store in=\mcThickness, 
pattern thickness = 0.3pt,
pattern radius/.store in=\mcRadius, 
pattern radius = 1pt}
\makeatletter
\pgfutil@ifundefined{pgf@pattern@name@_su7zzwc8a}{
\pgfdeclarepatternformonly[\mcThickness,\mcSize]{_su7zzwc8a}
{\pgfqpoint{0pt}{0pt}}
{\pgfpoint{\mcSize+\mcThickness}{\mcSize+\mcThickness}}
{\pgfpoint{\mcSize}{\mcSize}}
{
\pgfsetcolor{\tikz@pattern@color}
\pgfsetlinewidth{\mcThickness}
\pgfpathmoveto{\pgfqpoint{0pt}{0pt}}
\pgfpathlineto{\pgfpoint{\mcSize+\mcThickness}{\mcSize+\mcThickness}}
\pgfusepath{stroke}
}}
\makeatother

\tikzset{
pattern size/.store in=\mcSize, 
pattern size = 5pt,
pattern thickness/.store in=\mcThickness, 
pattern thickness = 0.3pt,
pattern radius/.store in=\mcRadius, 
pattern radius = 1pt}
\makeatletter
\pgfutil@ifundefined{pgf@pattern@name@_xxg34rr44}{
\pgfdeclarepatternformonly[\mcThickness,\mcSize]{_xxg34rr44}
{\pgfqpoint{0pt}{0pt}}
{\pgfpoint{\mcSize+\mcThickness}{\mcSize+\mcThickness}}
{\pgfpoint{\mcSize}{\mcSize}}
{
\pgfsetcolor{\tikz@pattern@color}
\pgfsetlinewidth{\mcThickness}
\pgfpathmoveto{\pgfqpoint{0pt}{0pt}}
\pgfpathlineto{\pgfpoint{\mcSize+\mcThickness}{\mcSize+\mcThickness}}
\pgfusepath{stroke}
}}
\makeatother

\tikzset{
pattern size/.store in=\mcSize, 
pattern size = 5pt,
pattern thickness/.store in=\mcThickness, 
pattern thickness = 0.3pt,
pattern radius/.store in=\mcRadius, 
pattern radius = 1pt}
\makeatletter
\pgfutil@ifundefined{pgf@pattern@name@_4aaxmu7dk}{
\pgfdeclarepatternformonly[\mcThickness,\mcSize]{_4aaxmu7dk}
{\pgfqpoint{0pt}{0pt}}
{\pgfpoint{\mcSize+\mcThickness}{\mcSize+\mcThickness}}
{\pgfpoint{\mcSize}{\mcSize}}
{
\pgfsetcolor{\tikz@pattern@color}
\pgfsetlinewidth{\mcThickness}
\pgfpathmoveto{\pgfqpoint{0pt}{0pt}}
\pgfpathlineto{\pgfpoint{\mcSize+\mcThickness}{\mcSize+\mcThickness}}
\pgfusepath{stroke}
}}
\makeatother
\tikzset{every picture/.style={line width=0.75pt}} 

\makeatletter
\pgfutil@ifundefined{pgf@pattern@name@_4dhvc4nk9}{
\pgfdeclarepatternformonly[\mcThickness,\mcSize]{_4dhvc4nk9}
{\pgfqpoint{0pt}{-\mcThickness}}
{\pgfpoint{\mcSize}{\mcSize}}
{\pgfpoint{\mcSize}{\mcSize}}
{
\pgfsetcolor{\tikz@pattern@color}
\pgfsetlinewidth{\mcThickness}
\pgfpathmoveto{\pgfqpoint{0pt}{\mcSize}}
\pgfpathlineto{\pgfpoint{\mcSize+\mcThickness}{-\mcThickness}}
\pgfusepath{stroke}
}}
\makeatother
\tikzset{every picture/.style={line width=0.75pt}} 

\newcommand{\be}{\begin{equation}}
\newcommand{\ee}{\end{equation}}
\newcommand{\beq}{\begin{eqnarray}}
\newcommand{\eeq}{\end{eqnarray}}
\newcommand{\ba}{\[\begin{aligned}}
\newcommand{\ea}{\end{aligned}\]}

\newcommand{\sgn}{{\rm sgn\,}}

\renewcommand{\vec}[1]{\boldsymbol{#1}}

\renewcommand{\phi}{\varphi}
\renewcommand{\epsilon}{\varepsilon}

\def\nn{\nonumber}

\def \q {\vec{q}}

\def \r {\vec{r}}
\def \ve{{\varepsilon}}
\def \el {\ell}

\def \w{{\omega}}

\def \k{{\vec{k}}}

\def \tn{\textnormal}

\def \ba{\begin{align*}}
\def \ea{\end{align*}}

\def \el{\ell}

\newcounter{indice}

\def \ph{\phi}
\def \o {\omega}
\newcommand*{\DC}[1]{\textcolor{orange}{ DC: {#1}}}
\newcommand*{\new}[1]{\textcolor{black}{{#1}}}

\newcommand*{\red}{\textcolor{black}}
\usepackage[english]{babel}
\begin{document}
\sffamily
\title{\textsf{Collective density fluctuations of strange metals with critical Fermi surfaces}}

\author{\textsf{Xuepeng Wang}}
\author{\textsf{Debanjan Chowdhury}}
\affiliation{\textsf{Department of Physics, Cornell University, Ithaca NY 14853}}
\begin{abstract}
Recent spectroscopic measurements in a number of strongly correlated metals that exhibit non-Fermi liquid like properties have observed evidence of anomalous frequency and momentum-dependent charge-density fluctuations. Specifically, in the strange metallic regime of the cuprate superconductors, there is a featureless particle-hole continuum exhibiting unusual power-laws, and experiments suggest that the plasmon mode decays into this continuum in a manner that is distinct from the expectations of conventional Fermi liquid theory. Inspired by these new experimental developments, we address the nature of low-energy collective modes and the particle-hole continua for different ``solvable'' lattice models of non-Fermi liquids that host a critical Fermi surface --- a sharp electronic Fermi surface without any low-energy electronic quasiparticles. We scrutinize theoretically the possible existence of a long-lived zero-sound mode, which is renormalized to the plasma frequency in the presence of long-ranged coulomb interactions, and its decay into the continuum over a wide range of frequencies and momenta. Quite remarkably, some of the models analyzed here can account for certain aspects of the universal experimental observations, that clearly lie beyond the purview of standard Fermi liquid theory.
\end{abstract}

\maketitle

\clearpage
\tableofcontents
\noindent\rule{\textwidth}{1pt}
\section{\textsf{Introduction}}
\label{sec:intro}

For decades the Landau-Boltzmann paradigm has been the edifice on which the theory of weakly interacting metals has been built \cite{AGD}. In addition to the electronic quasiparticle-like excitations that remain long-lived near the Fermi surface, a remarkable property associated with Landau Fermi liquids are their collective modes. These are the smooth long-wavelength fluctuations of the entire Fermi surface, perhaps most notable of which is the ``zero-sound'' in neutral Fermi liquids (associated with a uniform breathing mode in the $\ell=0$ angular momentum channel), which is a long-lived ``acoustic'' excitation at small frequency and momentum  \cite{pines}. For frequencies $\omega>\omega_\star(\q)$, the kinematics associated with the low-energy excitations near the Fermi surface in metals leads to an unavoidable decay of this mode into the particle-hole continuum via the standard mechanism of Landau-damping. In conventional metals, where electrons interact via long-ranged Coulomb interactions, all of these features remain broadly similar and the only modification is related to the zero-sound mode being renormalized to the well known plasmon excitation.

It is natural to ask whether strongly interacting metals with a critical Fermi surface --- a sharp electronic Fermi surface without any low-energy Landau quasiparticles \cite{senthil2008critical} --- that defy the standard expectations of Fermi liquid (FL) theory (i) continue to host a zero-sound (ZS) mode, and if yes, (ii) what is the nature of its decay into the possibly unconventional continuum. Clearly, there can be no ``universal'' answer to these questions. For instance, if the critical Fermi surface supports a ZS mode at asymptotically small frequency and momenta for a specific non-Fermi liquid (NFL), whether it is overdamped or not will be determined by the spectrum of low-energy excitations and the kinematics of decay processes. Therefore, it is important to have a concrete microscopic model where the full spectrum of single-particle excitations as well as collective modes can be computed reliably at strong coupling. In general, computing the dynamical correlation functions for generic lattice Hamiltonians (e.g. Hubbard or $t-J$ models \cite{hubbard_review,LNW}) is a difficult task without making somewhat uncontrolled approximations. This paper fills this gap by addressing the questions raised above and analyzing the frequency and momentum resolved density correlations in three distinct families of non-Fermi liquids, that are all solvable at strong coupling. Our explicit calculations for the lattice models will illustrate the relevant mechanisms that control the interplay between collective modes and their decay into the continua. We will focus exclusively on translationally invariant models without any spatial randomness --- momentum will then be a good quantum number and the Fermi surfaces will be sharp. We will find an interesting interplay of non-trivial momentum and frequency dependence associated with the density fluctuations across all the models analyzed here; see Sec.~\ref{sec:summary}. We note recent complementary theoretical work on $(0+1)-$dimensional disordered models, where the frequency dependence of density correlations have been analyzed \cite{joshi}, as well as computations of plasmon attenuation into an unconventional continuum in certain holographic models \cite{zaanen}.

The central questions of interest to us here are especially timely in light of recent experimental developments in the field. With the advent of momentum-resolved electron energy-loss spectroscopy (M-EELS) \cite{MEELS}, the dynamical charge response of numerous strongly correlated materials has been analyzed carefully over a broad range of frequencies and momenta \cite{Abbamonte1,Abbamonte2}. 
For instance in the cuprate strange metal, there is evidence for a featureless particle-hole continuum that spans much of the Brillouin zone (BZ), while being independent of temperature and doping. The results are quite remarkable, in that this is not a small effect; the continuum seemingly exists up to the highest measurable energies and contains more than $99\%$ of the total spectral weight \cite{Abbamonte1,Abbamonte2}. As a result, there is no evidence for the existence of a sharply dispersing plasmon beyond an exceptionally small momentum scale, $q\lesssim0.01$ r.l.u., near the $\Gamma-$point \cite{Marel16,Hayden20}. These results are broadly consistent with earlier Raman studies \cite{bozovic87,ginsberg91} that reported the existence of such an unconventional continuum. Interestingly, many of these features associated with the density correlations are not restricted to just the cuprate strange metal \cite{Abbamonte3}, suggesting a possible universal character of the anomalous behavior. The microscopic origin for the unconventional decay of the plasmon into the particle-hole continuum remains unclear, especially given the paucity of theoretical methods that can reliably obtain the detailed frequency and momentum-dependent correlations in the strong-coupling regime. Importantly, there are regimes where all three realizations of the non-Fermi liquids we study in this paper (possibly, with additional perturbations) exhibit the celebrated $T-$linear resistivity in transport \cite{DCrmp}; however, their momentum and energy-resolved density correlations appear vastly different. This clearly shows the importance of distinguishing between non-Fermi liquids with seemingly similar transport behavior on the basis of their detailed correlation functions.

The remainder of this paper is organized as follows: In Sec.~\ref{sec:summary}, we briefly summarize our theoretical results that highlight the important differences that can arise in the momentum and frequency-resolved density response of quantum critical strange metals. In Sec.~\ref{sec:prelimA}, we introduce three models for non-Fermi liquids that rely on generalizations of the purely electronic Sachdev-Ye-Kitaev (SYK) models \cite{SYK1,SYK2,SYK3,SYK4,SYK5,SYK6}, and the ``random'' Yukawa models of coupled electrons and critical bosons. Both classes of models are solvable in the limit of a large number of electron (and critical boson) flavors. Importantly, by construction we analyze models that have a notion of spatial locality and are translationally invariant; the ``randomness'' only arises in ``flavor-space''. In Sec.~\ref{sec:prelimB}, we review the two main technical tools used to obtain results for the density response, namely the Bethe-Salpeter equation for the charge-density vertex and the quantum Boltzmann equation, respectively. In Sec. ~\ref{sec:result}, we provide a more detailed exposition to the results summarized in Sec. ~\ref{sec:summary}. We end with a brief outlook in Sec.~\ref{sec:outlook}, where we also compare our theoretical results against some of the salient features observed in recent experiments upon including the effects of a Coulomb repulsion. The appendices \ref{app:SD} - \ref{app:equivalence} contain several technical details.

\section{\textsf{Summary of key results}}\label{sec:summary}

\begin{figure}[h!]
\includegraphics[width=150mm,scale=1.2]{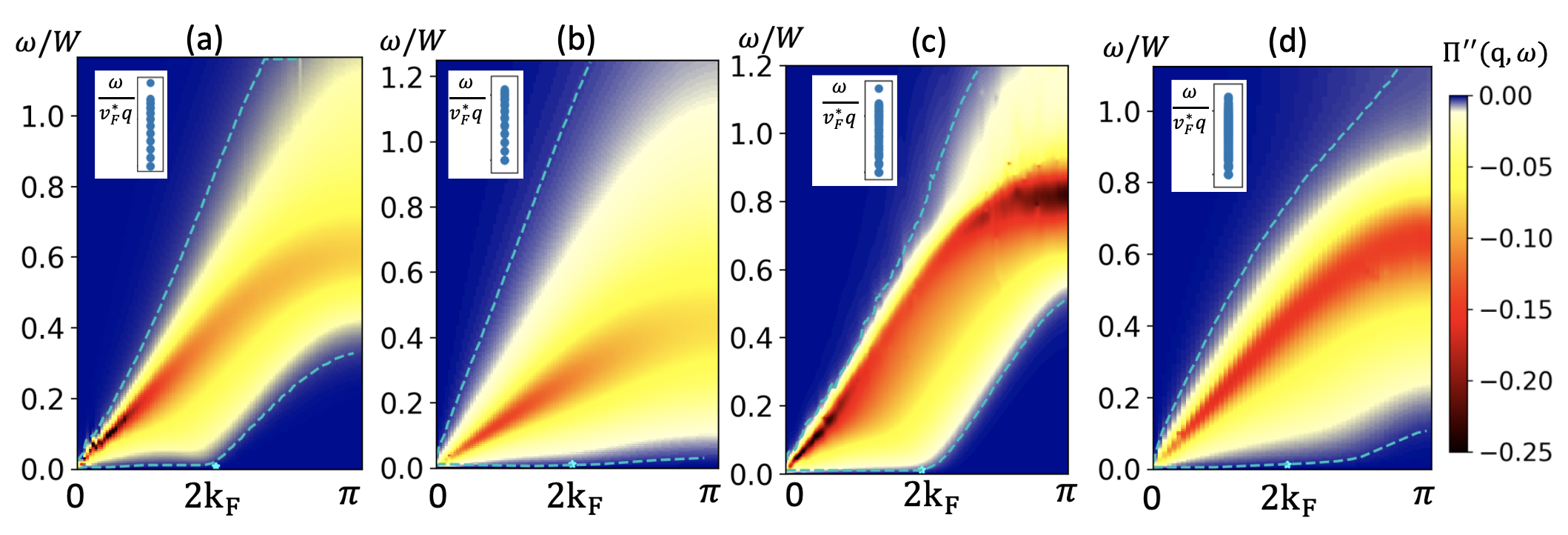}
\caption{\textsf{The frequency ($\omega$) and momentum ($\q$) dependent density-response function ($\Pi''(\q,\omega)$) for (a) model-A with $\new{t=0.75}~(\tn{i.e.}~W_f=6$ and $W_f^{*}=4), ~U=8, ~\mu_f=\new{-3t_f}$, (b) model-B with $\new{t_c=0.5}~(\tn{i.e.}~W=4),~J=8, ~U=8, ~\mu=\new{-3t_c}$, (c) model-C for ``weak'' interaction with $t_c=\new{0.5}, \lambda=2, \mu=\new{-3t_c}$, $\new{K=4}$, and (d) model-C for ``strong'' interaction with $t_c=\new{0.5}, \lambda=8, \mu=\new{-3t_c}$, $\new{K=4}$. The orange region represents the particle-hole continuum; the dashed lines denote the edge of this continuum where the spectral weight reaches $1\%$ of their maximum value. The hump in the dashed line and the black region at small $\omega,~q$ represent the region where zero-sound mode starts decaying into the continuum. Insets for all figures denote the spectrum of linearized eigen-energies plotted in units of $\frac{\omega}{v_F^{*} q}$, obtained from a solution of the quantum Boltzmann equation at small $\omega,~q$. The continuous part of the spectrum denotes the particle-hole continuum, and the well-separated eigenvalue (if it exists) is the zero-sound mode. In (a) and (c), there is an eigenvalue for an undamped zero-sound mode that is well separated from the particle-hole excitations. In (b) and (d), there is no such undamped ZS mode. Some of the irregular features in panel (c) arise from uncertainties associated with the analytical continuation via Pad\'e approximation.} 
}
\label{summary}
\end{figure}

We briefly summarize our key results here:
\begin{enumerate}
    \item In a solvable model of a strongly interacting ``heavy'' Fermi liquid, with renormalized quasiparticles and a reduced bandwidth, we find a long-lived ZS mode in the small $\omega,~\q-$limit that decays beyond $\omega_\star(\q)$ into a conventional particle-hole continuum via Landau damping (Fig.~\ref{summary}(a)). In this model, at low-energies the system remains a Fermi liquid \cite{Parcollet1,Balents} with a sharp Fermi surface at arbitrarily large $U/t$ \cite{DCsyk}, where $U$ and $t$ denote the typical strength of interactions and electron-hopping, respectively. We find that the ratio of zero-sound to the (renormalized) Fermi-velocity scales as $v_{S}/v_F^{*}\sim U/t$, implying that the collective mode disperses outside the continuum and remains undamped at small $\omega, \q$. At large frequency, as the heavy Fermi liquid crosses over into a non-Fermi liquid with a severely broadened Fermi surface and incoherent local excitations, the particle-hole continuum develops clear signatures that are at odds with the expectations of a Fermi liquid. Interestingly, even though the incoherent regime at high energies exhibits local criticality, the density correlations continue to retain some weak momentum dependence. The details appear in Sec.~\ref{sec:resultA}.
    
    \item In a solvable model of a marginal Fermi liquid (MFL) \cite{Varma89} with a momentum-independent (``local'') electron self-energy (i.e. $\Sigma(i\omega)\sim i\omega\ln(\omega)$) and a sharp Fermi surface \cite{DCsyk}, we find that there is {\it no} long-lived ZS mode even in the small $\omega,~\q-$limit. The collective mode is always burried deep inside the continuum, such that the ZS is overdamped at arbitrarily small $\omega,~\q$ (Fig.~\ref{summary}(b)). The particle-hole continuum associated with this MFL exhibits marked differences with a conventional Fermi liquid, especially as a function of $\q$ at small $\omega$. Contrasting Figs.~\ref{summary}(a) and (b), we note that there is a significant transfer of spectral weight down to the smallest $\omega$ for $|\q|>2k_F$, which is tied to the presence of local criticality. Finally, the density response also exhibits an unusual universal power law at large $(\omega/v_Fq)$ --- reminiscent of experiments \cite{Abbamonte1} --- that we discuss further in Sec.~\ref{sec:resultB}.

    \item In a solvable model of a non-Fermi liquid, where the singular nature of the self-energy (i.e. $\Sigma(i\omega)\sim (i\omega)^\alpha,~~\alpha<1$) is tied to the near vicinity of the sharp Fermi surface, we find that the existence of an undamped zero-sound mode depends on the underlying details of the low-energy physics. Specifically, depending on the effective retarded interaction between electrons mediated by the quantum critical boson (that drives the non-Fermi liquids physics in the first place), we can either obtain a long-lived zero-sound mode at low energy that decays beyond a $\omega_\star(\q)$, or that is overdamped altogether even at the smallest $\omega,~\q$ (Fig.~\ref{summary}(c) and (d)). Once again, there are clear differences between the particle-hole continuum associated with the locally critical marginal Fermi liquid in Fig.~\ref{summary}(b) and the non-Fermi liquid with $\alpha<1$  in Fig.~\ref{summary}(c) for small $\omega$ and $|\q|>2k_F$. The details appear in Sec.~\ref{sec:resultC}.
    
\end{enumerate}

In Sec.~\ref{sec:outlook}, we will briefly revisit these results and the plots in Fig.~\ref{summary} after including the effects of a long-range Coulomb repulsion, especially with an eye for contrasting the results with those observed in the experiments \cite{Abbamonte1,Abbamonte2}. 

\section{\textsf{Preliminaries}}

In this section, we will first review the single-particle properties of the three different models of interest to us \cite{DCrmp}. We will then summarize the basic computational setup for obtaining the full momentum and frequency dependence of the two-particle density response, as well as the quantum Boltzmann equation that directly obtains the linearized spectrum of collective modes and particle-hole continua at low energies. 

\subsection{\textsf{Models}}\label{sec:prelimA}
\subsubsection{\textsf{Model-A: Locally critical one-band electronic model}}
\label{prelim:A}

Consider a model consisting of just a single electron species with $i=1,..,N$ orbitals, with a Hamiltonian
\beq
H_f = \sum_{\r,\r'} \sum_{\el}(-t_{\r\r'}  - \mu_f \delta_{\r\r'}) f_{\r \el}^\dagger f^{\phantom\dagger}_{\r'\el}  + \frac{1}{(2N)^{3/2}}\sum_{\r}\sum_{ijk\el} U_{ijk\el} f^\dagger_{\r i}f^\dagger_{\r j} f_{\r k} f_{\r \el}, 
\label{hf}
\eeq
where the hopping matrix element, $t_{\r\r'}$, connects sites $\r,~\r'$ and is orbital diagonal. The interaction strength, $U_{ijk\el}$, is on-site and assumed to be independent of the site-label, $\r$, making it perfectly translationally invariant \cite{DCsyk}, even though they are drawn from a ``disorder'' distribution with $\overline{U_{ijk\el}}=0$ and $\overline{(U_{ijk\el})^2} \equiv U^2$; antisymmetrization demands that $U_{ijk\el} = -U_{jik\el} = -U_{ij\el k}$ and $U_{ijkl} = U_{klij}$. The chemical potential, $\mu_f$, is used to control the conserved density, $Q_f = \sum_{\r,\el}\langle f^\dagger_{\r\el}f^{\phantom\dagger}_{\r\el}\rangle/NV$. The above model builds on previous constructions without spatial locality \cite{Parcollet1} and partial translational invariance \cite{Balents,Zhang17,shenoy,Jose}.

The model can be solved at arbitrary strength of interactions in the limit $N\rightarrow\infty$ \cite{DCrmp}. At frequencies and temperatures smaller than $W^*\sim W^2/U$, the ground state is a renormalized (heavy) Fermi liquid for any ratio of $U/t\gg 1$. For $\omega,T\gg W^*$, there is a cross over to an incoherent local non-Fermi liquid without any singular momentum dependence. The single electron (Matsubara) Green's function, $G_f(\k,i\omega)$, can be obtained from the usual ``melonic'' series at strong coupling,
\begin{equation}
G_f(\k,i\omega) \sim 
\begin{cases}
\frac{Z}{i\omega - Z \tilde\varepsilon_\k + i \alpha \nu_0^2 U|\omega|^2 \ln(\frac{U}{|\omega|})\mathrm{sgn}(\omega)}, ~~\omega \ll W^*,\\
\frac{i\mathrm{sgn}(\omega)}{\sqrt{U |\omega|}} -  B(\omega) \frac{\varepsilon_{\k}}{U |\omega|}, ~~~~~~~~W^* \ll \omega \ll U,
\end{cases}
\label{limits}
\end{equation}
where $Z \sim 1/(\nu_0 U)$ is the quasiparticle residue, $\nu_0$ is the non-interacting density of states (DoS), and $\alpha$ is a number of order unity. $B(\omega)$ is a constant independent of frequency for both $\omega > 0$ and $\omega < 0$ though its precise value is different for the two signs of $\omega$.

\subsubsection{\textsf{Model-B: Locally critical marginal Fermi liquid in two-band electronic model}}
\label{prelim:B}
In order to obtain a non-Fermi liquid with singular momentum dependent features (i.e. a critical Fermi surface), one way forward has been to leverage the locally critical fluctuations associated with the $f-$electrons in the one-band model and use it as a scattering bath for a different band of electrons \cite{SSmagneto,DCsyk}. Consider then an additional band of $c-$fermions with $\el=1,..,N$ orbitals and an associated conserved $U(1)$ charge density, $Q_c$, tuned by a chemical potential $\mu_c$. The modified Hamiltonian is 
\beq
H &=& H_c + H_f + H_{cf},\\
H_{cf} &=& \frac{1}{N^{3/2}}\sum_{\r}\sum_{ijk\el} J_{ijk\el} c^\dagger_{\r i}f^\dagger_{\r j} c_{\r k} f_{\r \el},
\label{Htot}
\eeq
where $H_f$ is identical to Eq.~(\ref{hf}), and $H_c$ can be taken to be a translationally invariant free electron model in the absence of any coupling to the $f-$electrons. The coefficients, $J_{ijkl}$, are also chosen to be identical at every site with $\overline{J_{ijkl}} =  0$ and  $\overline{(J_{ijk\el})^2}=J^2$. It is convenient to set the $f-$electron bandwidth, $t_{\r\r'}=0$, and study the model as a function of $J/U$. \red{However, our analysis is also valid for a generic $t_f$ for energy scales above $t_f^2/U$, when $f-$electrons are in their local incoherent metallic regime \cite{DCsyk}.}

In the $N\rightarrow\infty$ limit, a set of melonic diagrams that now include the mutual feedback of the $c$ and $f-$electrons on each other leads to a wide range of energy scales where the $c-$electrons develop marginal Fermi liquid-like correlations \cite{Varma89,DCsyk}. Importantly, the $c-$electrons have a sharply defined Fermi surface, $G^{-1}_c(\k,\omega=0) = 0$ for $|\k|=k_F$, that satisfies Luttinger's theorem. In explicit form, the leading singular piece associated with the Green's function is given by,
\beq
G_c(\k,i\omega) &=& \frac{1}{-\Sigma_{cf}(i\omega) - \ve_\k},~\tn{where}\\
\Sigma_{cf}(i\omega) &\sim& -\frac{\nu_0J^2}{ U}i\omega\log\frac{U}{|\omega|}.
\eeq
Here $\nu_0$ is the c-electron DoS and the effective dimensionless coupling strength is $\tilde{J}=\nu_0 J^2/U$. Note that $\Sigma_{cf}$ is momentum-independent, i.e. the singular $\omega-$dependence is not tied just to the vicinity of the Fermi surface, but is rather present everywhere in the BZ. This is due to the local criticality in the $f-$sector, which serves as a momentum sink for the $c-$electrons. \red{The same local criticality plays a crucial role in giving rise to an unconventional particle-hole continuum in the density-response, leading to the absence of long-lived ZS mode, as we will show in Section \ref{sec:resultB}.}

\subsubsection{Model-C: Non Fermi liquid from quantum critical boson }
\label{prelim:C}
A different route to a non-Fermi liquid with a critical Fermi surface relies on coupling an electronic Fermi surface to a quantum critical boson. Consider the Hamiltonian \cite{largeN,DCrmp},
\beq
H &=& \sum_{i=1}^N\sum_{\r,\r'}(-t_{\r\r'} - \mu~\delta_{\r\r'})c^\dagger_{\r i} c^{\phantom\dagger}_{\r' i},\nn\\
&+& \frac{1}{2}\sum_{\alpha=1}^M \sum_\k [\pi^2_{\k\alpha} + \o_\k^2\ph_{\k\alpha}^2], \nn\\
&+& \sqrt{\frac{2}{MN}}\sum_{\alpha=1}^M  \sum_{\{i_1,i_2\}=1}^N \lambda_{i_1i_2;\alpha}~ \ph_{\r\alpha} c^\dagger_{\r i_1}c_{\r i_2}, \label{hi}
\eeq 
with the couplings $\lambda$ drawn from a random distribution with $\overline{\lambda_{ijk\el}}=0$ and $\overline{(\lambda_{ijk\el})^2} \equiv \lambda^2$, but while preserving exact translational invariance. See Refs. \cite{Fu:2016vas,Murugan:2017eto,Patel:2018zpy,Marcus:2018tsr,Wang:2019bpd,JS,Wang:2020dtj,Kim:2020jpz,Adalpe20,WangMeng21} for earlier studies of related random Yukawa-like models. The boson dispersion on a square lattice is taken to be $\omega^2_k = \omega_0^2 + K(2-\cos(k_xa)-\cos(k_ya))$, where $\omega_0^2$ is the bare mass, $K$ is related to the bare stiffness and $a$ is the lattice spacing (set to $a=1$ from here onwards). For the renormalized boson mass $\widetilde{\omega}_0^2\neq0$, the $c-$electrons form a heavy Fermi liquid with $Z^{-1}\sim\nu_0\lambda^2/\widetilde{\omega}_0^2$. In the $N,~M\rightarrow \infty$ limit (with $N/M$ fixed and set to $1$), the non-Fermi liquid regime can be accessed by tuning  $\widetilde{\omega_0}^2$ to be zero, where
\beq
\begin{aligned}
&
G_c(\vec{k},i\omega)=\frac{1}{i\sgn(\omega)B|\omega|^{2/3}-\tilde{\epsilon}_{\vec{k}}},\\
&D(\vec{q},i\Omega)=\frac{1}{\Omega^2+\widetilde{\omega}_q^2+i\eta\frac{|\Omega|}{|\vec{q}|}}.
\end{aligned}
\label{yukawa}
\eeq
This is the classic RPA result \cite{PALee89,AIM,Polchinski:1993ii,metlitski1,sungsik1,mross}, but now obtained in a controlled large $N,~M$ setting, where the boson has $z_b=3$ dynamics while the fermions have $z_c=3/2$. The coefficient $\eta=\frac{\lambda^2}{v_F^2}, B=\frac{\sqrt{3}}{6\pi}\frac{\lambda^{4/3}}{K^{2/3} v_F^{1/3}}$. Once again, $c-$electrons have a sharply defined Fermi surface satisfying Luttinger's theorem. However, two key differences from model-B are that the singular frequency dependence in model-C is tied to the near vicinity of the Fermi surface, and that the Fermi surface volume counts the density of all electrons. We end by noting that in the presence of disorder, the nature of the non-Fermi liquid (i.e. both $z_b,~z_c$) changes \cite{SS22}. However, our intention in this paper is to focus exclusively on models where translational symmetry is preserved at every stage and the Fermi surface remains sharply defined without any disorder-induced ``smearing''. 

\subsection{\textsf{Methods}} \label{sec:prelimB}

In a Fermi liquid with a sharply defined Fermi surface, the ZS mode in the long-wavelength and low-energy limit is well described by the collisionless Boltzmann equation for the quasiparticle distribution function \cite{pines}. However, for the non-Fermi liquids of interest to us here, there are no (or only marginally) well-defined quasiparticles near the Fermi surface as a result of the singular frequency dependence of electronic self energy and the effective density-density interaction. Therefore, instead of the usual Boltzmann equation written on the basis of existence of quasiparticles, we will build on the classic work of Prange and Kadanoff \cite{PK} which expresses a closed set of equations of motion for a generalized Fermi surface displacement --- the quantum Boltzmann equation (QBE). This formalism has been utilized previously to analyze the low-energy collective excitations for the composite Fermi liquid \cite{qbe} and the U(1) quantum spin liquid with a spinon Fermi surface \cite{nave}. Building on these works, we will use the non-equilibrium Green's function technique \cite{Keldysh} to obtain the QBE for all three models of interest to us. The key components for this calculation will include the singular self-energy correction,  a generalized Landau-type interaction-function, and the collision integral. Looking ahead, whether the ZS mode will survive as an undamped excitation will depend on the competition between the ``classical'' momentum transfer process and the singular frequency-dependent piece in the density interaction.

Solving for the eigenvalues of the QBE correspond to a set of linearly-dispersing modes in the small $\omega,~\q-$limit. When there is a clearly separated eigenvalue from the tower of states corresponding to the continuum, it corresponds to a well-defined (umdamped) ZS mode that exists outside the continuum. On the other hand, if there is no such clear separation, it already indicates the absence of an undamped ZS outside the continuum. In order to complement our understanding for the detailed density fluctuations, not just in the low-energy limit but instead for the entire range of frequency and momentum, we will analyze the Bethe-Salpeter equations associated with the density response in the large$-N$ limit directly. 

\subsubsection{\textsf{Bethe-Salpeter equation for density vertex}}

We begin by summarizing the standard formalism that is used to evaluate the two-particle density response functions \cite{Andrey3} at leading order in the large$-N~(M)$ approximation for models A-C; see e.g. \cite{Andrey1,Andrey2} for a similar strategy applied to other two-particle response functions. The Bethe-Salpeter equation for the intra-orbital density vertex takes the form:
\begin{equation}
\label{BSE}
\begin{tikzpicture}[x=0.75pt,y=0.75pt,yscale=-1,xscale=1]

\draw    (229.15,153.42) .. controls (230.67,152.4) and (229,153.06) .. (243,139.06) ;
\draw    (229.15,153.42) .. controls (233.67,159.4) and (227.67,151.4) .. (242.67,168.4) ;
\draw  [fill={rgb, 255:red, 0; green, 0; blue, 0 }  ,fill opacity=1 ] (233.72,147.72) -- (235.69,139.88) -- (241.41,145.23) -- cycle ;
\draw  [fill={rgb, 255:red, 0; green, 0; blue, 0 }  ,fill opacity=1 ] (238.64,163.77) -- (232.11,161.76) -- (238.04,156.96) -- cycle ;
\draw    (219.67,153.73) .. controls (221.28,152.01) and (222.94,151.96) .. (224.66,153.57) -- (229.15,153.42) -- (229.15,153.42) ;
\draw  [pattern=_4dhvc4nk9,pattern size=6pt,pattern thickness=0.75pt,pattern radius=0pt, pattern color={rgb, 255:red, 0; green, 0; blue, 0}] (132.17,151.2) -- (148.93,136.88) -- (149.5,164.96) -- cycle ;
\draw    (149.5,164.96) -- (179,187.73) ;
\draw    (148.93,136.88) -- (179.89,113.06) ;
\draw  [fill={rgb, 255:red, 0; green, 0; blue, 0 }  ,fill opacity=1 ] (164.41,124.97) -- (167.17,118.99) -- (170.58,123.88) -- cycle ;
\draw  [fill={rgb, 255:red, 0; green, 0; blue, 0 }  ,fill opacity=1 ] (166.62,177.92) -- (160.08,177.14) -- (163.68,172.39) -- cycle ;
\draw  [pattern=_4dhvc4nk9,pattern size=6pt,pattern thickness=0.75pt,pattern radius=0pt, pattern color={rgb, 255:red, 0; green, 0; blue, 0}] (360,144.06) -- (376,144.06) -- (376,161.06) -- (360,161.06) -- cycle ;
\draw    (320,128.06) .. controls (331.59,119.16) and (347,126.06) .. (360,144.06) ;
\draw    (323.25,176.34) .. controls (336.25,182.68) and (347,174.06) .. (360,162.06) ;
\draw    (376,144.06) -- (398.49,130.96) ;
\draw    (376,161.06) -- (398.49,172.96) ;
\draw  [fill={rgb, 255:red, 0; green, 0; blue, 0 }  ,fill opacity=1 ] (384.77,138.89) -- (387.86,132.8) -- (391.58,139.45) -- cycle ;
\draw  [fill={rgb, 255:red, 0; green, 0; blue, 0 }  ,fill opacity=1 ] (389.68,168.46) -- (382.86,168.83) -- (386.76,162.28) -- cycle ;
\draw  [pattern=_4dhvc4nk9,pattern size=6pt,pattern thickness=0.75pt,pattern radius=0pt, pattern color={rgb, 255:red, 0; green, 0; blue, 0}] (291.17,151.2) -- (307.93,136.88) -- (308.5,164.96) -- cycle ;
\draw    (308.5,164.96) -- (323.25,176.34) ;
\draw    (307.93,136.88) -- (320,128.06) ;
\draw  [fill={rgb, 255:red, 0; green, 0; blue, 0 }  ,fill opacity=1 ] (317.77,129.83) -- (320.53,123.85) -- (323.94,128.74) -- cycle ;
\draw  [fill={rgb, 255:red, 0; green, 0; blue, 0 }  ,fill opacity=1 ] (323.25,176.34) -- (316.7,175.56) -- (320.3,170.81) -- cycle ;

\draw (255,143.4) node [anchor=north west][inner sep=0.75pt]    {$+$};
\draw (181,142.17) node [anchor=north west][inner sep=0.75pt]    {$=$};

\end{tikzpicture}
\end{equation}
where the shaded triangle and shaded rectangle represent the fully dressed particle-hole vertex and four-point (4pt) interaction, respectively. All the solid lines represent the fully dressed Green's function; the vertex functions are then clearly chosen to satisfy Ward identities \cite{Andrey3} associated with the underlying charge-conservation. In the $N(,~M)\rightarrow\infty$ limit, the above equation only involves the standard ``ladder'' series. The effective 4pt interaction for models-A, B and C involve the $f-$fermion polarizability, the $c,~f$-fermion polarizability and the fully dressed critical boson propagator, respectively. The detailed diagrammatic expansions appear in Appendix~\ref{app:SD} and we discuss the Ward-identity in Appendix~\ref{app:ward}.

In order to obtain the detailed dependence of the density response over the entire BZ and for energies comparable to the full bandwidth and beyond, we solve Eq.~\ref{BSE} numerically in a fully self-consistent fashion. We define the problem on a square lattice with a generic non-interacting Fermi sea and solve the problem in a discretized momentum space with mesh size of $51\times51$. The Matsubara frequency sums are similarly carried out with a discrete grid-size of $N_\omega=2^9$. We ensure numerical convergence of both the single and two-particle correlation functions up to $10^{-4}$, and check that the Ward identity is satisfied at all stages of our computation; see Appendix \ref{app:ward}. We also show results for the density response functions obtained using the Bethe-Salpeter equations in the (theoretically inconsistent) limit where vertex corrections are ignored in Appendix \ref{app:ward}. The results are dramatically different, which clearly demonstrates the importance of the momentum-dependent vertex corrections and the need to go beyond purely local models to address the experimental puzzles. Note that since the computations are carried out in imaginary frequency, we use the standard Pad\'e approximation to analytically continue the response functions to real frequencies numerically \cite{pade1,pade2}. 

\subsubsection{\textsf{Quantum Boltzmann equation for critical Fermi surface displacement}}\label{qbesec}
 
 In this section we review the key steps involved in formulating the QBE at zero temperature, as the equation of motion for the Green's function $G^{<}$. This is re-expressed in terms of the generalized Fermi surface displacement: $f(\vec{\el},\omega,\theta,\Omega)\equiv\int d|\vec{L}| ~G^{<} (\vec{\el},\omega,\vec{L},\Omega)$, where $\theta$ is the angle between $\vec{L}$ and $\vec{\el}$. We use capital letters for center-of-mass frequency and momentum, and small letters for the relative frequency and momentum. After a standard set of manipulations \cite{PK,qbe}, the QBE takes the simplified form,
\begin{equation}
\label{QBE}
\begin{aligned}
    &[\omega-v_F \el \cos \theta]f(\vec{\el},\omega,\theta,\Omega)\\
     &=\nu_0 v_F^2 \int_{\theta^{'},\Omega^{'}} \tn{Re}~ V(k_F|\theta^{'}-\theta|,\Omega^{'}-\Omega)\bigg[f_0(\Omega^{'}+\frac{\w}{2})-f_0(\Omega^{'}-\frac{\w}{2})\bigg]f(\vec{\el},\omega,\theta,\Omega)\\
     &-\nu_0 v_F^2 \int_{\theta^{'},\Omega^{'}}\tn{Re}~ V(k_F|\theta^{'}-\theta|,\Omega^{'}-\Omega)\bigg[f_0(\Omega+\frac{\w}{2})-f_0(\Omega-\frac{\w}{2})\bigg]f(\vec{\el},\omega,\theta',\Omega').
\end{aligned}
\end{equation}
Here $V(k_F|\theta^{'}-\theta|,\Omega^{'}-\Omega)$ denotes the 4pt-interaction on the Fermi surface. The first term on the RHS of Eq.(\ref{QBE}) reduces to the self energy after integration, whereas the second term acts effectively as the analog of the Landau interaction in the classical Boltzmann equation. We now assume a rotationally invariant system and decompose the displacements into different angular momentum channels, $j$:
\begin{equation}
\label{angularQBE}
\begin{aligned}
    &[\omega-(\Sigma(\Omega+\omega/2)-\Sigma(\Omega-\omega/2))]f_j(\el,\omega,\Omega)-\frac{v_F \el}{2}[f_{j-1}(\el,\omega,\Omega)+f_{j+1}(\el,\omega,\Omega)]\\
    &=-\nu_0 v_F^2 \int_{\Omega'} V_j(\Omega^{'}-\Omega) f_j(\el,\omega,\Omega').
\end{aligned}
\end{equation}
Let us now focus on the $j=0$ channel, which will correspond to the ZS collective mode. Numerically, we treat Eq.(\ref{angularQBE}) as a hopping model on 1-D lattice, where the ``sites'' correspond to the different angular momentum channels \cite{qbe}, and compute its eigenvalue spectrum. We make the ansatz: $f_j=\lambda^{j-1}f_1$ for $j>1$ \cite{PhysRevB.97.115449,mandal}, where $\lambda_j$ can be understood as the probability amplitudes of the wavefunction on different sites of this 1-D lattice. For the ZS mode, we assume the $j=0$ channel contributes the largest probability amplitude. Due to normalizability of the eigenmode, we search for a solution with $|\lambda|<1$. The typical analytical treatment for the problem above proceeds by finding a characteristic $\theta_c$, and dividing the interaction in Eq.(\ref{QBE}) into two parts: when $\theta<\theta_c$ the frequency dependent piece dominates, and when $\theta>\theta_c$ the momentum transfer process dominates \cite{qbe}. However, as we will discuss below, for the examples with local criticality, this treatment needs a more careful consideration. See Appendix \ref{app:num} for additional numerical details.
 
\section{\textsf{Results}}\label{sec:result}
\subsection{\textsf{Model-A: Locally critical one-band electronic model}}\label{sec:resultA}

\begin{figure}[h!]
\includegraphics[width=150mm,scale=1.2]{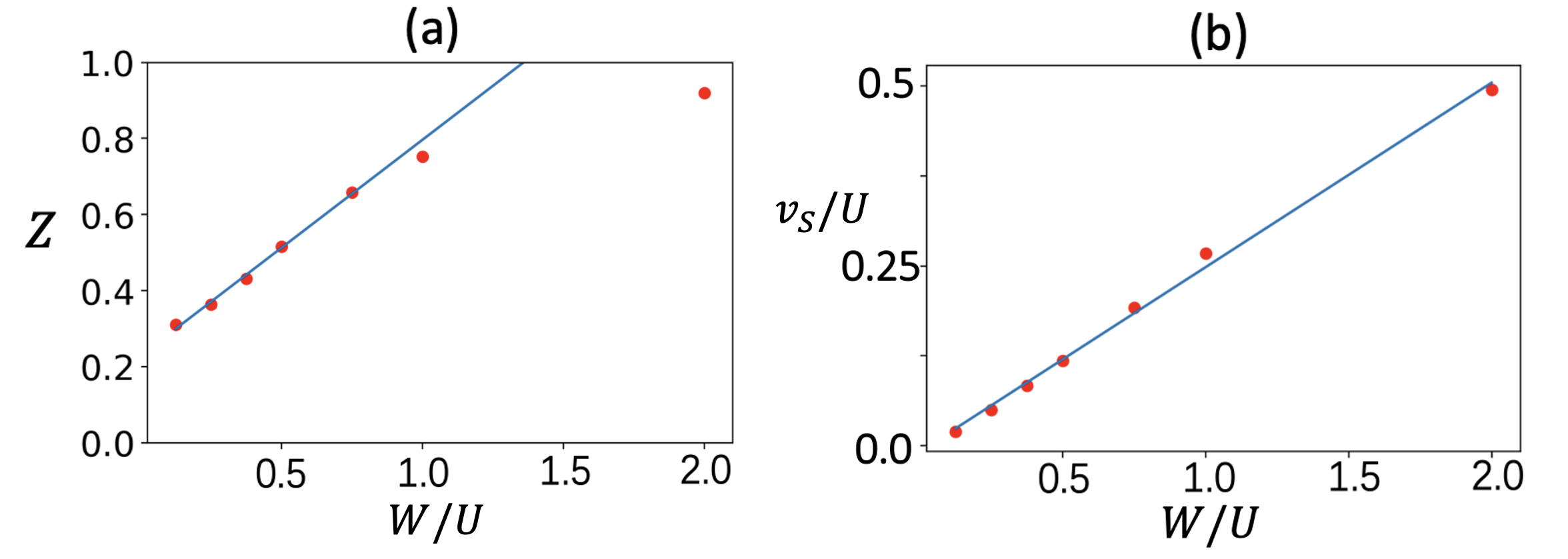}
\caption{\textsf{(a) The quasiparticle residue $Z$, extracted from the low-energy electron self-energy as a function of $t_f$ \red{by taking numerical derivative along the Matsubara axis}.  (b) The ZS velocity, $v_S$, obtained by fitting the peak of $\Pi''(\q,\omega)$ to a dispersing quasiparticle \red{(without invoking numerical derivative along the Matsubara axis) suggests the linear scaling as discussed in the main text}. Both are evaluated at a fixed $U=8,~ T=0.05,~ \mu=\new{-3}t_f$ \red{when varying the bandwidth $W=8t_f$ }. The solid line is a guide to the eye. }}
\label{Zvs}
\end{figure}

\begin{figure}[h!]
\includegraphics[width=120mm,scale=1.2]{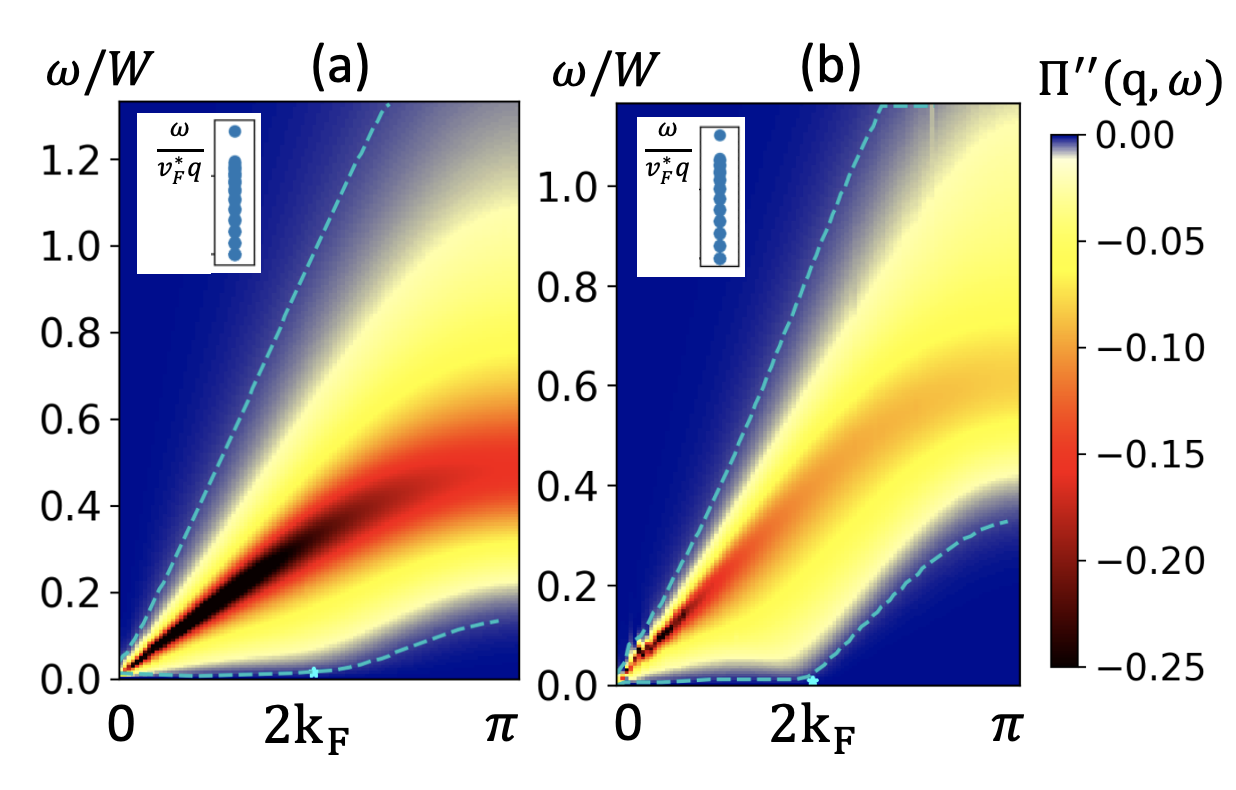}
\caption{\textsf{The density response, $\Pi''(\q,\omega)$, for model-A evaluated at $U=8,~ T=0.05,~ \mu=-3t_f$ and (a) \new{$t_f=0.375$}, (b) \new{$t_f=0.75$}. The dashed lines denote the edge of the particle-hole continuum where the spectral weight reaches $1\%$ of their maximum value. The inset shows the QBE eigenenergies in units of $\omega/v_F^*q$. At the lowest energy and momentum, there is always a well-separated eigenvalue associated with an undamped zero-sound mode in the heavy Fermi liquid at arbitrary ratio of $U/t_f\gg1$.}}
\label{one} 
\end{figure}

As outlined in Sec.~\ref{prelim:A}, the low-energy solution for model-A is a renormalized heavy Fermi liquid for all $U/t\gg1$ \cite{Parcollet1,Balents,DCsyk}; the quasiparticle residue $Z$ is extracted as a function of $t_f$ at fixed $U$ in Fig.~\ref{Zvs}(a). The result agrees with the asymptotic analytical result, $Z\sim t_f/U$, at large $U/t_f$; the saturation towards $Z\rightarrow1$ is also expected as $t_f\sim U$. We anticipate a long-lived ZS mode in the Fermi liquid regime, that is well described by the usual classical Boltzmann equation, as long as the Landau-parameter $F^s_0$ is large \cite{pines}. It has already been pointed out that $F^s_0 \sim (U/t_f)^2$ in this regime \cite{Balents,DCsyk}, which suggests  that the ZS speed, $v_{S}\sim Zv_F\sqrt{F_0^s} \sim t_f$. Indeed by fitting the peak of $\Pi''$ in Fig.~\ref{one}(a), (b), that we describe below, we can obtain the dispersing ZS mode in the small $q$ and $\omega$ limit. This allows us to extract $v_S$ for the undamped ZS, which agrees with the analytical expectation; see Fig.~\ref{Zvs}(b).  \red{Note that as $t_f\rightarrow0$ (i.e. when the physics is governed by an isolated SYK island), $Z$ should also go to zero. The extapolated residue $Z_{t_f\rightarrow0}\neq0$ in Fig.~\ref{Zvs}(a) is due to a numerical uncertainty associated with evaluating frequency derivatives of the self-energy along the Matsubara axis at a finite temperature.}

The full density response, $\Pi''(\q,\omega)$, is shown for two different values of the bare bandwidth (keeping $U$ fixed) in Fig.~\ref{one}(a), (b). The relatively sharp onset of the continuum, $\omega_\star(q)$, is clearly visible, and so is the feature near $2k_F$. It is worth noting that for the case with smaller bare bandwidth in panel (a), there is a more pronounced transfer of spectral weight down to lower energies near the $2k_F$ vectors. To illustrate that the ZS mode is indeed outside the continuum, we also diagonalized Eq.~(\ref{angularQBE}) numerically to obtain the eigenvalue spectrum, and find a well-separated eigenvalue corresponding to the undamped ZS; see insets of Fig.~\ref{one} (a) and (b).

Let us now turn out attention to the regime $\omega \gg W^{*}$, where the system enters the incipient incoherent regime. At large $\omega,~\q$, we notice a finite spectral weight even in regions of the phase-space where nominally there should not be any response (e.g. for momenta near the BZ boundary, beyond any $2k_F$ region, and for frequencies larger than the renormalized bandwdith). At the same time, instead of a completely local and featureless (as a function of momentum) response associated with the expected SYK-like regime, we find a non-trivial momentum dependence in $\Pi''(\omega,\q)$ for $\omega>W^{*}$. This can be traced back to the momentum dependence that is generated from the momentum-dependent vertex correction, and a ``mixing'' between high and low-energy contributions in the ladder-sum that enters the Bethe-Salpeter equation. Such mixing will contribute a subleading term with $\frac{\omega}{v_F^* q}$ scaling to the fully dressed density response (see Appendix \ref{app:modA}), which leads to the $q$-dependent boundary for $\omega>W^*$.

\subsection{\textsf{Model-B: Locally critical marginal Fermi liquid in two-band electronic model}}\label{sec:resultB}

\begin{figure}[h!]
\includegraphics[width=175mm,scale=1]{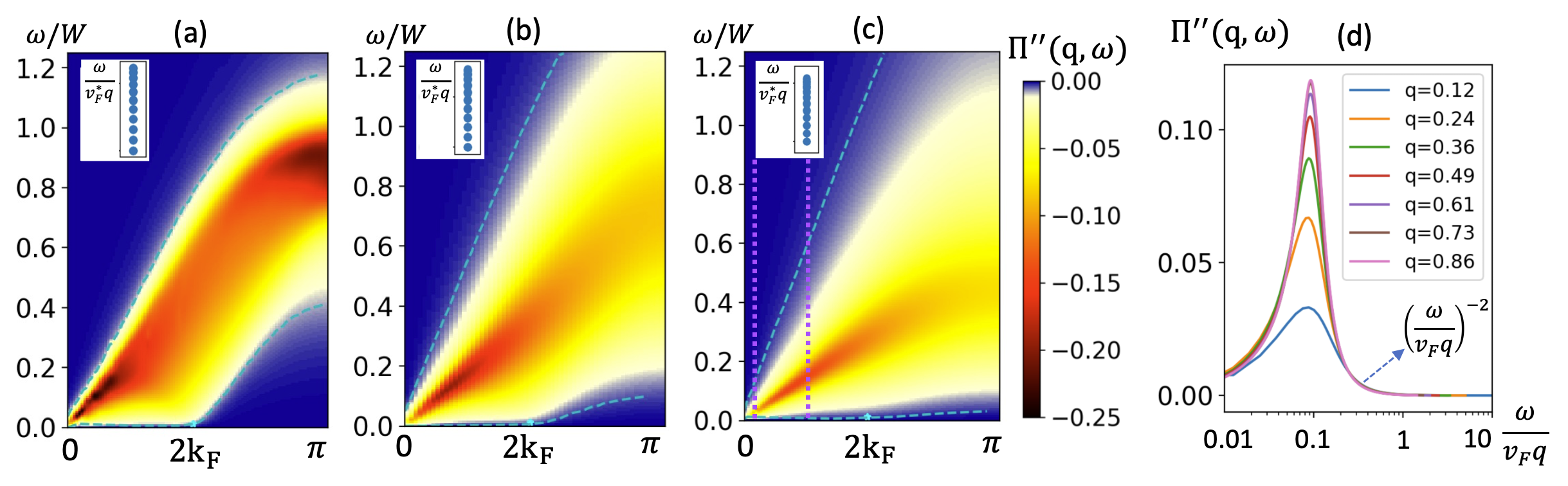}
\caption{\textsf{$\Pi''(q,\omega)$ evaluated for model-B with $\new{t_c=0.5},U=8, T=0.05, \mu=-3t_c$ and (a) $J=2$, (b) $J=4$, and (c) $J=8$. The insets show the eigenenergies in units of $\omega/v_F^*q$ obtained from the QBE; there is no isolated eigenvalue corresponding to an undamped ZS mode at any value of $J$. The blue dashed lines denote the edge of the particle-hole continuum where the spectral weight reaches $1\%$ of their maximum value. (d) Scans of $\Pi^{''}(q,\omega)$ at fixed $q$ (for a range of $q-$values along purple dotted lines) in (c) plotted as a function of $\omega/v_F q$.}}
\label{two} 
\end{figure}

As outlined in Sec.~\ref{prelim:B}, model-B realizes a marginal Fermi liquid with a critical Fermi surface. {\it A priori}, it is far from being obvious whether the sharp Fermi surface can sustain an undamped ZS mode. The numerical results for the full momentum and frequency dependent density response for the $c-$fermions (in the two-band model) are shown in Fig.~\ref{two}. The computations based on the Bethe-Salpeter equation carried out at a finite temperature show the ZS mode buried inside the continuum. To confirm whether the ZS exists outside the continuum at $T=0$ for any value of $J$ (the strength of inter-band interactions), we compute the eigenvalue spectrum from the QBE, which indeed shows no well-separated mode (see inset of Figs.~\ref{two} (a)-(c)). We will now further corroborate the absence of an undamped ZS mode by analyzing the problem analytically in the small $\omega,\q$ limit\red{, which can be understood from the $\k-$independence of the electronic self-energy (i.e. the singular self-energy is present in the entire BZ)}. As a matter of principle, this model demonstrates the absence of a long-lived collective mode associated with a sharp Fermi surface as a result of its decay into an anomalous particle-hole continuum at arbitrarily small energies.

\begin{figure}[h!]
\includegraphics[width=150mm,scale=1]{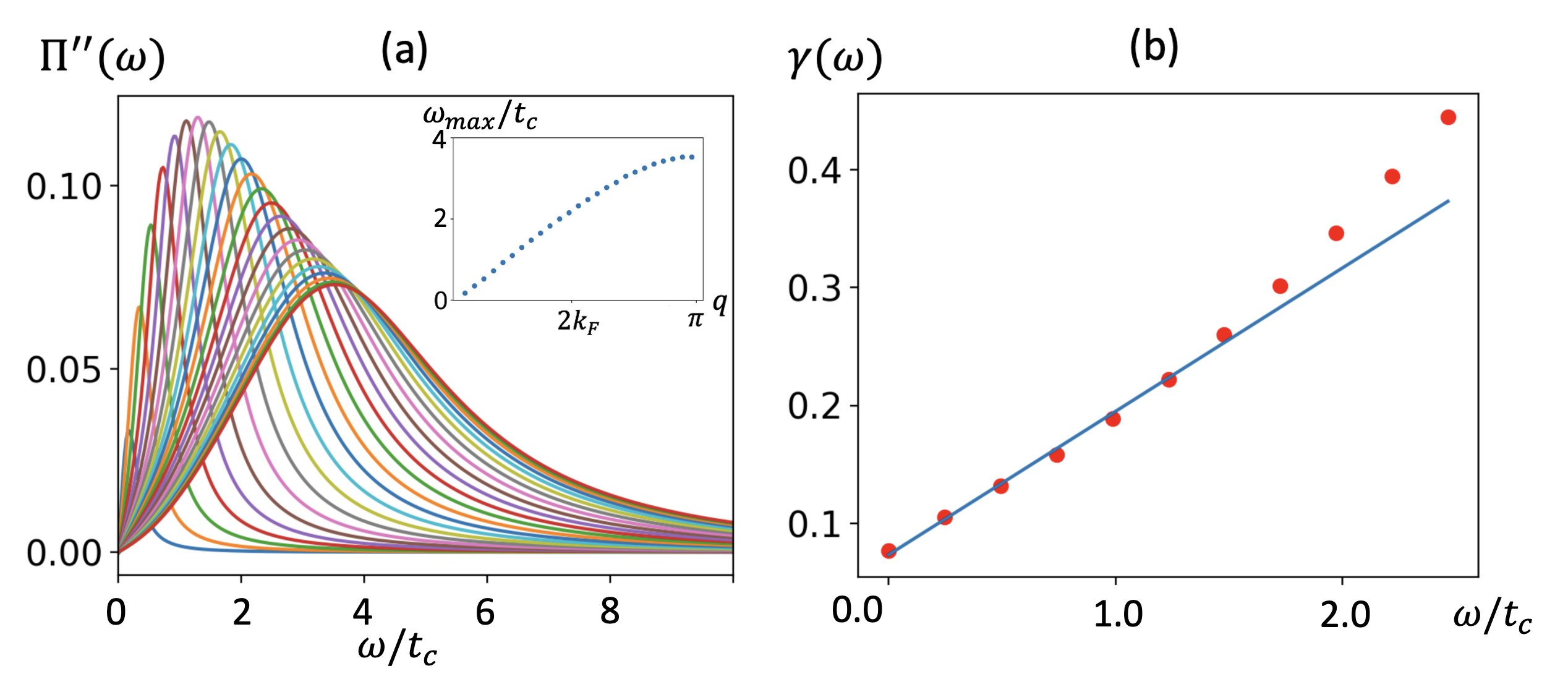}
\caption{\textsf{(a) Slices of $\Pi^{''}(q,\omega)$ at different $q$ from $q=0$ to $q=\pi$ in Fig.(\ref{two}) and a fit of the peak frequency $\omega_{\tn{max}}(q)$. Different colors denote scans at different $q$. (b) The ZS scattering rate, $\gamma(\omega)$, obtained from a Lorentzian fit to the numerical result for $\Pi^{''}(q,\omega)$.}}
\label{two2} 
\end{figure}


The self-consistent Bethe-Salpeter equation for the $c-$fermion density vertex (Eq.~\ref{BSE}) is given by,
\begin{equation}\label{bbse}
\begin{aligned}
    &\Gamma(\vec{q},i\Omega,i\omega)=1+\tilde{J}\int_{-\Omega}^{0} d\tilde{\omega} \frac{\Gamma(\vec{q},i\Omega,i\tilde{\omega})}{\sqrt{|\Xi(\tilde{i\omega},i\Omega)|^2+(v_F q)^2}} \log \bigg(\frac{U}{|\omega-\tilde{\omega}|}\bigg),\\
    &\Xi(i\omega,i\Omega)=i\Omega+\Sigma_{cf}(i\omega)-\Sigma_{cf}(i\omega+i\Omega),
\end{aligned}
\end{equation}
where $\tilde{J}=J^2 \nu_0/U$ is the effective interaction strength introduced as before, and $\vec{q}, \Omega$ and $k, \omega$ denote center-of-mass momentum/frequency and relative momentum/frequency, respectively. Here we have used the completely local character of the $f-$fermions, that contribute to the shaded box in Eq.(\ref{BSE}), to integrate over the momentum of internal lines. Note that the solution for the vertex in the limit $\vec{q}\rightarrow 0$ limit is already constrained by the Ward identity: $\Gamma(\vec{q}=0,i\Omega,i\omega)=\Xi(i\omega,i\Omega)/i\Omega$; see Appendix \ref{app:ward}. Let us now consider the limit $q \lesssim \Omega \ll U$ and make several approximations to Eq.~(\ref{bbse}) in order to proceed analytically. We will expand to leading order in $(v_F q/\Xi(i\omega,i\Omega))$ and express the self-energy piece in $\Xi(i\omega,i\Omega)$ in terms of a derivative, $\Lambda(i\Omega)$. Without affecting \new{the position of} the pole/branch-cut of the integrand, we get a modified equation for the vertex:
\begin{equation}
\begin{aligned}
    &\Gamma(\vec{q},i\Omega,i\omega)=1+\tilde{J}\int_{-\Omega}^{0} d\tilde{\omega} \frac{\Gamma(\vec{q},i\Omega,i\tilde{\omega})\Lambda(i\Omega)}{\new{-}\Lambda(i\Omega)\Xi(\tilde{i\omega},i\Omega)+(v_F q)^2} \log \bigg(\frac{U}{|\omega-\tilde{\omega}|}\bigg),\\
    &\Lambda(i\Omega)=i\Omega\bigg(1-\frac{\partial \Sigma_{cf}(i\Omega)}{\partial \Omega}\bigg)=i\Omega\bigg[1+\tilde{J}\log\bigg(\frac{U}{|\Omega|}\bigg)\bigg].\label{chargevertex}
\end{aligned}
\end{equation}
The solution to the vertex in this limit is $\Gamma(\vec{q},i\Omega,i\omega)=[-\Lambda(i\Omega)\Xi(i\omega,i\Omega)+(v_F q)^2]/[-i\Omega \Lambda(i\Omega)+(v_F q)^2]$. The full density response is then
\begin{equation}
\begin{aligned}
\label{fullpi2}
    \Pi(\vec{q},\Omega)
    =\frac{\new{\nu_0}(v_F q)^2}{[\Omega^2(1+\tilde{J}\log \frac{U}{|\Omega|})-(v_F q)^2]+i\frac{\pi}{2}\tilde{J}\Omega^2},
\end{aligned}
\end{equation}
and the imaginary part is given by,
\begin{equation}
\begin{aligned}
\label{pi2}
    \Pi''(\vec{q},\Omega)
    =\frac{\new{\nu_0}(\frac{v_F |q|}{\Omega})^2}{(\frac{\pi \tilde{J}}{2})^2+[(\frac{v_F |q|}{\Omega})^2-(1+\tilde{J} \log (\frac{U}{\Omega}))]^2}.
\end{aligned}
\end{equation}
Our analysis yields a $\Pi''(\vec{q},\Omega)\sim (v_F |q|/\Omega)^2$ scaling at large $\Omega$, which is observed in the full numerical analysis (Fig.~\ref{two} d). For large $\tilde{J}$, $\Pi''(\vec{q},\Omega)$ has an additional $1/\log^2(\frac{U}{\Omega})$ prefactor, which cures the apparent divergence arising from a purely $1/\Omega^2$ behavior for the ``f-sum rule'' \cite{Abbamonte1,Abbamonte2}. Interestingly, the recent experiments also report a similar behavior \cite{Abbamonte1}. The full $\omega-$dependence of $\Pi''(\omega,\q)$ for different $q$ is shown in Fig.~\ref{two2}(a); the peak ($\omega_{\tn{max}}$) corresponds to the dispersing, but damped, ZS mode.  

In the long-wavelength limit, we can extract a speed for the damped ZS mode from Eq.~(\ref{pi2})
\beq
v_{S}=\frac{v_F}{\sqrt{1+\tilde{J}\log(U/|v_F q|)}},
\eeq
which lies inside the continuum for any $\tilde{J}$. Let us rewrite Eq.~(\ref{fullpi2}) into a conventional Lorentzian form, $\Pi(\q,\Omega)=\frac{A(\q,\Omega)}{[\Omega^2-\omega(\q)^2]+i\Omega \gamma(\Omega)}$. Then $\omega(\q)$ is the dispersion of the ZS mode with velocity $v_S$ while the ZS decay rate is $\gamma(\Omega) \sim \Omega/\log|\frac{U}{\Omega}|$; see Fig.~\ref{two2}(b). The amplitude $A(\q,\Omega) \sim \new{1/}[1+\tilde{J}\log(\frac{U}{\Omega})]$ sets the quasiparticle residue for the ZS mode (and is controlled in part by the $\Lambda(i\Omega)$ term in Eq.~\ref{chargevertex}). The broadened and dispersing ZS mode can be seen in the plots of $\Pi''(\omega,\q)$ in Fig.~\ref{two2}(a).

We now turn our attention to analyzing the collective mode directly using the QBE in Eq.~(\ref{angularQBE}). The effective interaction is mediated by the $f-$fermion polarizability, which is completely independent of any spatial structure (including angle); thus only the $j=0$ piece survives in the decomposition in terms of angular momenta:
\begin{equation}
\begin{aligned}
\label{2qbe}
    &[\omega-(\Sigma(\Omega+\omega/2)-\Sigma(\Omega-\omega/2))]f_j(\el,\omega,\Omega)-\frac{v_F \el}{2}[f_{j-1}(\el,\omega,\Omega)+f_{j+1}(\el,\omega,\Omega)]\\
    &=-\nu_0 \new{J^2} \int_{\Omega'} \Pi_f(\Omega^{'}-\Omega) f_j(\el,\omega,\Omega')\delta_{j,0}.
\end{aligned}
\end{equation}
To obtain the ZS speed $\frac{\omega}{v_F \el}$, we seek a frequency eigenvalue $\alpha(\omega,\Omega)$ of $\int_{\Omega'} \Pi_f(\Omega^{'}-\Omega) f_{j=0}(\el,\omega,\Omega')=\alpha f_{j=0}(l,\omega,\Omega)$. In the long-wavelength and low-frequency limit, we would find $\alpha(\omega,\Omega) \sim \omega \log \frac{U}{\Omega} $ which should cancel with the self energy according to the analysis of Ref.~\cite{qbe}.
For $j>1$, one can get $f_j=\lambda f_{j-1}$ with \new{$\lambda=\frac{|\Lambda(\Omega)|}{v_F \el}-\sqrt{(\frac{|\Lambda(\Omega)|}{v_F \el})^2-1}$}. The eigenvalue problem only involves the $j=0$ and $j=1$ channels; in the limit of $\tilde{J}\gg 1$,  we arrive at the parametrically similar result $\frac{\omega}{v_F \el}\sim \frac{1}{\sqrt{1+2\tilde{J}\log\frac{U}{|\Omega|}}}$.

\subsection{\textsf{Model-C: Non Fermi liquid from quantum critical boson}}\label{sec:resultC}

As outlined in Sec.\ref{prelim:C}, model-C also realizes a non Fermi liquid with a critical Fermi surface. However, in contrast to model-B, the singular frequency dependence of electron self energy is only tied to the vicinity of the Fermi surface (i.e. the singular frequency dependence does not appear everywhere in the BZ). Such momentum dependence gives rise to dramatic difference for the density correlation between model-B and model-C. The qualitatively distinct numerical results for the fully dressed density correlation were already shown in Fig.~\ref{summary}(c) and Fig.~\ref{summary}(d) for weak and strong (Yukawa) coupling limit, respectively, along with their corresponding eigenvalue spectra in the insets. These results suggest that for model-C, the undamped ZS mode exists in the weak-coupling limit, whereas in the strong-coupling limit, the overdamped ZS mode is always buried inside the continuum.  These differences can be traced back to the momentum dependence of the Green's function for the critical $\phi$-boson, as we demonstrate below by analyzing the QBE \cite{qbe}. It is important to address the role of the bare $i\omega$ term in the electron Green's function on the stability of the ZS mode, even though there is a more singular self-energy; see Appendix \ref{app:barew} for a discussion.

As is well known, the interaction between electrons is mediated by the critical $\phi$-boson, with a propagator: $D(q,\Omega)\sim \frac{q}{i\eta|\Omega|+Kq^3}$. The effective Landau interaction, which serves as the input of Eq.(\ref{angularQBE}), is the real part of $D(q,\Omega)$, 
\begin{equation}
\label{boson}
\tn{Re}~ D(q,\Omega)=\frac{Kq^4}{\eta^2\Omega^2+K^2q^6},
\end{equation}
where $K,~\eta$ were introduced in Sec.~\ref{prelim:C}.

We consider the initial and final electronic states near the Fermi surface separated by $q=k_F|\theta-\theta'|$. As reviewed in Sec.~\ref{qbesec}, we define the critical angle $\theta_c=\new{k_F^{-1}}(\frac{\eta}{K}\Omega)^{1/3}$ at which the two contribution in the denominator of Eq.~(\ref{boson}) are comparable, and divide the interaction into two parts, 
\begin{equation}
\label{realboson}
\tn{Re}~ D(q,\Omega) = 
\begin{cases}
\frac{1}{K^{1/3}(\eta \Omega)^{2/3}}, ~~\theta < \theta_c,\\
\frac{1}{K q^2}, ~~~~~~~~\theta > \theta_c.
\end{cases}
\end{equation}
As highlighted in Ref.~\cite{qbe}, for $\theta>\theta_c$, approximating the boson mediated interaction by an instantaneous interaction inevitably leads to a Fermi liquid-like response. On the other hand, for $\theta<\theta_c$, the singular frequency-dependent interaction is important for the low-energy non-Fermi liquid behavior. Recall that the ZS mode favors a large momentum but small frequency transfer process. Associated with the critical angle, $\theta_c$, we can define a critical angular momentum $l_c\sim1/\theta_c$. For $l<l_c$ (i.e. $\theta>\theta_c$), the effective 4pt interaction involves pure momentum transfer and the density correlation behaves ``classically'' (i.e. similar to a conventinal Fermi liquid). On the other hand, for $l>l_c$ (i.e. $\theta<\theta_c$), the frequency dependent piece dominates over the classical momentum transfer process, and the singular interaction leads to density correlations that display a ``local'' non-Fermi liquid behavior (e.g. with some features that are reminiscent of model-B).

The ZS mode remains undamped if $\theta_c$ is ``small'', i.e. if $l_c=k_F (\frac{\eta}{K}\Omega)^{-1/3}\gg1$. For model-C, we estimate $l_c\sim k_F \sqrt{\frac{K}{\eta v_F^{*}}}$, which suggests that in the weak coupling regime, the ZS stays undamped, whereas in the strong coupling regime, the ZS decays into the particle-hole continuum, as illustrated in the detailed computations in Figs.~\ref{summary}(c)-(d). To conclude, the existence of an undamped ZS mode in model-C depends on the coupling strength and the bandwidth.  

\section{\textsf{Outlook}}\label{sec:outlook}

In this work, we have examined a fundamental question associated with non-Fermi liquids, namely what basic properties determine the fate of the zero-sound associated with a critical Fermi surface as it decays into the particle-hole continuum. In general, this is a difficult question to analyze reliably using any controlled analytical or numerical method for a generic microscopic model in the regime of strong interactions. Here we have focused on three families of translationally invariant non-Fermi liquids that host different forms of quantum criticality. We have demonstrated through explicit computations for solvable lattice models that the frequency and momentum-resolved density correlations associated with a critical Fermi surface displaying local criticality are distinct from those where the criticality has non-trivial space-time correlations. For the specific models considered here, a marginal Fermi liquid with local criticality does not host a long-lived zero-sound mode. On the other hand, in a model where the criticality has singular space-time correlations, the fate of zero-sound depends on the specific aspects of the effective interactions mediated by the critical bosonic degree of freedom. We provide a pedagogical summary of the connections between the QBE and the Bethe-Salpeter equation in Appendix \ref{app:equivalence}.

\begin{figure}[h!]
\includegraphics[width=160mm,scale=1.2]{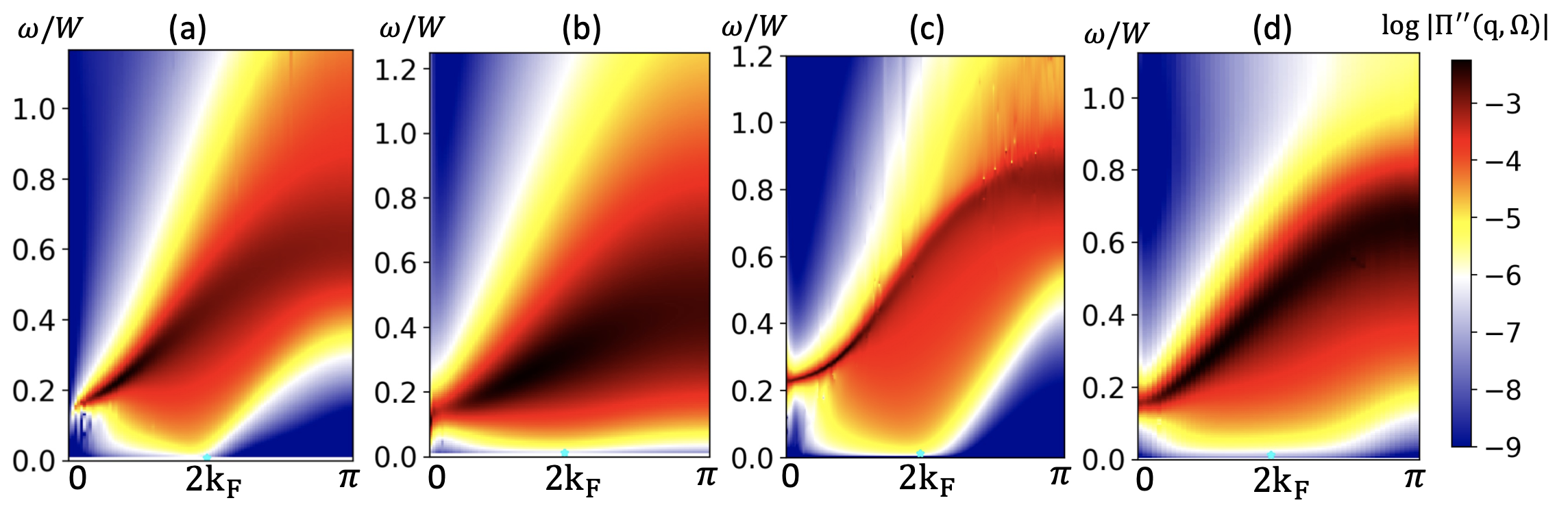}
\caption{\textsf{A re-plot of the results in Fig.~(\ref{summary}) (with an identical set of parameter values) after including the effects of Coulomb interactions within RPA for (a) model-A, (b) model-B, and model-C in (c) weak, and (d) strong-coupling regimes, respectively.}}
\label{exp1} 
\end{figure}

We end by making a few remarks about the possible connections between our results and recent experiments \cite{Abbamonte1,Abbamonte2}, with the obvious caveat that neither the models analyzed here nor the specific large$-N$ limits describe the microscopic models relevant for the real materials. It is useful to include the Coulomb interaction between electrons explicitly and treat it within the standard random-phase approximation (RPA), building on the full polarizability ($\Pi''(\q,\omega)$) that has been analyzed up until now. \red{Note that the problem is no longer ``solvable'' in the SYK-sense after including the Coulomb interaction, and the RPA computations on top of the strongly renormalized (non-)Fermi liquid states are only meant to provide an approximate picture for the charge-dynamics.} We choose to include the three-dimensional Coulomb interaction $V_{c}(q)\sim1/q^2$, as is appropriate for any correlated quasi two-dimensional layered material (e.g. the cuprates). Within RPA, this will have the standard effect of renormalizing the zero-sound up to the plasma frequency, as shown in Fig.~(\ref{exp1}). As expected, for models-A and C (the latter in the ``weak'' coupling regime), the low-energy undamped ZS mode is renormalized to an undamped plasmon mode at the smallest $q$ and finite $\omega$; see Figs.~\ref{exp1}(a) and (c). Once the plasmon enters the continuum, it undergoes a significant broadening due to a strong damping. For models-B and C (the latter in the ``strong'' coupling regime), the gapped plasmon mode exhibits an anomalous decay starting with significantly lower $q$; see Figs.~\ref{exp1}(b) and (d). This is expected given the strongly overdamped character of the parent ZS mode. Perhaps more interestingly, the $(\omega/v_Fq)^{-2}$ scaling collapse we find for the marginal Fermi liquid in model-B (Fig.~\ref{two}d) is reminiscent of the recent experiment \cite{Abbamonte1,Abbamonte2}.  
On the other hand, the interesting frequency independent (and rather universal) ``plateau'' observed in the experiment is absent in our models and explaining its origin remains a worthwhile challenge for future theoretical work. Studying the role of both short and long-wavelength disorder in addition to the non-trivial effects analyzed here is also left for future studies.

\acknowledgements

We thank P. Abbamonte, A. Chubukov, \new{E. Altman, A. A. Patel} and S. Sachdev for a number of illuminating discussions. DC thanks J.P. Sethna and S.J. Thornton for a related collaboration analyzing the density response of strange metals from a complementary point of view. DC is supported by faculty startup grants at Cornell University. This work was performed in part at the Aspen Center for Physics, which is supported by National Science Foundation grant PHY-1607611.

\appendix

\section{\textsf{Diagrammatic expansion in the large$-N$ limit}}
\label{app:SD}
In this section, we provide additional details for our computations of the Bethe-Salpeter equation starting with the ``$G-\Sigma$'' action. It is useful to formulate the method directly for model-B. To recover the results for model-A, we can set $J=0$ and simply set $t_f$ back to being non-zero. The $G-\Sigma$ action has the following standard form for model-B \cite{DCsyk,SSmagneto},
\begin{equation}
\label{gs}
\begin{aligned}
    &F/N=\tn{Tr} \log [-i\omega_n+\epsilon_k-\mu_c+\Sigma_{cf}] +\tn{Tr} \log [-i\omega_n-\mu_f+\Sigma_{cf}^{'}]\\
    &+\tn{Tr}[\Sigma_{cf} G_c]+\tn{Tr}[(\Sigma_{cf}^{'}+\Sigma_f) G_f]-\Phi_{LW}(G_c,G_f),
\end{aligned}    
\end{equation}
where the Luttinger-Ward functional is given by (red and black lines denote dressed $f$ and $c$ Green's functions, respectively)

\begin{equation}
\begin{aligned}
\begin{tikzpicture}[x=0.75pt,y=0.75pt,yscale=-1,xscale=1]

\draw  [fill={rgb, 255:red, 0; green, 0; blue, 0 }  ,fill opacity=1 ] (316.9,142.21) -- (316.69,149.04) -- (310.49,144.59) -- cycle ;
\draw    (343.06,123.32) .. controls (335.06,152.32) and (296.67,150.06) .. (286.67,124.06) ;
\draw    (343.06,123.32) .. controls (331.67,96.06) and (296.67,97.06) .. (286.67,124.06) ;
\draw  [fill={rgb, 255:red, 0; green, 0; blue, 0 }  ,fill opacity=1 ] (319.53,103.27) -- (313.95,107.23) -- (313.76,99.61) -- cycle ;
\draw [color={rgb, 255:red, 208; green, 2; blue, 5 }  ,draw opacity=1 ]   (343.06,123.32) .. controls (338.73,129.8) and (325.46,132.5) .. (312.63,131.9) .. controls (301.78,131.39) and (291.25,128.54) .. (286.67,123.63) ;
\draw [color={rgb, 255:red, 208; green, 2; blue, 5 }  ,draw opacity=1 ]   (343.06,123.32) .. controls (331.67,112.08) and (296.67,112.49) .. (286.67,123.63) ;
\draw  [fill={rgb, 255:red, 0; green, 0; blue, 0 }  ,fill opacity=1 ] (319.53,115.27) -- (313.95,119.23) -- (313.76,111.61) -- cycle ;
\draw  [fill={rgb, 255:red, 0; green, 0; blue, 0 }  ,fill opacity=1 ] (317.04,129.52) -- (316.82,136.35) -- (310.63,131.9) -- cycle ;
\draw  [fill={rgb, 255:red, 0; green, 0; blue, 0 }  ,fill opacity=1 ] (426.9,142.21) -- (426.69,149.04) -- (420.49,144.59) -- cycle ;
\draw [color={rgb, 255:red, 208; green, 9; blue, 2 }  ,draw opacity=1 ]   (453.06,123.32) .. controls (445.06,152.32) and (406.67,150.06) .. (396.67,124.06) ;
\draw [color={rgb, 255:red, 208; green, 9; blue, 2 }  ,draw opacity=1 ]   (453.06,123.32) .. controls (441.67,96.06) and (406.67,97.06) .. (396.67,124.06) ;
\draw  [fill={rgb, 255:red, 0; green, 0; blue, 0 }  ,fill opacity=1 ] (429.53,103.27) -- (423.95,107.23) -- (423.76,99.61) -- cycle ;
\draw [color={rgb, 255:red, 208; green, 2; blue, 5 }  ,draw opacity=1 ]   (453.06,123.32) .. controls (448.73,129.8) and (435.46,132.5) .. (422.63,131.9) .. controls (411.78,131.39) and (401.25,128.54) .. (396.67,123.63) ;
\draw [color={rgb, 255:red, 208; green, 2; blue, 5 }  ,draw opacity=1 ]   (453.06,123.32) .. controls (441.67,112.08) and (406.67,112.49) .. (396.67,123.63) ;
\draw  [fill={rgb, 255:red, 0; green, 0; blue, 0 }  ,fill opacity=1 ] (429.53,115.27) -- (423.95,119.23) -- (423.76,111.61) -- cycle ;
\draw  [fill={rgb, 255:red, 0; green, 0; blue, 0 }  ,fill opacity=1 ] (427.04,129.52) -- (426.82,136.35) -- (420.63,131.9) -- cycle ;

\draw (153,113.4) node [anchor=north west][inner sep=0.75pt]  [font=\normalsize]  {$\Phi _{LW}( G_{c} ,G_{f}) =\frac{J^{2}}{2}$};
\draw (354,109.4) node [anchor=north west][inner sep=0.75pt]    {$+\frac{U_{f}^{2}}{4}$};
\end{tikzpicture}.
\end{aligned}  
\end{equation}

The Schwinger-Dyson equations follow directly from Eq.(\ref{gs}):
\begin{equation}
\label{self_con}
\begin{aligned}
G_c(k,i\omega)&=\frac{1}{i\omega_n-\epsilon_k-\Sigma_{cf}(k,i\omega)}\\
G_f(k,i\omega)&=\frac{1}{i\omega_n-\Sigma_f(k,i\omega)-\Sigma_{cf}^{'}(k,i\omega)}
\end{aligned}    
\end{equation}
where the self-energies are given by:
\begin{equation}
\label{se_2}
\begin{aligned}
\begin{tikzpicture}[x=0.75pt,y=0.75pt,yscale=-1,xscale=1]

\draw  [fill={rgb, 255:red, 0; green, 0; blue, 0 }  ,fill opacity=1 ] (190.83,84.6) -- (185.21,88.49) -- (185.09,80.87) -- cycle ;
\draw    (193.67,84.63) .. controls (173.09,84.87) and (196.09,84.87) .. (182.09,84.87) ;
\draw    (250.06,84.32) .. controls (238.67,57.06) and (203.67,58.06) .. (193.67,85.06) ;
\draw  [fill={rgb, 255:red, 0; green, 0; blue, 0 }  ,fill opacity=1 ] (226.53,64.27) -- (220.95,68.23) -- (220.76,60.61) -- cycle ;
\draw [color={rgb, 255:red, 208; green, 2; blue, 5 }  ,draw opacity=1 ]   (250.06,84.32) .. controls (245.73,90.8) and (232.46,93.5) .. (219.63,92.9) .. controls (208.78,92.39) and (198.25,89.54) .. (193.67,84.63) ;
\draw [color={rgb, 255:red, 208; green, 2; blue, 5 }  ,draw opacity=1 ]   (250.06,84.32) .. controls (238.67,73.08) and (203.67,73.49) .. (193.67,84.63) ;
\draw  [fill={rgb, 255:red, 0; green, 0; blue, 0 }  ,fill opacity=1 ] (226.53,76.27) -- (220.95,80.23) -- (220.76,72.61) -- cycle ;
\draw  [fill={rgb, 255:red, 0; green, 0; blue, 0 }  ,fill opacity=1 ] (224.04,90.52) -- (223.82,97.35) -- (217.63,92.9) -- cycle ;
\draw    (261.64,84.07) .. controls (241.06,84.32) and (264.06,84.32) .. (250.06,84.32) ;
\draw  [dash pattern={on 0.84pt off 2.51pt}]  (250.06,84.32) .. controls (242.06,113.32) and (203.67,111.06) .. (193.67,85.06) ;
\draw  [fill={rgb, 255:red, 0; green, 0; blue, 0 }  ,fill opacity=1 ] (258.64,84.07) -- (253.02,87.97) -- (252.91,80.35) -- cycle ;
\draw [color={rgb, 255:red, 208; green, 2; blue, 27 }  ,draw opacity=1 ]   (472.67,81.63) .. controls (452.09,81.87) and (475.09,81.87) .. (461.09,81.87) ;
\draw [color={rgb, 255:red, 208; green, 2; blue, 27 }  ,draw opacity=1 ]   (529.06,81.32) .. controls (517.67,54.06) and (482.67,55.06) .. (472.67,82.06) ;
\draw  [fill={rgb, 255:red, 0; green, 0; blue, 0 }  ,fill opacity=1 ] (505.53,61.27) -- (499.95,65.23) -- (499.76,57.61) -- cycle ;
\draw [color={rgb, 255:red, 208; green, 2; blue, 5 }  ,draw opacity=1 ]   (529.06,81.32) .. controls (524.73,87.8) and (511.46,90.5) .. (498.63,89.9) .. controls (487.78,89.39) and (477.25,86.54) .. (472.67,81.63) ;
\draw [color={rgb, 255:red, 208; green, 2; blue, 5 }  ,draw opacity=1 ]   (529.06,81.32) .. controls (517.67,70.08) and (482.67,70.49) .. (472.67,81.63) ;
\draw  [fill={rgb, 255:red, 0; green, 0; blue, 0 }  ,fill opacity=1 ] (505.53,73.27) -- (499.95,77.23) -- (499.76,69.61) -- cycle ;
\draw  [fill={rgb, 255:red, 0; green, 0; blue, 0 }  ,fill opacity=1 ] (503.04,87.52) -- (502.82,94.35) -- (496.63,89.9) -- cycle ;
\draw [color={rgb, 255:red, 208; green, 2; blue, 27 }  ,draw opacity=1 ]   (540.64,81.07) .. controls (520.06,81.32) and (543.06,81.32) .. (529.06,81.32) ;
\draw  [dash pattern={on 0.84pt off 2.51pt}]  (529.06,81.32) .. controls (521.06,110.32) and (482.67,108.06) .. (472.67,82.06) ;
\draw  [fill={rgb, 255:red, 0; green, 0; blue, 0 }  ,fill opacity=1 ] (537.64,81.07) -- (532.02,84.97) -- (531.91,77.35) -- cycle ;
\draw  [fill={rgb, 255:red, 0; green, 0; blue, 0 }  ,fill opacity=1 ] (469.83,81.6) -- (464.21,85.49) -- (464.09,77.87) -- cycle ;
\draw [color={rgb, 255:red, 208; green, 2; blue, 27 }  ,draw opacity=1 ]   (341.67,83.63) .. controls (321.09,83.87) and (344.09,83.87) .. (330.09,83.87) ;
\draw [color={rgb, 255:red, 208; green, 2; blue, 27 }  ,draw opacity=1 ]   (398.06,83.32) .. controls (386.67,56.06) and (351.67,57.06) .. (341.67,84.06) ;
\draw  [fill={rgb, 255:red, 0; green, 0; blue, 0 }  ,fill opacity=1 ] (374.53,63.27) -- (368.95,67.23) -- (368.76,59.61) -- cycle ;
\draw [color={rgb, 255:red, 0; green, 0; blue, 0 }  ,draw opacity=1 ]   (398.06,83.32) .. controls (393.73,89.8) and (380.46,92.5) .. (367.63,91.9) .. controls (356.78,91.39) and (346.25,88.54) .. (341.67,83.63) ;
\draw [color={rgb, 255:red, 0; green, 0; blue, 0 }  ,draw opacity=1 ]   (398.06,83.32) .. controls (386.67,72.08) and (351.67,72.49) .. (341.67,83.63) ;
\draw  [fill={rgb, 255:red, 0; green, 0; blue, 0 }  ,fill opacity=1 ] (374.53,75.27) -- (368.95,79.23) -- (368.76,71.61) -- cycle ;
\draw  [fill={rgb, 255:red, 0; green, 0; blue, 0 }  ,fill opacity=1 ] (372.04,89.52) -- (371.82,96.35) -- (365.63,91.9) -- cycle ;
\draw [color={rgb, 255:red, 208; green, 2; blue, 27 }  ,draw opacity=1 ]   (409.64,83.07) .. controls (389.06,83.32) and (412.06,83.32) .. (398.06,83.32) ;
\draw  [dash pattern={on 0.84pt off 2.51pt}]  (398.06,83.32) .. controls (390.06,112.32) and (351.67,110.06) .. (341.67,84.06) ;
\draw  [fill={rgb, 255:red, 0; green, 0; blue, 0 }  ,fill opacity=1 ] (406.64,83.07) -- (401.02,86.97) -- (400.91,79.35) -- cycle ;
\draw  [fill={rgb, 255:red, 0; green, 0; blue, 0 }  ,fill opacity=1 ] (338.83,83.6) -- (333.21,87.49) -- (333.09,79.87) -- cycle ;

\draw (134,74.4) node [anchor=north west][inner sep=0.75pt]  [font=\normalsize]  {$\Sigma _{cf} =$};
\draw (281,73.4) node [anchor=north west][inner sep=0.75pt]  [font=\normalsize]  {$\Sigma _{cf'} =$};
\draw (421,71.4) node [anchor=north west][inner sep=0.75pt]  [font=\normalsize]  {$\Sigma _{f} =$};

\end{tikzpicture}
\end{aligned}    
\end{equation}
To evaluate the dressed charge vertex, we need the effective 4pt interaction as an input of Bethe-Salpeter equation, which can be obtained from the $G-\Sigma$ action in Eq.(\ref{gs}). Specifically, we can consider the fluctuations of the $G-\Sigma$ action near the saddle point Eq.(\ref{se_2}), which is given by \cite{Gu2020NotesOT, linear}

\begin{equation}
\label{linres}
	\begin{gathered}
		\delta S=
		\begin{pmatrix}
			\delta \Sigma & \delta G 
		\end{pmatrix}
	\begin{pmatrix}
		W_\Sigma & -1 \\
		-1 & W_G
	\end{pmatrix}
	\begin{pmatrix}
		\delta \Sigma \\ \delta G 
	\end{pmatrix} 
	+ \tn{tr}[\sigma \cdot \delta G],
	\end{gathered}
\end{equation}
where $W_G$ and $W_\Sigma$ are $2\times2$ matrices whose entries are described by $c$ and $f$ electrons. We have included an external source term, $\sigma$. The matrix elements of $W_G$ and $W_\Sigma$ can be expressed diagrammatically, with dashed lines as delta function in real space (for simplicity, we drop the disorder-average dotted line from now on):

\begin{equation}
\begin{aligned}
\includegraphics[width=100mm,scale=1.2]{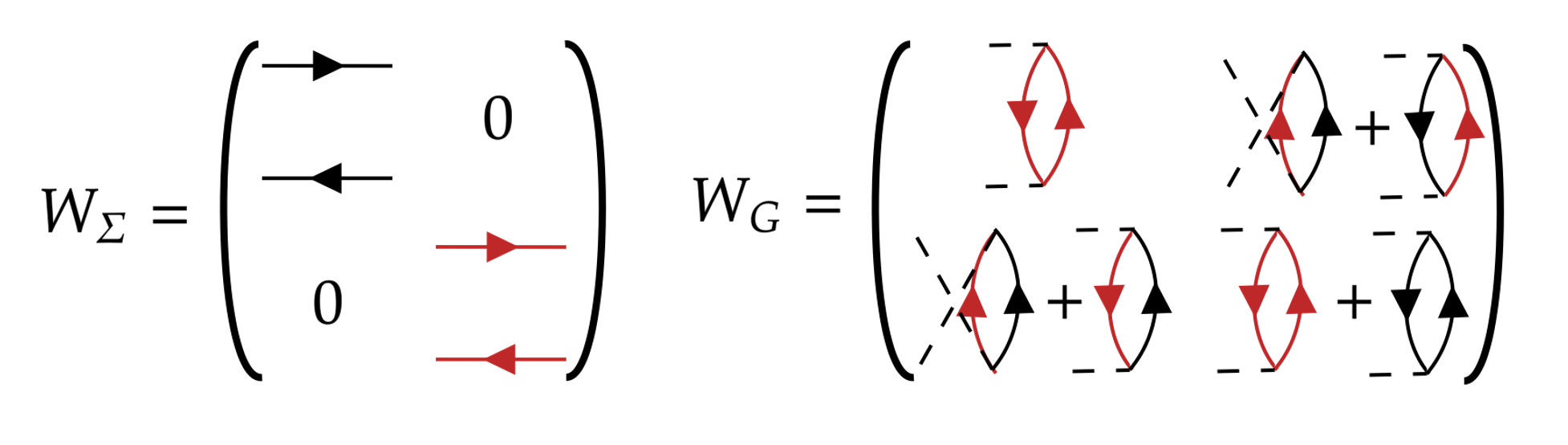}
\end{aligned}
\end{equation}
The 4pt interaction in terms of the bilocal field linear response theory is given by the linear response kernel of $\sigma$, i.e. the term in action that is quadratic to $\sigma$. After integrating out $\Sigma$, we have
\begin{equation}
	\begin{gathered}
		\delta S=
		\begin{pmatrix}
			\delta G_c & \delta G_f
		\end{pmatrix}
    W_\Sigma^{-1}(1-W_\Sigma W_{G})
	\begin{pmatrix}
		\delta G_c \\ \delta G_f
	\end{pmatrix} 
	+\tn{tr}[\sigma \cdot \delta G].
	\end{gathered}
\end{equation}

 The response of $G$ to the external source $\sigma$ is given by $\delta G=(1-W_\Sigma W_{G})^{-1} W_\Sigma \cdot \sigma $, where $(1-W_\Sigma W_{G})^{-1} W_\Sigma$ is the linear response kernel that generates the effective 4pt interaction. By expanding $(1-W_\Sigma W_{G})^{-1}$, one obtains the ladder summation for 4pt interaction. The charge vertex can be expressed as,

\begin{equation}
\label{ver1}
\begin{tikzpicture}[x=0.75pt,y=0.75pt,yscale=-0.8,xscale=0.8]

\draw [line width=1.5]    (172,82.06) .. controls (159,82.13) and (159,175.06) .. (173,174.06) ;
\draw [line width=1.5]    (257.02,173.99) .. controls (270.02,174.02) and (270.69,81.08) .. (256.68,81.98) ;
\draw  [pattern=_zkop8g48k,pattern size=6pt,pattern thickness=0.75pt,pattern radius=0pt, pattern color={rgb, 255:red, 0; green, 0; blue, 0}] (169.18,104.04) -- (181.4,95.41) -- (181.82,112.34) -- cycle ;
\draw    (181.82,112.34) -- (203.35,126.06) ;
\draw    (181.4,95.41) -- (204,81.06) ;
\draw  [fill={rgb, 255:red, 0; green, 0; blue, 0 }  ,fill opacity=1 ] (192.7,88.24) -- (194.72,84.63) -- (197.21,87.58) -- cycle ;
\draw  [fill={rgb, 255:red, 0; green, 0; blue, 0 }  ,fill opacity=1 ] (194.32,120.15) -- (189.54,119.68) -- (192.17,116.82) -- cycle ;
\draw  [pattern=_zkop8g48k,pattern size=6pt,pattern thickness=0.75pt,pattern radius=0pt, pattern color={rgb, 255:red, 0; green, 0; blue, 0}] (218.18,104.04) -- (230.4,95.41) -- (230.82,112.34) -- cycle ;
\draw [color={rgb, 255:red, 208; green, 2; blue, 27 }  ,draw opacity=1 ][fill={rgb, 255:red, 208; green, 2; blue, 27 }  ,fill opacity=1 ]   (230.82,112.34) -- (252.35,126.06) ;
\draw [color={rgb, 255:red, 208; green, 2; blue, 27 }  ,draw opacity=1 ]   (230.4,95.41) -- (253,81.06) ;
\draw  [color={rgb, 255:red, 208; green, 2; blue, 27 }  ,draw opacity=1 ][fill={rgb, 255:red, 208; green, 2; blue, 27 }  ,fill opacity=1 ] (241.7,88.24) -- (243.72,84.63) -- (246.21,87.58) -- cycle ;
\draw  [color={rgb, 255:red, 208; green, 2; blue, 27 }  ,draw opacity=1 ][fill={rgb, 255:red, 208; green, 2; blue, 27 }  ,fill opacity=1 ] (243.32,120.15) -- (238.54,119.68) -- (241.17,116.82) -- cycle ;
\draw  [color={rgb, 255:red, 208; green, 2; blue, 27 }  ,draw opacity=1 ][pattern=_66zr4cbsr,pattern size=6pt,pattern thickness=0.75pt,pattern radius=0pt, pattern color={rgb, 255:red, 208; green, 2; blue, 27}] (170.18,159.04) -- (182.4,150.41) -- (182.82,167.34) -- cycle ;
\draw    (182.82,167.34) -- (204.35,181.06) ;
\draw    (182.4,150.41) -- (205,136.06) ;
\draw  [fill={rgb, 255:red, 0; green, 0; blue, 0 }  ,fill opacity=1 ] (193.7,143.24) -- (195.72,139.63) -- (198.21,142.58) -- cycle ;
\draw  [fill={rgb, 255:red, 0; green, 0; blue, 0 }  ,fill opacity=1 ] (195.32,175.15) -- (190.54,174.68) -- (193.17,171.82) -- cycle ;
\draw  [color={rgb, 255:red, 208; green, 2; blue, 27 }  ,draw opacity=1 ][pattern=_66zr4cbsr,pattern size=6pt,pattern thickness=0.75pt,pattern radius=0pt, pattern color={rgb, 255:red, 208; green, 2; blue, 27}] (222.18,157.04) -- (234.4,148.41) -- (234.82,165.34) -- cycle ;
\draw [color={rgb, 255:red, 208; green, 2; blue, 27 }  ,draw opacity=1 ][fill={rgb, 255:red, 208; green, 2; blue, 27 }  ,fill opacity=1 ]   (234.82,165.34) -- (256.35,179.06) ;
\draw [color={rgb, 255:red, 208; green, 2; blue, 27 }  ,draw opacity=1 ]   (234.4,148.41) -- (257,134.06) ;
\draw  [color={rgb, 255:red, 208; green, 2; blue, 27 }  ,draw opacity=1 ][fill={rgb, 255:red, 208; green, 2; blue, 27 }  ,fill opacity=1 ] (245.7,141.24) -- (247.72,137.63) -- (250.21,140.58) -- cycle ;
\draw  [color={rgb, 255:red, 208; green, 2; blue, 27 }  ,draw opacity=1 ][fill={rgb, 255:red, 208; green, 2; blue, 27 }  ,fill opacity=1 ] (247.32,173.15) -- (242.54,172.68) -- (245.17,169.82) -- cycle ;
\draw    (302,101.08) -- (313.68,108.45) -- (335.26,122.06) ;
\draw    (302,101.08) -- (336,79.06) ;
\draw  [fill={rgb, 255:red, 0; green, 0; blue, 0 }  ,fill opacity=1 ] (319.22,89.99) -- (321.26,86.3) -- (323.78,89.32) -- cycle ;
\draw  [fill={rgb, 255:red, 0; green, 0; blue, 0 }  ,fill opacity=1 ] (320.38,112.54) -- (315.55,112.06) -- (318.21,109.13) -- cycle ;
\draw [line width=1.5]    (303,81.06) .. controls (290,81.13) and (290,174.06) .. (304,173.06) ;
\draw [line width=1.5]    (388.02,172.99) .. controls (401.02,173.02) and (401.69,80.08) .. (387.68,80.98) ;
\draw [color={rgb, 255:red, 208; green, 2; blue, 27 }  ,draw opacity=1 ][fill={rgb, 255:red, 208; green, 2; blue, 27 }  ,fill opacity=1 ]   (352,151.42) -- (363.68,158.9) -- (385.26,172.73) ;
\draw [color={rgb, 255:red, 208; green, 2; blue, 27 }  ,draw opacity=1 ][fill={rgb, 255:red, 208; green, 2; blue, 27 }  ,fill opacity=1 ]   (352,151.42) -- (386,129.06) ;
\draw  [color={rgb, 255:red, 208; green, 2; blue, 27 }  ,draw opacity=1 ][fill={rgb, 255:red, 208; green, 2; blue, 27 }  ,fill opacity=1 ] (369.22,140.16) -- (371.26,136.41) -- (373.78,139.48) -- cycle ;
\draw  [color={rgb, 255:red, 208; green, 2; blue, 27 }  ,draw opacity=1 ][fill={rgb, 255:red, 208; green, 2; blue, 27 }  ,fill opacity=1 ] (370.38,163.06) -- (365.55,162.57) -- (368.21,159.6) -- cycle ;

\draw (268,119.17) node [anchor=north west][inner sep=0.75pt]    {$=$};
\draw (317,141.4) node [anchor=north west][inner sep=0.75pt]    {$0$};
\draw (362,92.4) node [anchor=north west][inner sep=0.75pt]    {$0$};
\draw (405,120.4) node [anchor=north west][inner sep=0.75pt]    {$( 1-W_\Sigma W_{G})^{-1} W_{\Sigma }$};
\end{tikzpicture}
\end{equation}
The off-diagonal term in Eq.(\ref{ver1}) converts an $f-$electron ($c-$electron) particle-hole pair, to a $c-$electron ($f-$electron) particle-hole pair, respectively. However, these off-diagonal terms  vanish, as is also consistent with the respective Ward identities; see App.~\ref{app:ward}. We also demonstrate this explicitly by analyzing the first non-trivial correction in Eq.~(\ref{ver1}),
\begin{equation}
\includegraphics[width=80mm,scale=1.2]{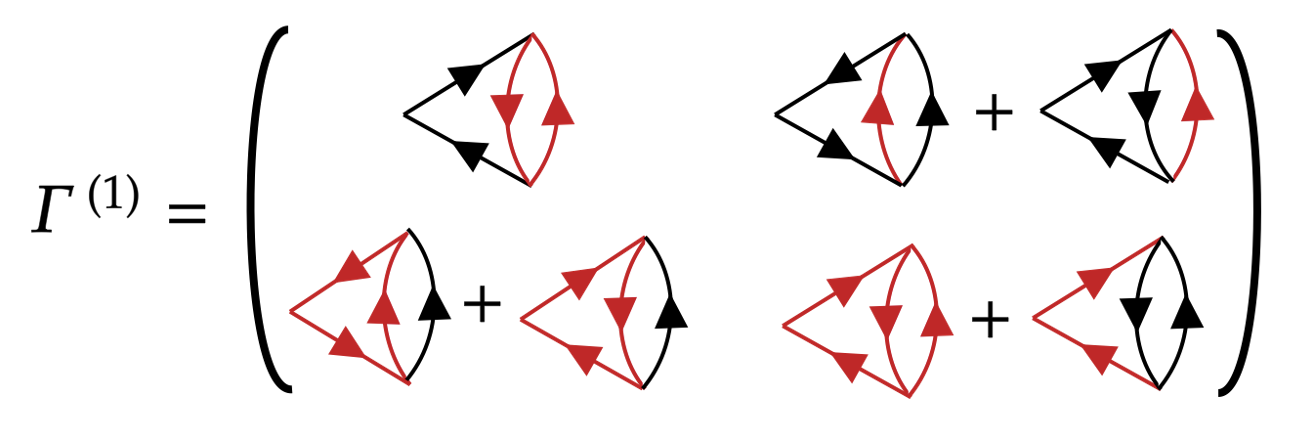}
\end{equation}
The off-diagonal terms consist of two triangular Fermi loops with opposite orientation which cancel each other; we defer a further discussion to the end of this section. The higher order off-diagonal corrections also vanish due to similar cancellations.

Thus, the vertex correction reads:
\begin{equation}
\begin{aligned}
\label{ver_corr2}
\begin{tikzpicture}[x=0.75pt,y=0.75pt,yscale=-0.8,xscale=0.8]

\draw  [pattern=_zkop8g48k,pattern size=6pt,pattern thickness=0.75pt,pattern radius=0pt, pattern color={rgb, 255:red, 0; green, 0; blue, 0}] (147.17,108.2) -- (163.93,93.88) -- (164.5,121.96) -- cycle ;
\draw    (164.5,121.96) -- (194,144.73) ;
\draw    (163.93,93.88) -- (194.89,70.06) ;
\draw  [fill={rgb, 255:red, 0; green, 0; blue, 0 }  ,fill opacity=1 ] (179.41,81.97) -- (182.17,75.99) -- (185.58,80.88) -- cycle ;
\draw    (234,107.73) -- (279,141.73) ;
\draw    (234,107.73) -- (280,72.06) ;
\draw  [pattern=_zkop8g48k,pattern size=6pt,pattern thickness=0.75pt,pattern radius=0pt, pattern color={rgb, 255:red, 0; green, 0; blue, 0}] (323.17,110.07) -- (337.51,96.55) -- (338,123.06) -- cycle ;
\draw    (338,123.06) -- (376,150.06) ;
\draw    (337.51,96.55) -- (375,66.06) ;
\draw [color={rgb, 255:red, 208; green, 2; blue, 27 }  ,draw opacity=1 ]   (363.01,75.06) .. controls (352.02,91.32) and (350.08,118.37) .. (363.01,140.06) ;
\draw [color={rgb, 255:red, 208; green, 2; blue, 27 }  ,draw opacity=1 ]   (363.01,75.06) .. controls (377.65,95.15) and (372.02,130.61) .. (363.01,140.06) ;
\draw  [fill={rgb, 255:red, 0; green, 0; blue, 0 }  ,fill opacity=1 ] (257.29,89.77) -- (260.06,83.78) -- (263.46,88.67) -- cycle ;
\draw  [fill={rgb, 255:red, 0; green, 0; blue, 0 }  ,fill opacity=1 ] (181.62,134.92) -- (175.08,134.14) -- (178.68,129.39) -- cycle ;
\draw  [fill={rgb, 255:red, 0; green, 0; blue, 0 }  ,fill opacity=1 ] (258.87,126.31) -- (252.33,125.52) -- (255.93,120.78) -- cycle ;
\draw  [fill={rgb, 255:red, 0; green, 0; blue, 0 }  ,fill opacity=1 ] (346.41,89.91) -- (348.79,83.76) -- (352.5,88.42) -- cycle ;
\draw  [fill={rgb, 255:red, 0; green, 0; blue, 0 }  ,fill opacity=1 ] (348.87,131.31) -- (342.33,130.52) -- (345.93,125.78) -- cycle ;
\draw  [fill={rgb, 255:red, 0; green, 0; blue, 0 }  ,fill opacity=1 ] (353.69,111.84) -- (350.39,106.14) -- (356.34,106.17) -- cycle ;
\draw  [fill={rgb, 255:red, 0; green, 0; blue, 0 }  ,fill opacity=1 ] (372.45,105.15) -- (375.58,110.95) -- (369.62,110.74) -- cycle ;
\draw  [fill={rgb, 255:red, 0; green, 0; blue, 0 }  ,fill opacity=1 ] (367.01,73.06) -- (369.39,66.91) -- (373.1,71.58) -- cycle ;
\draw  [fill={rgb, 255:red, 0; green, 0; blue, 0 }  ,fill opacity=1 ] (372,147.06) -- (365.42,146.61) -- (368.78,141.69) -- cycle ;
\draw  [color={rgb, 255:red, 208; green, 2; blue, 27 }  ,draw opacity=1 ][pattern=_su7zzwc8a,pattern size=6pt,pattern thickness=0.75pt,pattern radius=0pt, pattern color={rgb, 255:red, 208; green, 2; blue, 27}] (101.17,205.2) -- (117.93,190.88) -- (118.5,218.96) -- cycle ;
\draw [color={rgb, 255:red, 208; green, 2; blue, 27 }  ,draw opacity=1 ]   (118.5,218.96) -- (148,241.73) ;
\draw [color={rgb, 255:red, 208; green, 2; blue, 27 }  ,draw opacity=1 ]   (117.93,190.88) -- (148.89,167.06) ;
\draw  [fill={rgb, 255:red, 0; green, 0; blue, 0 }  ,fill opacity=1 ] (133.41,178.97) -- (136.17,172.99) -- (139.58,177.88) -- cycle ;
\draw [color={rgb, 255:red, 208; green, 2; blue, 27 }  ,draw opacity=1 ]   (188,204.73) -- (233,238.73) ;
\draw [color={rgb, 255:red, 208; green, 2; blue, 27 }  ,draw opacity=1 ]   (188,204.73) -- (234,169.06) ;
\draw  [color={rgb, 255:red, 208; green, 2; blue, 27 }  ,draw opacity=1 ][pattern=_xxg34rr44,pattern size=6pt,pattern thickness=0.75pt,pattern radius=0pt, pattern color={rgb, 255:red, 208; green, 2; blue, 27}] (277.17,207.07) -- (291.51,193.55) -- (292,220.06) -- cycle ;
\draw [color={rgb, 255:red, 208; green, 2; blue, 27 }  ,draw opacity=1 ]   (292,220.06) -- (330,247.06) ;
\draw [color={rgb, 255:red, 208; green, 2; blue, 27 }  ,draw opacity=1 ]   (291.51,193.55) -- (329,163.06) ;
\draw [color={rgb, 255:red, 208; green, 2; blue, 27 }  ,draw opacity=1 ]   (317.01,172.06) .. controls (306.02,188.32) and (304.08,215.37) .. (317.01,237.06) ;
\draw [color={rgb, 255:red, 208; green, 2; blue, 27 }  ,draw opacity=1 ]   (317.01,172.06) .. controls (331.65,192.15) and (326.02,227.61) .. (317.01,237.06) ;
\draw  [fill={rgb, 255:red, 0; green, 0; blue, 0 }  ,fill opacity=1 ] (211.29,186.77) -- (214.06,180.78) -- (217.46,185.67) -- cycle ;
\draw  [fill={rgb, 255:red, 0; green, 0; blue, 0 }  ,fill opacity=1 ] (135.62,231.92) -- (129.08,231.14) -- (132.68,226.39) -- cycle ;
\draw  [fill={rgb, 255:red, 0; green, 0; blue, 0 }  ,fill opacity=1 ] (212.87,223.31) -- (206.33,222.52) -- (209.93,217.78) -- cycle ;
\draw  [fill={rgb, 255:red, 0; green, 0; blue, 0 }  ,fill opacity=1 ] (300.41,186.91) -- (302.79,180.76) -- (306.5,185.42) -- cycle ;
\draw  [fill={rgb, 255:red, 0; green, 0; blue, 0 }  ,fill opacity=1 ] (302.87,228.31) -- (296.33,227.52) -- (299.93,222.78) -- cycle ;
\draw  [fill={rgb, 255:red, 0; green, 0; blue, 0 }  ,fill opacity=1 ] (307.69,208.84) -- (304.39,203.14) -- (310.34,203.17) -- cycle ;
\draw  [fill={rgb, 255:red, 0; green, 0; blue, 0 }  ,fill opacity=1 ] (326.45,202.15) -- (329.58,207.95) -- (323.62,207.74) -- cycle ;
\draw  [fill={rgb, 255:red, 0; green, 0; blue, 0 }  ,fill opacity=1 ] (321.01,170.06) -- (323.39,163.91) -- (327.1,168.58) -- cycle ;
\draw  [fill={rgb, 255:red, 0; green, 0; blue, 0 }  ,fill opacity=1 ] (326,244.06) -- (319.42,243.61) -- (322.78,238.69) -- cycle ;
\draw  [color={rgb, 255:red, 208; green, 2; blue, 27 }  ,draw opacity=1 ][pattern=_4aaxmu7dk,pattern size=6pt,pattern thickness=0.75pt,pattern radius=0pt, pattern color={rgb, 255:red, 208; green, 2; blue, 27}] (381.17,205.07) -- (395.51,191.55) -- (396,218.06) -- cycle ;
\draw [color={rgb, 255:red, 208; green, 2; blue, 27 }  ,draw opacity=1 ]   (396,218.06) -- (434,245.06) ;
\draw [color={rgb, 255:red, 208; green, 2; blue, 27 }  ,draw opacity=1 ]   (395.51,191.55) -- (433,161.06) ;
\draw [color={rgb, 255:red, 0; green, 0; blue, 0 }  ,draw opacity=1 ]   (421.01,170.06) .. controls (410.02,186.32) and (408.08,213.37) .. (421.01,235.06) ;
\draw [color={rgb, 255:red, 0; green, 0; blue, 0 }  ,draw opacity=1 ]   (421.01,170.06) .. controls (435.65,190.15) and (430.02,225.61) .. (421.01,235.06) ;
\draw  [fill={rgb, 255:red, 0; green, 0; blue, 0 }  ,fill opacity=1 ] (404.41,184.91) -- (406.79,178.76) -- (410.5,183.42) -- cycle ;
\draw  [fill={rgb, 255:red, 0; green, 0; blue, 0 }  ,fill opacity=1 ] (406.87,226.31) -- (400.33,225.52) -- (403.93,220.78) -- cycle ;
\draw  [fill={rgb, 255:red, 0; green, 0; blue, 0 }  ,fill opacity=1 ] (411.69,206.84) -- (408.39,201.14) -- (414.34,201.17) -- cycle ;
\draw  [fill={rgb, 255:red, 0; green, 0; blue, 0 }  ,fill opacity=1 ] (430.45,200.15) -- (433.58,205.95) -- (427.62,205.74) -- cycle ;
\draw  [fill={rgb, 255:red, 0; green, 0; blue, 0 }  ,fill opacity=1 ] (425.01,168.06) -- (427.39,161.91) -- (431.1,166.58) -- cycle ;
\draw  [fill={rgb, 255:red, 0; green, 0; blue, 0 }  ,fill opacity=1 ] (430,242.06) -- (423.42,241.61) -- (426.78,236.69) -- cycle ;
\draw  [dash pattern={on 0.84pt off 2.51pt}]  (363.01,75.06) .. controls (389,81.06) and (395,135.06) .. (363.01,140.06) ;
\draw  [dash pattern={on 0.84pt off 2.51pt}]  (317.01,172.06) .. controls (343,178.06) and (349,232.06) .. (317.01,237.06) ;
\draw  [dash pattern={on 0.84pt off 2.51pt}]  (421.01,170.06) .. controls (447,176.06) and (453,230.06) .. (421.01,235.06) ;

\draw (196,99.17) node [anchor=north west][inner sep=0.75pt]    {$=$};
\draw (284,98.17) node [anchor=north west][inner sep=0.75pt]    {$+$};
\draw (150,196.17) node [anchor=north west][inner sep=0.75pt]    {$=$};
\draw (238,195.17) node [anchor=north west][inner sep=0.75pt]    {$+$};
\draw (355,195.17) node [anchor=north west][inner sep=0.75pt]    {$+$};
\end{tikzpicture}
\end{aligned}
\end{equation}
The first (second) line of Eq.~(\ref{ver_corr2}) represents the Bethe-Salpeter equation for $c-$electron ($f-$electron), which will form the basis for the computations of the density response for $c-$electrons in model-B.

It is straightforward to obtain the results for model-A, where we set $J=0$ and focus only on the one-band model with a finite $t_f$ for the $f-$electrons. From the Schwinger-Dyson equation (Eq.~(\ref{self_con})), we obtain
\begin{equation}
\begin{aligned}
G_f(k,i\omega)=\frac{1}{i\omega_n-\epsilon_k-\Sigma_{f}(k,i\omega)},\\
\end{aligned}    
\end{equation}
with the self energy $\Sigma_f$,
\begin{equation}
\begin{aligned}
\begin{tikzpicture}[x=0.75pt,y=0.75pt,yscale=-1,xscale=1]

\draw [color={rgb, 255:red, 0; green, 0; blue, 0 }  ,draw opacity=1 ]   (309.67,136.63) .. controls (289.09,136.87) and (312.09,136.87) .. (298.09,136.87) ;
\draw [color={rgb, 255:red, 0; green, 0; blue, 0 }  ,draw opacity=1 ]   (366.06,136.32) .. controls (354.67,109.06) and (319.67,110.06) .. (309.67,137.06) ;
\draw  [color={rgb, 255:red, 0; green, 0; blue, 0 }  ,draw opacity=1 ][fill={rgb, 255:red, 0; green, 0; blue, 0 }  ,fill opacity=1 ] (342.53,116.27) -- (336.95,120.23) -- (336.76,112.61) -- cycle ;
\draw [color={rgb, 255:red, 0; green, 0; blue, 0 }  ,draw opacity=1 ]   (366.06,136.32) .. controls (361.73,142.8) and (348.46,145.5) .. (335.63,144.9) .. controls (324.78,144.39) and (314.25,141.54) .. (309.67,136.63) ;
\draw [color={rgb, 255:red, 0; green, 0; blue, 0 }  ,draw opacity=1 ]   (366.06,136.32) .. controls (354.67,125.08) and (319.67,125.49) .. (309.67,136.63) ;
\draw  [color={rgb, 255:red, 0; green, 0; blue, 0 }  ,draw opacity=1 ][fill={rgb, 255:red, 0; green, 0; blue, 0 }  ,fill opacity=1 ] (342.53,128.27) -- (336.95,132.23) -- (336.76,124.61) -- cycle ;
\draw  [color={rgb, 255:red, 0; green, 0; blue, 0 }  ,draw opacity=1 ][fill={rgb, 255:red, 0; green, 0; blue, 0 }  ,fill opacity=1 ] (340.04,142.52) -- (339.82,149.35) -- (333.63,144.9) -- cycle ;
\draw [color={rgb, 255:red, 0; green, 0; blue, 0 }  ,draw opacity=1 ]   (377.64,136.07) .. controls (357.06,136.32) and (380.06,136.32) .. (366.06,136.32) ;
\draw [color={rgb, 255:red, 0; green, 0; blue, 0 }  ,draw opacity=1 ] [dash pattern={on 0.84pt off 2.51pt}]  (366.06,136.32) .. controls (358.06,165.32) and (319.67,163.06) .. (309.67,137.06) ;
\draw  [color={rgb, 255:red, 0; green, 0; blue, 0 }  ,draw opacity=1 ][fill={rgb, 255:red, 0; green, 0; blue, 0 }  ,fill opacity=1 ] (374.64,136.07) -- (369.02,139.97) -- (368.91,132.35) -- cycle ;
\draw  [color={rgb, 255:red, 0; green, 0; blue, 0 }  ,draw opacity=1 ][fill={rgb, 255:red, 0; green, 0; blue, 0 }  ,fill opacity=1 ] (306.83,136.6) -- (301.21,140.49) -- (301.09,132.87) -- cycle ;

\draw (258,126.4) node [anchor=north west][inner sep=0.75pt]  [font=\normalsize]  {$\Sigma _{f} =$};
\end{tikzpicture}
\end{aligned}    
\end{equation}
The Bethe-Salpeter equation for the charge vertex takes the simplified form:
\begin{equation}
    \begin{aligned}
    \label{ver_corr1}
        \begin{tikzpicture}[x=0.75pt,y=0.75pt,yscale=-1,xscale=1]

\draw  [color={rgb, 255:red, 0; green, 0; blue, 0 }  ,draw opacity=1 ][pattern=_zkop8g48k,pattern size=6pt,pattern thickness=0.75pt,pattern radius=0pt, pattern color={rgb, 255:red, 0; green, 0; blue, 0}] (147.17,108.2) -- (163.93,93.88) -- (164.5,121.96) -- cycle ;
\draw [color={rgb, 255:red, 0; green, 0; blue, 0 }  ,draw opacity=1 ]   (164.5,121.96) -- (194,144.73) ;
\draw [color={rgb, 255:red, 0; green, 0; blue, 0 }  ,draw opacity=1 ]   (163.93,93.88) -- (194.89,70.06) ;
\draw  [color={rgb, 255:red, 0; green, 0; blue, 0 }  ,draw opacity=1 ][fill={rgb, 255:red, 0; green, 0; blue, 0 }  ,fill opacity=1 ] (179.41,81.97) -- (182.17,75.99) -- (185.58,80.88) -- cycle ;
\draw [color={rgb, 255:red, 0; green, 0; blue, 0 }  ,draw opacity=1 ]   (234,107.73) -- (279,141.73) ;
\draw [color={rgb, 255:red, 0; green, 0; blue, 0 }  ,draw opacity=1 ]   (234,107.73) -- (280,72.06) ;
\draw  [color={rgb, 255:red, 0; green, 0; blue, 0 }  ,draw opacity=1 ][pattern=_zkop8g48k,pattern size=6pt,pattern thickness=0.75pt,pattern radius=0pt, pattern color={rgb, 255:red, 0; green, 0; blue, 0}] (323.17,110.07) -- (337.51,96.55) -- (338,123.06) -- cycle ;
\draw [color={rgb, 255:red, 0; green, 0; blue, 0 }  ,draw opacity=1 ]   (338,123.06) -- (376,150.06) ;
\draw [color={rgb, 255:red, 0; green, 0; blue, 0 }  ,draw opacity=1 ]   (337.51,96.55) -- (375,66.06) ;
\draw [color={rgb, 255:red, 0; green, 0; blue, 0 }  ,draw opacity=1 ]   (363.01,75.06) .. controls (352.02,91.32) and (350.08,118.37) .. (363.01,140.06) ;
\draw [color={rgb, 255:red, 0; green, 0; blue, 0 }  ,draw opacity=1 ]   (363.01,75.06) .. controls (377.65,95.15) and (372.02,130.61) .. (363.01,140.06) ;
\draw  [color={rgb, 255:red, 0; green, 0; blue, 0 }  ,draw opacity=1 ][fill={rgb, 255:red, 0; green, 0; blue, 0 }  ,fill opacity=1 ] (257.29,89.77) -- (260.06,83.78) -- (263.46,88.67) -- cycle ;
\draw  [color={rgb, 255:red, 0; green, 0; blue, 0 }  ,draw opacity=1 ][fill={rgb, 255:red, 0; green, 0; blue, 0 }  ,fill opacity=1 ] (181.62,134.92) -- (175.08,134.14) -- (178.68,129.39) -- cycle ;
\draw  [color={rgb, 255:red, 0; green, 0; blue, 0 }  ,draw opacity=1 ][fill={rgb, 255:red, 0; green, 0; blue, 0 }  ,fill opacity=1 ] (258.87,126.31) -- (252.33,125.52) -- (255.93,120.78) -- cycle ;
\draw  [color={rgb, 255:red, 0; green, 0; blue, 0 }  ,draw opacity=1 ][fill={rgb, 255:red, 0; green, 0; blue, 0 }  ,fill opacity=1 ] (346.41,89.91) -- (348.79,83.76) -- (352.5,88.42) -- cycle ;
\draw  [color={rgb, 255:red, 0; green, 0; blue, 0 }  ,draw opacity=1 ][fill={rgb, 255:red, 0; green, 0; blue, 0 }  ,fill opacity=1 ] (348.87,131.31) -- (342.33,130.52) -- (345.93,125.78) -- cycle ;
\draw  [color={rgb, 255:red, 0; green, 0; blue, 0 }  ,draw opacity=1 ][fill={rgb, 255:red, 0; green, 0; blue, 0 }  ,fill opacity=1 ] (353.69,111.84) -- (350.39,106.14) -- (356.34,106.17) -- cycle ;
\draw  [color={rgb, 255:red, 0; green, 0; blue, 0 }  ,draw opacity=1 ][fill={rgb, 255:red, 0; green, 0; blue, 0 }  ,fill opacity=1 ] (372.45,105.15) -- (375.58,110.95) -- (369.62,110.74) -- cycle ;
\draw  [color={rgb, 255:red, 0; green, 0; blue, 0 }  ,draw opacity=1 ][fill={rgb, 255:red, 0; green, 0; blue, 0 }  ,fill opacity=1 ] (367.01,73.06) -- (369.39,66.91) -- (373.1,71.58) -- cycle ;
\draw  [color={rgb, 255:red, 0; green, 0; blue, 0 }  ,draw opacity=1 ][fill={rgb, 255:red, 0; green, 0; blue, 0 }  ,fill opacity=1 ] (372,147.06) -- (365.42,146.61) -- (368.78,141.69) -- cycle ;
\draw [color={rgb, 255:red, 0; green, 0; blue, 0 }  ,draw opacity=1 ] [dash pattern={on 0.84pt off 2.51pt}]  (363.01,75.06) .. controls (389,81.06) and (395,135.06) .. (363.01,140.06) ;

\draw (196,99.17) node [anchor=north west][inner sep=0.75pt]    {$=$};
\draw (284,98.17) node [anchor=north west][inner sep=0.75pt]    {$+$};
\end{tikzpicture}
    \end{aligned}
\end{equation}

Let us now turn to model-C, where the coupled set of Schwinger-Dyson equations for the $c-$electron and $\varphi-$boson are given by,
\begin{equation}
\begin{aligned}
\label{sde_yukawa}
G_c(k,i\omega)=\frac{1}{i\omega_n-\epsilon_k-\Sigma_{c}(k,i\omega)},\\
D(q,i\Omega)=\frac{1}{\Omega^2+\omega_k^2-\Pi(q,i\Omega)}
\end{aligned}    
\end{equation}
with their respective self-energies (solid lines denote $c-$Green's function and dashed lines denote $\varphi-$propagator)
\begin{equation}
\begin{aligned}
\begin{tikzpicture}[x=0.75pt,y=0.75pt,yscale=-1,xscale=1]

\draw [color={rgb, 255:red, 0; green, 0; blue, 0 }  ,draw opacity=1 ]   (219.67,118.63) .. controls (199.09,118.87) and (222.09,118.87) .. (208.09,118.87) ;
\draw [color={rgb, 255:red, 0; green, 0; blue, 0 }  ,draw opacity=1 ]   (276.06,118.32) .. controls (264.67,91.06) and (229.67,92.06) .. (219.67,119.06) ;
\draw  [color={rgb, 255:red, 0; green, 0; blue, 0 }  ,draw opacity=1 ][fill={rgb, 255:red, 0; green, 0; blue, 0 }  ,fill opacity=1 ] (252.53,98.27) -- (246.95,102.23) -- (246.76,94.61) -- cycle ;
\draw [color={rgb, 255:red, 0; green, 0; blue, 0 }  ,draw opacity=1 ]   (276.06,118.32) .. controls (275.96,120.87) and (274.68,121.9) .. (272.22,121.41) .. controls (270.33,120.21) and (268.71,120.51) .. (267.34,122.32) .. controls (265.55,123.99) and (263.84,123.94) .. (262.21,122.18) .. controls (260.79,120.39) and (259.17,120.24) .. (257.36,121.73) .. controls (255.47,123.22) and (253.75,123.07) .. (252.21,121.26) .. controls (250.67,119.49) and (249.02,119.41) .. (247.27,121) .. controls (245.64,122.65) and (244.01,122.66) .. (242.38,121.02) .. controls (240.61,119.41) and (238.93,119.48) .. (237.34,121.23) .. controls (235.71,122.97) and (234.04,123.04) .. (232.34,121.44) .. controls (230.71,119.79) and (229.03,119.77) .. (227.3,121.38) .. controls (225.37,122.83) and (223.77,122.52) .. (222.5,120.46) -- (219.6,118.55) ;
\draw [color={rgb, 255:red, 0; green, 0; blue, 0 }  ,draw opacity=1 ]   (287.64,118.07) .. controls (267.06,118.32) and (290.06,118.32) .. (276.06,118.32) ;
\draw [color={rgb, 255:red, 0; green, 0; blue, 0 }  ,draw opacity=1 ] [dash pattern={on 0.84pt off 2.51pt}]  (276.06,118.32) .. controls (268.06,147.32) and (229.67,145.06) .. (219.67,119.06) ;
\draw  [color={rgb, 255:red, 0; green, 0; blue, 0 }  ,draw opacity=1 ][fill={rgb, 255:red, 0; green, 0; blue, 0 }  ,fill opacity=1 ] (284.64,118.07) -- (279.02,121.97) -- (278.91,114.35) -- cycle ;
\draw  [color={rgb, 255:red, 0; green, 0; blue, 0 }  ,draw opacity=1 ][fill={rgb, 255:red, 0; green, 0; blue, 0 }  ,fill opacity=1 ] (216.83,118.6) -- (211.21,122.49) -- (211.09,114.87) -- cycle ;
\draw    (342.67,113.56) .. controls (344.54,112.12) and (346.19,112.33) .. (347.63,114.2) .. controls (349.06,116.07) and (350.71,116.28) .. (352.58,114.84) .. controls (354.45,113.4) and (356.1,113.61) .. (357.54,115.48) -- (358.67,115.63) -- (358.67,115.63) ;
\draw    (411.85,115.32) .. controls (413.52,113.66) and (415.19,113.67) .. (416.85,115.34) .. controls (418.51,117.01) and (420.18,117.02) .. (421.85,115.36) -- (426,115.38) -- (426,115.38) ;
\draw [color={rgb, 255:red, 0; green, 0; blue, 0 }  ,draw opacity=1 ]   (411.85,115.32) .. controls (406,119.06) and (399.1,120.66) .. (387,120.06) .. controls (374.9,119.46) and (362.99,120.54) .. (358.67,115.63) ;
\draw [color={rgb, 255:red, 0; green, 0; blue, 0 }  ,draw opacity=1 ]   (411.85,115.32) .. controls (402.36,98.06) and (373.13,97.06) .. (358.67,115.63) ;
\draw  [color={rgb, 255:red, 0; green, 0; blue, 0 }  ,draw opacity=1 ][fill={rgb, 255:red, 0; green, 0; blue, 0 }  ,fill opacity=1 ] (390.53,102.27) -- (384.95,106.23) -- (384.76,98.61) -- cycle ;
\draw  [color={rgb, 255:red, 0; green, 0; blue, 0 }  ,draw opacity=1 ][fill={rgb, 255:red, 0; green, 0; blue, 0 }  ,fill opacity=1 ] (388.66,117.76) -- (388.44,124.59) -- (382.25,120.14) -- cycle ;
\draw [color={rgb, 255:red, 0; green, 0; blue, 0 }  ,draw opacity=1 ] [dash pattern={on 0.84pt off 2.51pt}]  (415.06,114.88) .. controls (407.06,143.88) and (368.67,141.63) .. (358.67,115.63) ;

\draw (167,109.4) node [anchor=north west][inner sep=0.75pt]  [font=\normalsize]  {$\Sigma _{c} =$};
\draw (307,109.4) node [anchor=north west][inner sep=0.75pt]  [font=\normalsize]  {$\Pi =$};
\end{tikzpicture}
\end{aligned}    
\end{equation}
The Bethe-Salpeter equation for model-C can be derived in a fashion that is analogous to the derivation for model-B (including the cancellations, that we address below) and is given by 

\begin{equation}
\begin{aligned}
\label{ver_corry}
\begin{tikzpicture}[x=0.75pt,y=0.75pt,yscale=-1,xscale=1]

\draw  [color={rgb, 255:red, 0; green, 0; blue, 0 }  ,draw opacity=1 ][pattern=_zkop8g48k,pattern size=6pt,pattern thickness=0.75pt,pattern radius=0pt, pattern color={rgb, 255:red, 0; green, 0; blue, 0}] (147.17,108.2) -- (163.93,93.88) -- (164.5,121.96) -- cycle ;
\draw [color={rgb, 255:red, 0; green, 0; blue, 0 }  ,draw opacity=1 ]   (164.5,121.96) -- (194,144.73) ;
\draw [color={rgb, 255:red, 0; green, 0; blue, 0 }  ,draw opacity=1 ]   (163.93,93.88) -- (194.89,70.06) ;
\draw  [color={rgb, 255:red, 0; green, 0; blue, 0 }  ,draw opacity=1 ][fill={rgb, 255:red, 0; green, 0; blue, 0 }  ,fill opacity=1 ] (179.41,81.97) -- (182.17,75.99) -- (185.58,80.88) -- cycle ;
\draw [color={rgb, 255:red, 0; green, 0; blue, 0 }  ,draw opacity=1 ]   (234,107.73) -- (279,141.73) ;
\draw [color={rgb, 255:red, 0; green, 0; blue, 0 }  ,draw opacity=1 ]   (234,107.73) -- (280,72.06) ;
\draw  [color={rgb, 255:red, 0; green, 0; blue, 0 }  ,draw opacity=1 ][pattern=_zkop8g48k,pattern size=6pt,pattern thickness=0.75pt,pattern radius=0pt, pattern color={rgb, 255:red, 0; green, 0; blue, 0}] (323.17,110.07) -- (337.51,96.55) -- (338,123.06) -- cycle ;
\draw [color={rgb, 255:red, 0; green, 0; blue, 0 }  ,draw opacity=1 ]   (338,123.06) -- (376,150.06) ;
\draw [color={rgb, 255:red, 0; green, 0; blue, 0 }  ,draw opacity=1 ]   (337.51,96.55) -- (375,66.06) ;
\draw  [color={rgb, 255:red, 0; green, 0; blue, 0 }  ,draw opacity=1 ][fill={rgb, 255:red, 0; green, 0; blue, 0 }  ,fill opacity=1 ] (257.29,89.77) -- (260.06,83.78) -- (263.46,88.67) -- cycle ;
\draw  [color={rgb, 255:red, 0; green, 0; blue, 0 }  ,draw opacity=1 ][fill={rgb, 255:red, 0; green, 0; blue, 0 }  ,fill opacity=1 ] (181.62,134.92) -- (175.08,134.14) -- (178.68,129.39) -- cycle ;
\draw  [color={rgb, 255:red, 0; green, 0; blue, 0 }  ,draw opacity=1 ][fill={rgb, 255:red, 0; green, 0; blue, 0 }  ,fill opacity=1 ] (258.87,126.31) -- (252.33,125.52) -- (255.93,120.78) -- cycle ;
\draw  [color={rgb, 255:red, 0; green, 0; blue, 0 }  ,draw opacity=1 ][fill={rgb, 255:red, 0; green, 0; blue, 0 }  ,fill opacity=1 ] (346.41,89.91) -- (348.79,83.76) -- (352.5,88.42) -- cycle ;
\draw  [color={rgb, 255:red, 0; green, 0; blue, 0 }  ,draw opacity=1 ][fill={rgb, 255:red, 0; green, 0; blue, 0 }  ,fill opacity=1 ] (348.87,131.31) -- (342.33,130.52) -- (345.93,125.78) -- cycle ;
\draw  [color={rgb, 255:red, 0; green, 0; blue, 0 }  ,draw opacity=1 ][fill={rgb, 255:red, 0; green, 0; blue, 0 }  ,fill opacity=1 ] (367.01,73.06) -- (369.39,66.91) -- (373.1,71.58) -- cycle ;
\draw  [color={rgb, 255:red, 0; green, 0; blue, 0 }  ,draw opacity=1 ][fill={rgb, 255:red, 0; green, 0; blue, 0 }  ,fill opacity=1 ] (372,147.06) -- (365.42,146.61) -- (368.78,141.69) -- cycle ;
\draw [color={rgb, 255:red, 0; green, 0; blue, 0 }  ,draw opacity=1 ] [dash pattern={on 0.84pt off 2.51pt}]  (363.01,75.06) .. controls (389,81.06) and (395,135.06) .. (363.01,140.06) ;
\draw    (363.01,75.06) .. controls (364.68,76.73) and (364.68,78.39) .. (363.01,80.06) .. controls (361.34,81.73) and (361.34,83.39) .. (363.01,85.06) .. controls (364.68,86.73) and (364.68,88.39) .. (363.01,90.06) .. controls (361.34,91.73) and (361.34,93.39) .. (363.01,95.06) .. controls (364.68,96.73) and (364.68,98.39) .. (363.01,100.06) .. controls (361.34,101.73) and (361.34,103.39) .. (363.01,105.06) .. controls (364.68,106.73) and (364.68,108.39) .. (363.01,110.06) .. controls (361.34,111.73) and (361.34,113.39) .. (363.01,115.06) .. controls (364.68,116.73) and (364.68,118.39) .. (363.01,120.06) .. controls (361.34,121.73) and (361.34,123.39) .. (363.01,125.06) .. controls (364.68,126.73) and (364.68,128.39) .. (363.01,130.06) .. controls (361.34,131.73) and (361.34,133.39) .. (363.01,135.06) .. controls (364.68,136.73) and (364.68,138.39) .. (363.01,140.06) -- (363.01,140.06) ;

\draw (196,99.17) node [anchor=north west][inner sep=0.75pt]    {$=$};
\draw (284,98.17) node [anchor=north west][inner sep=0.75pt]    {$+$};

\end{tikzpicture}
\end{aligned}    
\end{equation}

\begin{figure}[h!]
\includegraphics[width=90mm,scale=1]{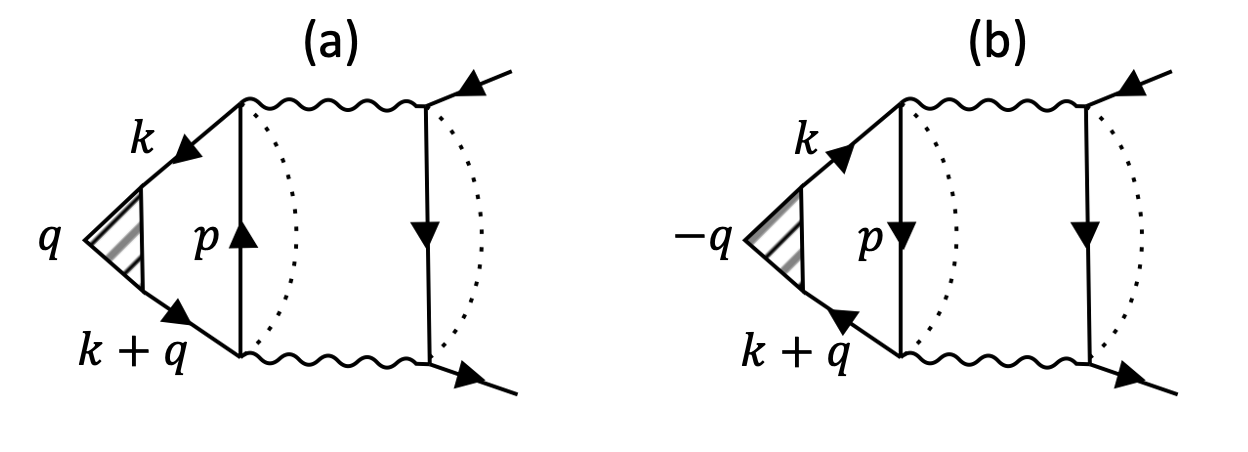}
\caption{\textsf{The two Aslamazov-Larking diagrams that cancel each other mutually}}
\label{AL}
\end{figure}

Let us now address the important cancellations that arise in the Bethe-Salpeter equation for the charge vertex --- these include the Aslamov-Larkin (AL) diagrams in model-C (Fig.~\ref{AL}) and the off-diagonal terms (Eq.~\ref{ver1}) in model-B. Both cancellations arise from the two diagrams with a reverse orientation of the triangular fermion loop. Note that the cancellations occur specifically for the charge vertex, which is an even function of the center-of-mass momentum/frequency. These nature of these cancellations are self-evident also from the observation that the Ward-identities for the charge vertex are only satisfied when these are not included in the Bethe-Salpeter equation, as discussed in Appendix \ref{app:ward} 

\section{\textsf{Ward identity for density vertex}}
\label{app:ward}
In all of the models studied in this paper, the density of electrons is conserved. Associated with this conservation law and the $U(1)$ charge, there is a Ward identity which relates the dressed electron self-energy to the density vertex \cite{Andrey3}. In this section, as a consistency check for our numerical computations of the Bethe-Salpeter equation, we demonstrate that the Ward identity is indeed satisfied. Moreover, this exercise also helps highlight the importance of the momentum-dependence of the vertex corrections when evaluating the density response.

\begin{figure}[h!]
\includegraphics[width=160mm,scale=1]{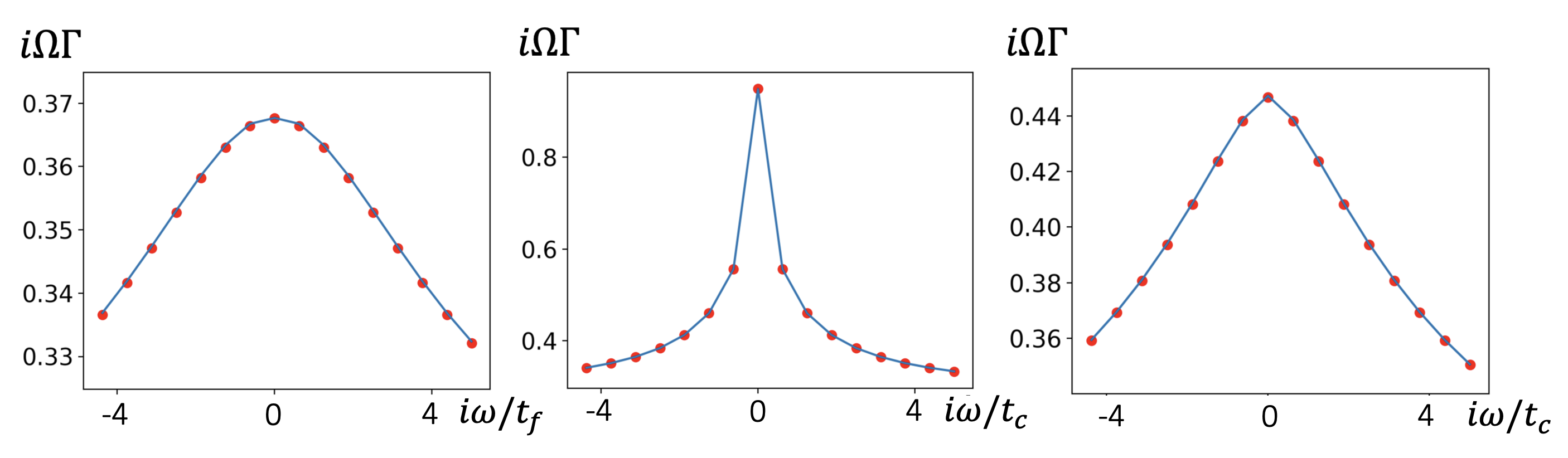}
\caption{\label{wardfig} \textsf{Plot of the LHS (line) and RHS (dots) of Eq.~(\ref{ward}) evaluated at $i\Omega=i\Omega_0, \vec{q}=0, \vec{k}=0$ as a function of $i\omega$ for models A, B and C, respectively, where $\Omega_0=2 \pi T$ with $T=0.01W$. }}
\end{figure}

We denote the fully dressed density vertex as $\Gamma(\q,i\Omega,\k,i\omega)$, as drawn on the left of Eqs.~(\ref{ver_corr1}), (\ref{ver_corr2}) and (\ref{ver_corry})) for models-A, B and C, respectively. Here $(\q,i\Omega)$ denote the momentum, Matsubara frequency for the incoming electron-hole pairs, and $(\k,i\omega)$ for the outgoing hole. The Ward identity for the respective U(1) conservation is given by:
\begin{equation}
    \begin{aligned}
    \label{ward}
        i\Omega \Gamma(\q=0,i\Omega,\k,i\omega)=i\Omega+\Sigma(\k,i\omega)-\Sigma(\k,i\omega+i\Omega),
    \end{aligned}
\end{equation}
with $\Sigma(\k,i\omega)$ as the corresponding self-energy for each model, respectively. We check the Ward identity for each model numerically, by comparing the left and right-hand sides of Eq.~\ref{ward} in Fig. (\ref{wardfig}).

For model B, there are two additional Ward identities, which correspond to the off diagonal density vertex in Eq.~(\ref{ver1}):
\beq
\begin{aligned}
\label{ward2}
&
i\Omega \Gamma_{cf}(\q=0,i\Omega,\k,i\omega)=0,\\
&i\Omega \Gamma_{fc}(\q=0,i\Omega,\k,i\omega)=0,\\
\end{aligned}
\eeq
where $\Gamma_{cf}$ and $\Gamma_{fc}$ represent the off-diagonal terms, respectively. The RHS of Eq.~(\ref{ward2}) vanishes identically as a result of the expectation values, $\langle c^{\dagger}f\rangle = \langle f^{\dagger}c\rangle = 0$. This also suggests that for $\Omega\neq0$, the off-diagonal density vertex should vanish non-perturbatively, as discussed in Appendix~ \ref{app:SD}.

\begin{figure}[h!]
\includegraphics[width=160mm,scale=1]{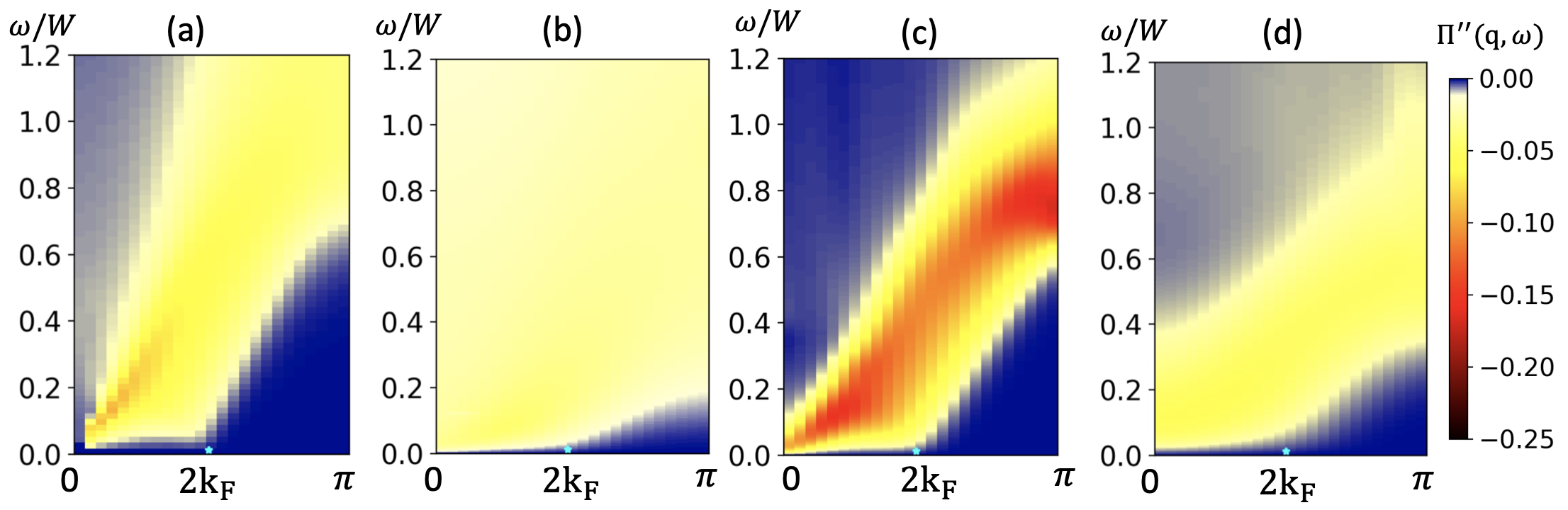}
\caption{\textsf{A replot of the density correlation functions in Fig.~\ref{summary}, but without including the vertex corrections. The density response here is inconsistent with the Ward identity.}}
\label{bare_verfig} 
\end{figure}

To illustrate the importance of the vertex correction, and its non-trivial momentum dependence, in determining the full density response, we show the results for $\Pi''(\q,\Omega)$ without these corrections in Fig.~(\ref{bare_verfig}). Comparison these results against the full computation in Fig.~\ref{two}, we find that the density response is finite even at $\q=0$ and a finite $\Omega$, and is tied to the (intentionally) incorrect computation of a simple convolution of two Green's functions. The contrast is most apparent for model-B, comparing Figs.~\ref{summary}(b) and \ref{bare_verfig}(b) --- while the full (and correct) computation continues to have momentum-dependent features due to $\Gamma$, the computation without vertex corrections becomes almost entirely momentum-independent.

It is useful to write down the equation of motion for both the charge, $\Pi(q,\Omega)$, and current vertex, $\Pi_{x,y}(\q,\Omega)$, as
\begin{equation}\label{eomward}
    \begin{aligned}
    \Omega \Pi(\q,\Omega)+q_x \Pi_x(\q,\Omega)+q_y\Pi_y(\q,\Omega)=0.
    \end{aligned}
\end{equation}
Thus, at $q=0$ and $\Omega \neq 0$, $\Pi(q,\Omega)$ is guaranteed to vanish by charge conservation. Qualitatively, the effect of vertex correction is to transfer the excess unphysical spectral weight above the edge of continuum to the low $q$ and low $\Omega$ region, and thereby also satisfy the Ward identity.

\section{\textsf{Further details of numerical methods}}
\label{app:num}

In Sec.~\ref{sec:prelimB}, we already provided the key steps involved in the computation of the density response using the Bethe-Salpeter equation. Here we provide some additional details associated with these and the QBE computations. We begin by noting that as a result of the coarse momentum-space grid, while generating the color plots for $\Pi''(\q,\Omega)$ in the main text, we have performed a numerical interpolation along the momentum axis on the original grid of 51$\times$51 using the quintic method. This interpolation is carried out before the analytical continuation (using Pad\'e). 

We solve for the eigenvalues of the QBE in Eq.~(\ref{QBE}) numerically in a self-consistent fashion. First, we use the Pad\'e approximation to get the effective 4pt interaction in real frequency as an input for the QBE. The cutoff for Pad\'e approximation can be read off by noting that the RHS of Eq.(\ref{QBE}) vanishes when $|\omega|>2|\Omega|$, where $\omega$ is comparable to the physical Fermi velocity $v_F^{*}$. We adjust the real frequency cutoff in Pad\'e approximation, using $v_F^{*}$ as an intial guess for the cutoff, and discretize the real frequency until the eigenvalue converges. 

\section{\textsf{Incoherent regime of model-A}}
\label{app:modA}

\begin{figure}[h!]
\includegraphics[width=160mm,scale=1.0]{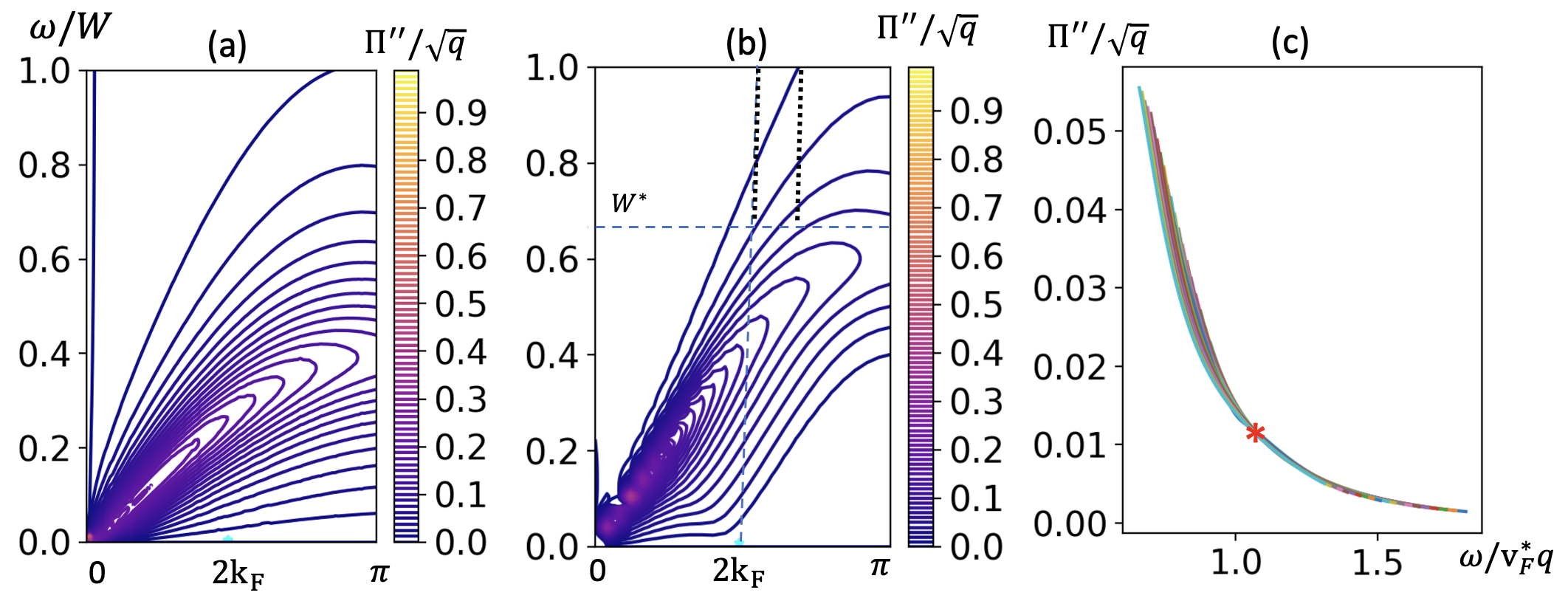}
\caption{\textsf{(a) Contour plot of $\Pi''(q,\omega)$ for model-A evaluated at $U=8,~ T=0.05,~ \mu=-3t_f$ and $\new{t_f=0.125}$. Nominally, the ZS mode appears overdamped and the singular edge near $2k_F$ is almost lost since the temperature is comparable to $t_f$. The QBE analysis at $T=0$ reveals the undamped ZS mode. (b) Contour plot of $\Pi''(q,\omega)$ for model-A with the same parameters as Fig.~\ref{one}(b). The contours beyond $\omega=W^{*}$ and $q=2k_F$ (i.e. upper right region enclosed by the dashed blue lines) show a linear dependence on $q$. (c) Scans of $\Pi''/\sqrt{q}$ along the regions enclosed inside dotted line in (a), which show scaling collapse for $\omega/v_F^{*}q$ starting from the red star, i.e. along the edge of continuum for the heavy FL.}}
\label{one2} 
\end{figure}

We begin this section by pointing out an interesting contrasting aspect of the conclusions from Bethe-Salpeter equations vs. the QBE for model-A in the $t_f\ll U$ regime. When we analyze the Bethe-Salpeter equations for $t_f\sim U/\new{64}$, the renormalized bandwidth $W^{*}$ becomes comparable to the low temperature at which the computations have been performed. In this regime, nominally we do not find an undamped ZS mode using the full two-particle response as the system is already dominated in part by the response from the incoherent regime at energies above $W^*$ (see Fig.~\ref{one2} (a)); however, the QBE analysis carried out at zero-temperature clearly indicates the presence of the undamped ZS with $v_S>v_F$ even for this extreme strong coupling regime.

We now provide details for the mixing between high and low-energy contribution, that was introduced in the main text as a possible explanation for the momentum-dependent features in the incoherent regime of model-A. To make analytical progress, we treat the crossover from the low-energy heavy Fermi liquid to the incoherent metal as sharp and located precisely at $\omega=W^{*}$. This crude model will already turn out to lead to interesting insights into the unusual numerical results.  

For both the Green's function, $G_f$, and the polarizability, $\Pi_f^0$, we make the ansatz,
\beq
\begin{aligned}
\label{sharp}
&
G_f(\k,i\omega)=\frac{\theta(W^{*}-\omega)}{i Z^{-1} \omega-\tilde{\epsilon}_{\k}}+\theta(\omega-W^{*})\frac{i \tn{sgn}(\omega)}{\sqrt{U|\omega|}},\\
&
\Pi_f^{0}(|\q| \ll k_F,i\Omega)=\theta(W^{*}-\Omega)\frac{1}{U}\bigg[1+\frac{|\Omega|}{v_F^{*}|\q|}+\tn{ln} (\frac{U}{W^{*}})\bigg]+\theta(\Omega-W^{*})\frac{1}{U} \ln\bigg(\frac{U}{|\Omega|}\bigg),
\end{aligned}
\eeq
with $Z\sim\frac{t_f}{U}$. Note that the expression of $\Pi_f^{0}$ in Eq.~(\ref{sharp}) is only valid when the center of mass momentum is much smaller than $2k_F$. To include the $2k_F$ physics, we use antipodal patches and expand the center-of-mass momentum near $2k_F$ in terms of $q$ along the direction perpendicular to the Fermi surface. 

We consider the momentum dependent contribution to the full polarizability for $\Omega>W^*$ order by order in $t_f/U$. The zeroth-order momentum-dependent contribution comes from Fermi-liquid regime, which is clearly not responsible for the residual momentum-dependence in the incoherent regime, and we ignore it henceforth. 

\begin{figure}[h!]
\includegraphics[width=160mm,scale=1]{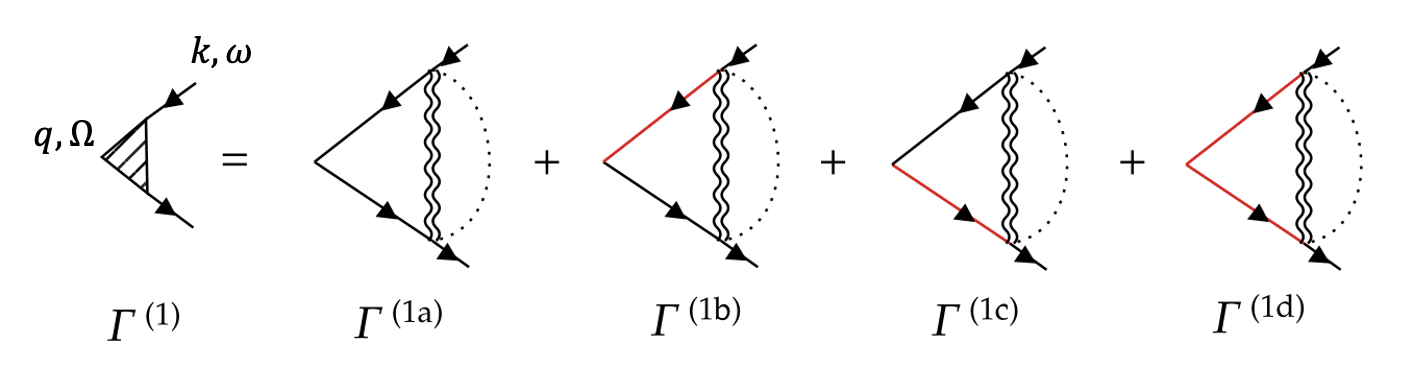}
\caption{\textsf{Contribution to the density vertex from different energy regimes, where the contributions from the high-energy and low-energy electrons are denoted by red and black lines, respectively.}}
\label{UVIR} 
\end{figure}

 To obtain the $\q$ dependence beyond the renormalized bandwidth $W^{*}$, we now consider the leading order corrections to the charge vertex, and classify the contributions from the high-energy and low-energy electrons, denoted by red and black in Fig.~\ref{UVIR}, respectively. To avoid confusion, we use double wavy line to denote $\Pi_f^{0}$ in Eq.(\ref{sharp}).

Note that both of the external legs in Fig.~\ref{UVIR} must be in the FL regime, as otherwise, contracting the external legs of $\Gamma^{(1)}$ with the incoherent local Green's function (that serve as a momentum sink) will lead to a density response independent of $\q$. In fact, $\Gamma^{(1d)}$ is $\q$-independent due to the same reason.

Consider now the limit $t_f \ll U$ and compare the relative contributions $\Gamma^{(1a)}$, $\Gamma^{(1b)}$ and $\Gamma^{(1c)}$. In what follows, we make very simple power-counting estimates and find, 
\beq
\begin{aligned}
&
\Gamma^{(1a)}\sim U^2\int_{0}^{W^{*}} d\tilde{\omega} ~d\vec{p} \frac{Z}{i\tilde{\omega}-Z\tilde{\epsilon}_{\vec{p}}} \frac{Z}{i(\tilde{\omega}+\Omega)-Z\tilde{\epsilon}_{\vec{p}+\q}} \Pi_f^0(\vec{p}-\k,\tilde{\omega}-\omega)\sim \frac{t_f^2}{U^2},\\&
\Gamma^{(1b)}\sim \Gamma^{(1c)}\sim U^2\int_{0}^{W^{*}} d\omega ~d\vec{p} \frac{Z}{i\tilde{\omega}-Z\tilde{\epsilon}_{\vec{p}}}\frac{1}{\sqrt{U|\tilde{\omega}+\Omega|}}\Pi_f^0(\vec{p}-\k,\tilde{\omega}-\omega)\sim{\sqrt{\frac{t_f}{U}}}.
\end{aligned}
\eeq
Note that $\Pi_f^0(...)$ is not momentum-independent in the above integrands, as their momentum dependence is controlled by the frequency transfer. 

The leading momentum dependence arises from $\Gamma^{(1b)}$, and is given by
\beq
\begin{aligned}
\label{gamma1b}
    &
    \Gamma^{(1b)}(\vec{q},\Omega,\vec{k},\omega)=U\int_{p_x,p_y,\tilde{\omega}} \frac{1}{iZ^{-1}\tilde{\omega}-v_F p_x-\kappa p_y^2} \frac{\sgn(\tilde{\omega}+\Omega)}{\sqrt{U|\tilde{\omega}+\Omega}|} \frac{|\tilde{\omega}-\omega|}{v_F|p_x-k_x|}\\
    &=\frac{\sqrt{U}}{v_F}\int_{p_y,\tilde{\omega}} \frac{\sgn(\tilde{\omega}+\Omega)\sgn(\tilde{\omega})}{iZ^{-1}\tilde{\omega}-v_F k_x-\kappa p_y^2} \frac{|\tilde{\omega}-\omega|}{\sqrt{|\tilde{\omega}+\Omega}|}\\
    &\sim\frac{t^2}{v_F\sqrt{\kappa U}}\frac{|\omega|}{\sqrt{|\Omega v_F k_x|}}.
\end{aligned}
\eeq
Note that the $k_x$ branch cut in the denominator of Eq.(\ref{gamma1b}) is important for obtaining non-vanishing response beyond the renormalized bandwidth when contracting with the external FL Green's function.  
This leads to a correction to the density response of the form,
\beq
\begin{aligned}
\label{pi1f}
    &\Pi^{(1)}(q,\Omega)=\frac{t^2}{v_F\sqrt{\kappa U}}\int_{\vec{p},\omega}\frac{1}{i Z^{-1}\omega+v_F p_x-\kappa p_y^2}\frac{1}{iZ^{-1}(\omega+\Omega)-v_F(p_x+q_x)-\kappa(p_y)^2}\frac{|\omega|}{\sqrt{v_F|\Omega p_x|}}\\
    &=\frac{t^2}{v_F^2\sqrt{\kappa U}}\int_{p_y,\omega}\frac{1}{i Z^{-1}(2\omega+\Omega)-v_F q_x -2\kappa p_y^2}\frac{|\omega|}{\sqrt{|\Omega|}}[\frac{\alpha_1}{\sqrt{i Z^{-1}\omega-\kappa p_y^2}}+\frac{\alpha_2}{\sqrt{i Z^{-1}(\omega+\Omega)-v_F q_x-\kappa p_y^2}}],\\
\end{aligned}
\eeq
where $\alpha_1$ and $\alpha_2$ are O(1) complex numbers. Note that Eq.(\ref{pi1f}) can be expressed as a scaling form $\Pi^{(1)}(q,\Omega)=\frac{t^2\new{Z}}{v_F^2\sqrt{\kappa U}}\sqrt{\frac{v_F^{*} q}{\kappa}}F(\frac{\Omega}{W^{*} q_x})$ where $F(x)$ is a scaling function. 
Our numerical results in this incoherent regime exhibiting a mixing between the high and low energy regions does exhibit a similar scaling collapse as can be clearly seen in Fig.~\ref{one2}(b) and (c).
Thus, the density response for $\Omega>W^{*}$ and $q>2k_F$, can be described order by order in $\frac{t_f}{U}$, with a leading momentum independent piece from usual SYK particle-hole pairs and a subleading contribution involving mixing between high/low-energy regimes ($\sim \frac{1}{(U)^{\frac{3}{2}}}$) satisfying $\frac{\Omega}{W^{*} q_x}$ scaling.

\section{\textsf{Role of ``bare'' electron dynamics: case of model-C}}
\label{app:barew}

\begin{figure}[h!]
\includegraphics[width=160mm,scale=1]{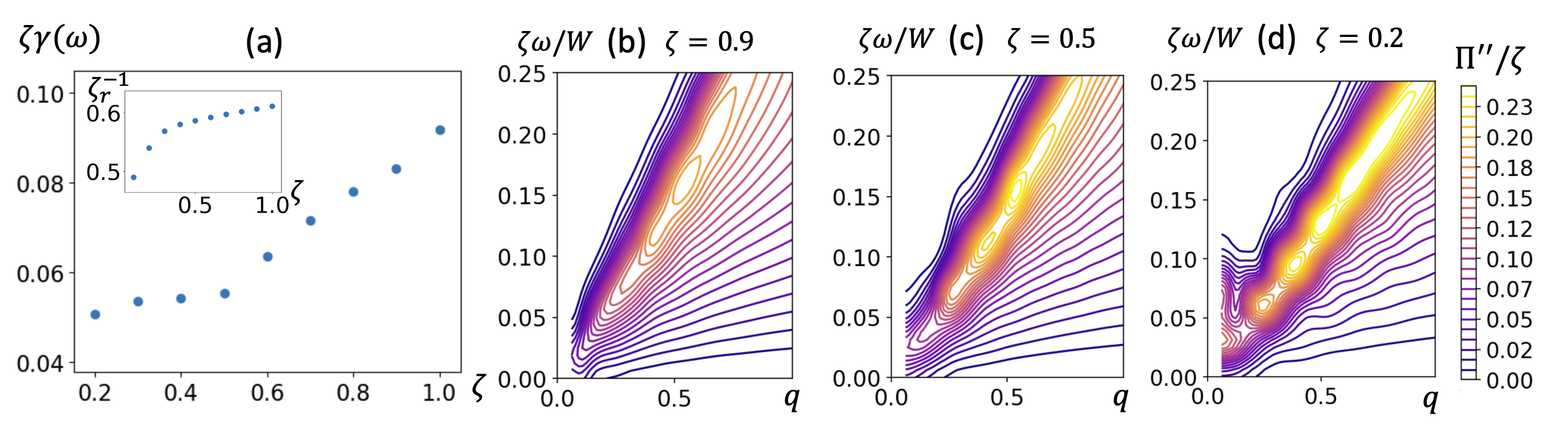}
\caption{\label{fig:nobare} \textsf{(a) The ZS decay rate $\gamma(\omega)$ obtained from a lorentzian fit to the peak of $\Pi''(\q,\omega)$ in (b)-(d) at the smallest $q$. The procedure is as described in Sec.~ \ref{sec:resultB}. The inset shows the renormalized $\zeta_r^{-1}$ as a function of the bare $\zeta$. (b)-(d) Fully dressed density response, $\Pi''(\q,\omega)$ for model-C starting with the artificial $\zeta$ factor.}}
\end{figure}

In this section, we study the effect of the ``bare'' electron dynamics, as encoded in the $i\omega$ term in the inverse propagator on the stability of the ZS mode. However, in the present context, the term ``bare'' does not necessarily imply the microscopic $i\omega$ term that we start with at the outset; instead such a term will always be generated self-consistently as one integrates out high-energy modes. It is conceivable that when non-Fermi liquids host an undamped ZS mode, it can be traced back to the presence of this $i\omega$ term (that is nevertheless weaker compared to the more singular self-energy).  Specifically, we focus on model-C in the ``weak'' coupling regime, where our previous calculations have found evidence of an undamped ZS mode that exists outside the continuum at low energies.
Consider now an artificial limit, where an explicit coefficient $\zeta$ is introduced for the $i\omega$ term. The resulting Schwinger-Dyson equation in Eq.~(\ref{sde_yukawa}) is then modified to
\begin{equation}
\begin{aligned}
\label{Zscf}
G_c(k,i\omega)=\frac{1}{i \zeta\omega_n-\epsilon_k-\Sigma_{c}(k,i\omega)},\\
D(q,i\Omega)=\frac{1}{\Omega^2+\omega_k^2-\Pi(q,i\Omega)}.
\end{aligned}
\end{equation}
In the non-Fermi liquid regime with a singular self-energy, the term proportional to $\zeta$ does not play an important role in as far as the single-particle properties at low energy are concerned. It nevertheless does play an important role for the density response, and in determining the structure of the onset of the particle-hole continuum at low-energy and long wavelength limit. We have numerically obtained the density response with the above modification as a function of varying $\zeta$, focusing on the low-energy limit. It is worth noting that interactions lead to a  self-consistent renormalization of $\zeta$, such that even a vanishingly small initial choice of $\zeta$ will generate a finite renormalized $\zeta_r$ as shown in the inset of Fig.~\ref{fig:nobare}(a). Importantly, at the smallest values of $q$ and $\omega$, the ZS mode remains long-lived even for small $\zeta$, as demonstrated in the decay rate in Fig.~\ref{fig:nobare}(a).

In Fig.~\ref{fig:nobare}(b)-(d), we show the fully dressed density response for several values of $\zeta$. The result shows no qualitative difference in as far as the existence of undamped ZS mode is concerned.  

\section{\textsf{Relationship between quantum Boltzmann equation and Bethe-Salpeter equation}}\label{app:equivalence}

In this section, we provide a pedagogical overview of the equivalence of the two methods that have been used extensively in this paper --- quantum Boltzmann equation and Bethe-Salpeter equation. Within Fermi liquid theory, the ZS mode can be obtained from the pole of the fully dressed polarizability \cite{RevModPhys.66.129}. This is still the case for a non-Fermi liquid, as we discuss below.

Let us define a ``pre-vertex'', $\gamma_{\theta,\omega}(q,p)$, as shown in Eq.~(\ref{prevertex}) and investigate its singular structure. Here $\theta$ denotes the angle between $\vec{q}$ and $\vec{v_F}$, and $\omega$ denotes the energy of outgoing particle, and the shaded square denotes the fully dressed 4pt interaction. The full density vertex can be obtained from the ``pre-''vertex after integrating over $\theta$. For simplicity, we use the 3-momentum convention with $q\equiv(\Omega,\vec{q})$ and $p\equiv(\omega,\vec{p})$.
\begin{equation}
\label{prevertex}
\begin{aligned}
\begin{tikzpicture}[x=0.75pt,y=0.75pt,yscale=-1,xscale=1]

\draw  [pattern=_4dhvc4nk9,pattern size=6pt,pattern thickness=0.75pt,pattern radius=0pt, pattern color={rgb, 255:red, 0; green, 0; blue, 0}] (315,106.06) -- (331,106.06) -- (331,123.06) -- (315,123.06) -- cycle ;
\draw    (264.15,112.42) .. controls (278.15,85.42) and (304.15,89.42) .. (315,106.06) ;
\draw    (264.15,112.42) .. controls (275.15,139.42) and (303.15,138.42) .. (315,123.06) ;
\draw    (331,106.06) -- (353.49,92.96) ;
\draw    (331,123.06) -- (353.49,134.96) ;
\draw  [fill={rgb, 255:red, 0; green, 0; blue, 0 }  ,fill opacity=1 ] (296.01,93.2) -- (290.34,97.02) -- (290.33,89.4) -- cycle ;
\draw  [fill={rgb, 255:red, 0; green, 0; blue, 0 }  ,fill opacity=1 ] (289.34,133.19) -- (295.04,129.41) -- (294.99,137.04) -- cycle ;
\draw  [fill={rgb, 255:red, 0; green, 0; blue, 0 }  ,fill opacity=1 ] (344.66,98.03) -- (341.82,104.24) -- (337.83,97.74) -- cycle ;
\draw  [fill={rgb, 255:red, 0; green, 0; blue, 0 }  ,fill opacity=1 ] (339.78,127.61) -- (346.59,127.09) -- (342.83,133.73) -- cycle ;
\draw    (254.67,112.73) .. controls (256.28,111.01) and (257.94,110.96) .. (259.66,112.57) -- (264.15,112.42) -- (264.15,112.42) ;

\draw (276,74.4) node [anchor=north west][inner sep=0.75pt]    {$\theta,\omega $};
\draw (242,99.4) node [anchor=north west][inner sep=0.75pt]    {$q$};
\draw (356,79.4) node [anchor=north west][inner sep=0.75pt]    {$p-q$};
\draw (357,126.4) node [anchor=north west][inner sep=0.75pt]    {$p$};
\draw (162,101.4) node [anchor=north west][inner sep=0.75pt]    {$\gamma _{\theta, \omega }( p,q) =$};

\end{tikzpicture}
\end{aligned}
\end{equation}

The equation of motion for $\gamma_{\theta,\omega}(q,p)$ in a Fermi liquid has been derived in classic works \cite{Abrikosov_1959, AGD} by investigating the singular structure of the density vertex. We generalize these computations now to our examples of non-Fermi liquids. Let us decompose the fully dressed 4pt-vertex into components that are singular and regular as a function of the momentum, $\vec{q}$, where $\vec{q}$ is as shown in Eq.(\ref{Gamma}). 

\begin{equation}
\label{Gamma}
\begin{tikzpicture}[x=0.75pt,y=0.75pt,yscale=-1,xscale=1]

\draw  [pattern=_4dhvc4nk9,pattern size=6pt,pattern thickness=0.75pt,pattern radius=0pt, pattern color={rgb, 255:red, 0; green, 0; blue, 0}] (304.13,68.7) -- (317.38,68.7) -- (317.38,82.23) -- (304.13,82.23) -- cycle ;
\draw    (317.38,68.7) -- (336,58.26) ;
\draw    (317.38,82.23) -- (336,91.71) ;
\draw  [fill={rgb, 255:red, 0; green, 0; blue, 0 }  ,fill opacity=1 ] (324.64,64.59) -- (327.17,59.72) -- (330.29,65.01) -- cycle ;
\draw  [fill={rgb, 255:red, 0; green, 0; blue, 0 }  ,fill opacity=1 ] (328.76,88.03) -- (323.13,88.6) -- (326.09,83.23) -- cycle ;
\draw    (286.72,55.52) -- (303.73,68.25) ;
\draw  [fill={rgb, 255:red, 0; green, 0; blue, 0 }  ,fill opacity=1 ] (293.31,60.58) -- (298.97,60.7) -- (295.32,65.67) -- cycle ;
\draw    (285.51,92.67) -- (304.13,82.23) ;
\draw  [fill={rgb, 255:red, 0; green, 0; blue, 0 }  ,fill opacity=1 ] (296.82,86.27) -- (294.47,91.22) -- (291.17,86.05) -- cycle ;
\draw  [color={rgb, 255:red, 0; green, 0; blue, 0 }  ,draw opacity=1 ][fill={rgb, 255:red, 0; green, 0; blue, 0 }  ,fill opacity=1 ] (304.13,136.7) -- (317.38,136.7) -- (317.38,150.23) -- (304.13,150.23) -- cycle ;
\draw    (317.38,136.7) -- (336,126.26) ;
\draw    (317.38,150.23) -- (336,159.71) ;
\draw  [fill={rgb, 255:red, 0; green, 0; blue, 0 }  ,fill opacity=1 ] (324.67,132.63) -- (327.08,127.71) -- (330.32,132.93) -- cycle ;
\draw  [fill={rgb, 255:red, 0; green, 0; blue, 0 }  ,fill opacity=1 ] (328.82,155.9) -- (323.24,156.81) -- (325.87,151.27) -- cycle ;
\draw    (287.12,123.97) -- (304.13,136.7) ;
\draw  [fill={rgb, 255:red, 0; green, 0; blue, 0 }  ,fill opacity=1 ] (293.65,129.12) -- (299.31,128.98) -- (295.89,134.11) -- cycle ;
\draw    (285.51,160.67) -- (304.13,150.23) ;
\draw  [fill={rgb, 255:red, 0; green, 0; blue, 0 }  ,fill opacity=1 ] (296.82,154.27) -- (294.47,159.22) -- (291.17,154.05) -- cycle ;
\draw    (414.33,128.06) .. controls (405.33,147.06) and (405.33,147.06) .. (415,162.06) ;
\draw    (414.33,128.06) .. controls (422.21,141.2) and (428.21,149.2) .. (415,162.06) ;
\draw    (400.21,128.2) -- (414.33,128.06) ;
\draw    (414.33,128.06) -- (428.45,127.92) ;
\draw    (400.88,162.2) -- (415,162.06) ;
\draw    (415,162.06) -- (429.12,161.92) ;

\draw (166,66.4) node [anchor=north west][inner sep=0.75pt]    {$\Gamma ( p_{1} ,p_{2} ,q) =$};
\draw (337,82.4) node [anchor=north west][inner sep=0.75pt]    {$p_{2}$};
\draw (267,84.4) node [anchor=north west][inner sep=0.75pt]    {$p_{1}$};
\draw (341,39.4) node [anchor=north west][inner sep=0.75pt]    {$p_{2} -q$};
\draw (166,133.4) node [anchor=north west][inner sep=0.75pt]    {$\Gamma ^{\Omega }( p_{1} ,p_{2} ,\vec{q}) =$};
\draw (337,150.4) node [anchor=north west][inner sep=0.75pt]    {$p_{2}$};
\draw (274,156.4) node [anchor=north west][inner sep=0.75pt]    {$p_{1}$};
\draw (341,107.4) node [anchor=north west][inner sep=0.75pt]    {$p_{2} -q$};
\draw (370,135.4) node [anchor=north west][inner sep=0.75pt]    {$=$};
\end{tikzpicture}
\end{equation}
Here $\Gamma^{\Omega}(p_1,p_2,\vec{q})$ denotes the regular piece, and we can take the limit of $\vec{q}=0$ to consider the long wavelength electron-hole excitations. 
The Schwinger-Dyson equation for $\Gamma(p_1,p_2,q)$ reads:
\begin{equation}
\label{4ptsd}
\begin{tikzpicture}[x=0.75pt,y=0.75pt,yscale=-1,xscale=1]

\draw  [pattern=_4dhvc4nk9,pattern size=6pt,pattern thickness=0.75pt,pattern radius=0pt, pattern color={rgb, 255:red, 0; green, 0; blue, 0}] (190.13,88.7) -- (203.38,88.7) -- (203.38,102.23) -- (190.13,102.23) -- cycle ;
\draw    (203.38,88.7) -- (222,78.26) ;
\draw    (203.38,102.23) -- (222,111.71) ;
\draw  [fill={rgb, 255:red, 0; green, 0; blue, 0 }  ,fill opacity=1 ] (214.69,82.3) -- (212.33,87.25) -- (209.03,82.07) -- cycle ;
\draw  [fill={rgb, 255:red, 0; green, 0; blue, 0 }  ,fill opacity=1 ] (210.64,105.86) -- (216.29,105.45) -- (213.17,110.73) -- cycle ;
\draw    (172.72,75.52) -- (189.73,88.25) ;
\draw  [fill={rgb, 255:red, 0; green, 0; blue, 0 }  ,fill opacity=1 ] (183.11,83.23) -- (177.46,82.97) -- (181.22,78.1) -- cycle ;
\draw    (171.51,112.67) -- (190.13,102.23) ;
\draw  [fill={rgb, 255:red, 0; green, 0; blue, 0 }  ,fill opacity=1 ] (182.82,106.27) -- (180.47,111.22) -- (177.17,106.05) -- cycle ;
\draw  [color={rgb, 255:red, 0; green, 0; blue, 0 }  ,draw opacity=1 ][fill={rgb, 255:red, 0; green, 0; blue, 0 }  ,fill opacity=1 ] (265.13,90.7) -- (278.38,90.7) -- (278.38,104.23) -- (265.13,104.23) -- cycle ;
\draw    (278.38,90.7) -- (297,80.26) ;
\draw    (278.38,104.23) -- (297,113.71) ;
\draw  [fill={rgb, 255:red, 0; green, 0; blue, 0 }  ,fill opacity=1 ] (289.69,84.3) -- (287.33,89.25) -- (284.03,84.07) -- cycle ;
\draw  [fill={rgb, 255:red, 0; green, 0; blue, 0 }  ,fill opacity=1 ] (285.64,107.86) -- (291.29,107.45) -- (288.17,112.73) -- cycle ;
\draw    (247.72,77.52) -- (264.73,90.25) ;
\draw  [fill={rgb, 255:red, 0; green, 0; blue, 0 }  ,fill opacity=1 ] (258.11,85.23) -- (252.46,84.97) -- (256.22,80.1) -- cycle ;
\draw    (246.51,114.67) -- (265.13,104.23) ;
\draw  [fill={rgb, 255:red, 0; green, 0; blue, 0 }  ,fill opacity=1 ] (257.82,108.27) -- (255.47,113.22) -- (252.17,108.05) -- cycle ;
\draw  [color={rgb, 255:red, 0; green, 0; blue, 0 }  ,draw opacity=1 ][fill={rgb, 255:red, 0; green, 0; blue, 0 }  ,fill opacity=1 ] (339.13,91.7) -- (352.38,91.7) -- (352.38,105.23) -- (339.13,105.23) -- cycle ;
\draw    (321.72,78.52) -- (338.73,91.25) ;
\draw  [fill={rgb, 255:red, 0; green, 0; blue, 0 }  ,fill opacity=1 ] (332.11,86.23) -- (326.46,85.97) -- (330.22,81.1) -- cycle ;
\draw    (320.51,115.67) -- (339.13,105.23) ;
\draw  [fill={rgb, 255:red, 0; green, 0; blue, 0 }  ,fill opacity=1 ] (331.82,109.27) -- (329.47,114.22) -- (326.17,109.05) -- cycle ;
\draw    (352.38,91.7) .. controls (355.55,89.33) and (358.45,87.62) .. (361.16,86.44) .. controls (371.82,81.81) and (379.64,85.48) .. (390,91.06) ;
\draw    (352.38,105.23) .. controls (366,112.06) and (374,112.6) .. (390,104.6) ;
\draw  [pattern=_4dhvc4nk9,pattern size=6pt,pattern thickness=0.75pt,pattern radius=0pt, pattern color={rgb, 255:red, 0; green, 0; blue, 0}] (390,91.06) -- (403.25,91.06) -- (403.25,104.6) -- (390,104.6) -- cycle ;
\draw  [fill={rgb, 255:red, 0; green, 0; blue, 0 }  ,fill opacity=1 ] (372.54,83.8) -- (367.91,87.05) -- (367.91,80.89) -- cycle ;
\draw  [fill={rgb, 255:red, 0; green, 0; blue, 0 }  ,fill opacity=1 ] (367.91,109.92) -- (372.61,106.77) -- (372.48,112.92) -- cycle ;
\draw    (403.38,90.7) -- (422,80.26) ;
\draw    (403.38,104.23) -- (422,113.71) ;
\draw  [fill={rgb, 255:red, 0; green, 0; blue, 0 }  ,fill opacity=1 ] (414.69,84.3) -- (412.33,89.25) -- (409.03,84.07) -- cycle ;
\draw  [fill={rgb, 255:red, 0; green, 0; blue, 0 }  ,fill opacity=1 ] (410.64,107.86) -- (416.29,107.45) -- (413.17,112.73) -- cycle ;

\draw (229,87.4) node [anchor=north west][inner sep=0.75pt]    {$=$};
\draw (299,89.4) node [anchor=north west][inner sep=0.75pt]    {$+$};

\end{tikzpicture}
\end{equation}
The singular structure and the $q-$dependence in $\Gamma(p_1,p_2,q)$ are generated entirely by the singularity in the bare bubble,  $\Pi(q)=\sum_{\k,\omega}G(\k,\omega)G(\k+\q,\omega+\Omega)$ in Eq.(\ref{4ptsd}), when summing over the ladder series. By focusing on the particle-hole excitations near the Fermi surface, we decompose $G(\k,\omega)G(\k+\q,\omega+\Omega)$ into terms proportional to $\delta(\vec{k}-\vec{k_F})$ and regular terms:
\begin{equation}
\label{singularplrz}
\begin{aligned}
    G(\k,\omega)G(\k+\q,\omega+\Omega)&=\frac{\new{\nu}\Omega}{(1-\frac{\partial \Sigma}{\partial \omega})\Omega-v_F |\q| cos \theta}\delta(|\vec{k}|-k_F)+\phi_{reg},
\end{aligned}
\end{equation}
where the regular piece, $\phi_{reg}$, does not affect the pole of $\gamma_{\theta,\omega}(q,p)$, and so we drop it. Plugging the singular part of Eq.(\ref{singularplrz}) into Eq.(\ref{4ptsd}), we have
\begin{equation}
\begin{aligned}
\label{4ptsdsing}
    \Gamma(p_1,p_3,q)
    &=\int_{\theta_2, \omega_2} \Gamma^{\Omega}(p_1,\theta_2,\omega_2)\times
    \bigg[\frac{\new{\nu}\Omega}{(1-\frac{\partial \Sigma}{\partial \omega_2})\Omega-v_F |\q| \tn{cos} \theta_2}\Gamma(\theta_2,\omega_2,p_3,q)\bigg],
\end{aligned}
\end{equation}
where the term inside $[...]$ in Eq.(\ref{4ptsdsing}) is just the pre-vertex defined in Eq.(\ref{prevertex}), with the bubble replaced by its singular part. The equation of motion for $\gamma_{\theta ,\omega}(q,p)$ can be obtained by multiplying both sides of Eq.(\ref{4ptsdsing}) by $\frac{\new{\nu}\Omega}{(1-\frac{\partial \Sigma}{\partial \omega_1})\Omega-v_F |\q| \tn{cos} \theta_1}\delta(|\vec{p_1}|-k_F)$ and integrating over $|\vec{p}_1|$. This yields,
\begin{equation}
\label{devqbe}
\begin{aligned}
    &\bigg[(1-\frac{\partial \Sigma}{\partial \omega})\Omega-v_F |\q| \cos \theta\bigg]\gamma_{\theta, \omega}(p,q)=\Omega\int_{\theta^{'},\omega^{'}} \Gamma^{\Omega}(\theta,\omega;\theta^{'},\omega^{'}) \gamma_{\theta^{'},\omega^{'}}(p,q),
\end{aligned}
\end{equation}
which is precisely the quantum Boltzmann equation in Eq.(\ref{QBE}). To analyze the fully dressed polarizability, $\Pi(\q,\Omega)$, we can similarly derive its equation of motion by introducing an analogous '`pre-polarizability'', shown in Eq.(\ref{preplrz}):
\begin{equation}
\label{preplrz}
    \begin{tikzpicture}[x=0.75pt,y=0.75pt,yscale=-1,xscale=1]

\draw  [pattern=_4dhvc4nk9,pattern size=6pt,pattern thickness=0.75pt,pattern radius=0pt, pattern color={rgb, 255:red, 0; green, 0; blue, 0}] (335,126.06) -- (351,126.06) -- (351,143.06) -- (335,143.06) -- cycle ;
\draw    (284.15,132.42) .. controls (298.15,105.42) and (324.15,109.42) .. (335,126.06) ;
\draw    (284.15,132.42) .. controls (295.15,159.42) and (323.15,158.42) .. (335,143.06) ;
\draw  [fill={rgb, 255:red, 0; green, 0; blue, 0 }  ,fill opacity=1 ] (316.01,113.2) -- (310.34,117.02) -- (310.33,109.4) -- cycle ;
\draw  [fill={rgb, 255:red, 0; green, 0; blue, 0 }  ,fill opacity=1 ] (309.34,153.19) -- (315.04,149.41) -- (314.99,157.04) -- cycle ;
\draw  [fill={rgb, 255:red, 0; green, 0; blue, 0 }  ,fill opacity=1 ] (378.9,152.21) -- (378.69,159.04) -- (372.49,154.59) -- cycle ;
\draw    (274.67,132.73) .. controls (276.28,131.01) and (277.94,130.96) .. (279.66,132.57) -- (284.15,132.42) -- (284.15,132.42) ;
\draw    (401.06,132.32) .. controls (393.06,161.32) and (364.06,158.32) .. (351,143.06) ;
\draw    (401.06,132.32) .. controls (389.06,109.32) and (368.06,108.32) .. (351,126.06) ;
\draw  [fill={rgb, 255:red, 0; green, 0; blue, 0 }  ,fill opacity=1 ] (379.53,113.27) -- (373.95,117.23) -- (373.76,109.61) -- cycle ;
\draw    (401.06,132.32) .. controls (402.67,130.6) and (404.34,130.55) .. (406.06,132.16) -- (410.54,132.01) -- (410.54,132.01) ;

\draw (296,92.4) node [anchor=north west][inner sep=0.75pt]    {$\theta $};
\draw (262,119.4) node [anchor=north west][inner sep=0.75pt]    {$q$};
\draw (194,121.4) node [anchor=north west][inner sep=0.75pt]    {$\Pi _{\theta }( q) =$};

\end{tikzpicture}
\end{equation}
The corresponding equation of motion is then given by,
\begin{equation}
\label{eomplrz}
    \bigg[(1-\frac{\partial \Sigma}{\partial \Omega})\Omega-v_F q \cos \theta\bigg]\Pi_{\theta}(q)=\Omega + \Omega\int_{\theta^{'}} \Gamma^{\Omega}(\theta,\theta^{'}), \Pi_{\theta^{'}}(q),
\end{equation}
where the $\Omega$ piece in the first term on the RHS of Eq.(\ref{eomplrz}) denotes the contribution from the bare 'pre-polarizability'. The $\Gamma^{\Omega}(\theta,\theta^{'})$ term in Eq.(\ref{eomplrz}) and Eq.(\ref{devqbe}) are the effective Landau interaction for QBE; we can also treat the effect of $\Gamma^{\Omega}(\theta,\theta^{'})$ perturbatively and obtain a diagrammatic expansion for the pre-polarizability and pre-vertex to obtain the associated Bethe-Salpeter equation.

\bibliographystyle{apsrev4-1_custom}
\bibliography{NFL.bib}

\begin{thebibliography}{63}%
\makeatletter
\providecommand \@ifxundefined [1]{%
 \@ifx{#1\undefined}
}%
\providecommand \@ifnum [1]{%
 \ifnum #1\expandafter \@firstoftwo
 \else \expandafter \@secondoftwo
 \fi
}%
\providecommand \@ifx [1]{%
 \ifx #1\expandafter \@firstoftwo
 \else \expandafter \@secondoftwo
 \fi
}%
\providecommand \natexlab [1]{#1}%
\providecommand \enquote  [1]{``#1''}%
\providecommand \bibnamefont  [1]{#1}%
\providecommand \bibfnamefont [1]{#1}%
\providecommand \citenamefont [1]{#1}%
\providecommand \href@noop [0]{\@secondoftwo}%
\providecommand \href [0]{\begingroup \@sanitize@url \@href}%
\providecommand \@href[1]{\@@startlink{#1}\@@href}%
\providecommand \@@href[1]{\endgroup#1\@@endlink}%
\providecommand \@sanitize@url [0]{\catcode `\\12\catcode `\$12\catcode
  `\&12\catcode `\#12\catcode `\^12\catcode `\_12\catcode `\%12\relax}%
\providecommand \@@startlink[1]{}%
\providecommand \@@endlink[0]{}%
\providecommand \url  [0]{\begingroup\@sanitize@url \@url }%
\providecommand \@url [1]{\endgroup\@href {#1}{\urlprefix }}%
\providecommand \urlprefix  [0]{URL }%
\providecommand \Eprint [0]{\href }%
\providecommand \doibase [0]{http://dx.doi.org/}%
\providecommand \selectlanguage [0]{\@gobble}%
\providecommand \bibinfo  [0]{\@secondoftwo}%
\providecommand \bibfield  [0]{\@secondoftwo}%
\providecommand \translation [1]{[#1]}%
\providecommand \BibitemOpen [0]{}%
\providecommand \bibitemStop [0]{}%
\providecommand \bibitemNoStop [0]{.\EOS\space}%
\providecommand \EOS [0]{\spacefactor3000\relax}%
\providecommand \BibitemShut  [1]{\csname bibitem#1\endcsname}%
\let\auto@bib@innerbib\@empty
\bibitem [{\citenamefont {Abrikosov}\ \emph {et~al.}(1975)\citenamefont
  {Abrikosov}, \citenamefont {Dzyaloshinskii}, \citenamefont {Gorkov},\ and\
  \citenamefont {Silverman}}]{AGD}%
  \BibitemOpen
  \bibfield  {author} {\bibinfo {author} {\bibfnamefont {A.~A.}\ \bibnamefont
  {Abrikosov}}, \bibinfo {author} {\bibfnamefont {I.}~\bibnamefont
  {Dzyaloshinskii}}, \bibinfo {author} {\bibfnamefont {L.~P.}\ \bibnamefont
  {Gorkov}}, \ and\ \bibinfo {author} {\bibfnamefont {R.~A.}\ \bibnamefont
  {Silverman}},\ }\href {https://cds.cern.ch/record/107441} {\emph {\bibinfo
  {title} {{Methods of quantum field theory in statistical physics}}}}\
  (\bibinfo  {publisher} {Dover},\ \bibinfo {address} {New York, NY},\ \bibinfo
  {year} {1975})\BibitemShut {NoStop}%
\bibitem [{\citenamefont {Pines}\ and\ \citenamefont
  {Nozi\`eres}(2018)}]{pines}%
  \BibitemOpen
  \bibfield  {author} {\bibinfo {author} {\bibfnamefont {D.}~\bibnamefont
  {Pines}}\ and\ \bibinfo {author} {\bibfnamefont {P.}~\bibnamefont
  {Nozi\`eres}},\ }\href@noop {} {\emph {\bibinfo {title} {Theory of Quantum
  Liquids: Normal Fermi Liquids}}}\ (\bibinfo  {publisher} {CRC Press},\
  \bibinfo {year} {2018})\BibitemShut {NoStop}%
\bibitem [{\citenamefont {Senthil}(2008)}]{senthil2008critical}%
  \BibitemOpen
  \bibfield  {author} {\bibinfo {author} {\bibfnamefont {T.}~\bibnamefont
  {Senthil}},\ }\bibfield  {title} {\enquote {\bibinfo {title} {Critical fermi
  surfaces and non-fermi liquid metals},}\ }\href@noop {} {\bibfield  {journal}
  {\bibinfo  {journal} {Physical Review B}\ }\textbf {\bibinfo {volume} {78}},\
  \bibinfo {pages} {035103} (\bibinfo {year} {2008})}\BibitemShut {NoStop}%
\bibitem [{\citenamefont {Arovas}\ \emph {et~al.}(2022)\citenamefont {Arovas},
  \citenamefont {Berg}, \citenamefont {Kivelson},\ and\ \citenamefont
  {Raghu}}]{hubbard_review}%
  \BibitemOpen
  \bibfield  {author} {\bibinfo {author} {\bibfnamefont {D.~P.}\ \bibnamefont
  {Arovas}}, \bibinfo {author} {\bibfnamefont {E.}~\bibnamefont {Berg}},
  \bibinfo {author} {\bibfnamefont {S.~A.}\ \bibnamefont {Kivelson}}, \ and\
  \bibinfo {author} {\bibfnamefont {S.}~\bibnamefont {Raghu}},\ }\bibfield
  {title} {\enquote {\bibinfo {title} {The hubbard model},}\ }\href {\doibase
  10.1146/annurev-conmatphys-031620-102024} {\bibfield  {journal} {\bibinfo
  {journal} {Annual Review of Condensed Matter Physics}\ }\textbf {\bibinfo
  {volume} {13}},\ \bibinfo {pages} {239} (\bibinfo {year} {2022})}\BibitemShut
  {NoStop}%
\bibitem [{\citenamefont {Lee}\ \emph {et~al.}(2006)\citenamefont {Lee},
  \citenamefont {Nagaosa},\ and\ \citenamefont {Wen}}]{LNW}%
  \BibitemOpen
  \bibfield  {author} {\bibinfo {author} {\bibfnamefont {P.~A.}\ \bibnamefont
  {Lee}}, \bibinfo {author} {\bibfnamefont {N.}~\bibnamefont {Nagaosa}}, \ and\
  \bibinfo {author} {\bibfnamefont {X.-G.}\ \bibnamefont {Wen}},\ }\bibfield
  {title} {\enquote {\bibinfo {title} {Doping a mott insulator: Physics of
  high-temperature superconductivity},}\ }\href {\doibase
  10.1103/RevModPhys.78.17} {\bibfield  {journal} {\bibinfo  {journal} {Rev.
  Mod. Phys.}\ }\textbf {\bibinfo {volume} {78}},\ \bibinfo {pages} {17}
  (\bibinfo {year} {2006})}\BibitemShut {NoStop}%
\bibitem [{\citenamefont {Joshi}\ and\ \citenamefont {Sachdev}(2020)}]{joshi}%
  \BibitemOpen
  \bibfield  {author} {\bibinfo {author} {\bibfnamefont {D.~G.}\ \bibnamefont
  {Joshi}}\ and\ \bibinfo {author} {\bibfnamefont {S.}~\bibnamefont
  {Sachdev}},\ }\bibfield  {title} {\enquote {\bibinfo {title} {Anomalous
  density fluctuations in a random $t\text{\ensuremath{-}}j$ model},}\ }\href
  {\doibase 10.1103/PhysRevB.102.165146} {\bibfield  {journal} {\bibinfo
  {journal} {Phys. Rev. B}\ }\textbf {\bibinfo {volume} {102}},\ \bibinfo
  {pages} {165146} (\bibinfo {year} {2020})}\BibitemShut {NoStop}%
\bibitem [{\citenamefont {Romero-Berm\'udez}\ \emph {et~al.}(2019)\citenamefont
  {Romero-Berm\'udez}, \citenamefont {Krikun}, \citenamefont {Schalm},\ and\
  \citenamefont {Zaanen}}]{zaanen}%
  \BibitemOpen
  \bibfield  {author} {\bibinfo {author} {\bibfnamefont {A.}~\bibnamefont
  {Romero-Berm\'udez}}, \bibinfo {author} {\bibfnamefont {A.}~\bibnamefont
  {Krikun}}, \bibinfo {author} {\bibfnamefont {K.}~\bibnamefont {Schalm}}, \
  and\ \bibinfo {author} {\bibfnamefont {J.}~\bibnamefont {Zaanen}},\
  }\bibfield  {title} {\enquote {\bibinfo {title} {Anomalous attenuation of
  plasmons in strange metals and holography},}\ }\href {\doibase
  10.1103/PhysRevB.99.235149} {\bibfield  {journal} {\bibinfo  {journal} {Phys.
  Rev. B}\ }\textbf {\bibinfo {volume} {99}},\ \bibinfo {pages} {235149}
  (\bibinfo {year} {2019})}\BibitemShut {NoStop}%
\bibitem [{\citenamefont {Vig}\ \emph {et~al.}(2017)\citenamefont {Vig},
  \citenamefont {Kogar}, \citenamefont {Mitrano}, \citenamefont {Husain},
  \citenamefont {Mishra}, \citenamefont {Rak}, \citenamefont {Venema},
  \citenamefont {Johnson}, \citenamefont {Gu}, \citenamefont {Fradkin},
  \citenamefont {Norman},\ and\ \citenamefont {Abbamonte}}]{MEELS}%
  \BibitemOpen
  \bibfield  {author} {\bibinfo {author} {\bibfnamefont {S.}~\bibnamefont
  {Vig}}, \bibinfo {author} {\bibfnamefont {A.}~\bibnamefont {Kogar}}, \bibinfo
  {author} {\bibfnamefont {M.}~\bibnamefont {Mitrano}}, \bibinfo {author}
  {\bibfnamefont {A.~A.}\ \bibnamefont {Husain}}, \bibinfo {author}
  {\bibfnamefont {V.}~\bibnamefont {Mishra}}, \bibinfo {author} {\bibfnamefont
  {M.~S.}\ \bibnamefont {Rak}}, \bibinfo {author} {\bibfnamefont
  {L.}~\bibnamefont {Venema}}, \bibinfo {author} {\bibfnamefont {P.~D.}\
  \bibnamefont {Johnson}}, \bibinfo {author} {\bibfnamefont {G.~D.}\
  \bibnamefont {Gu}}, \bibinfo {author} {\bibfnamefont {E.}~\bibnamefont
  {Fradkin}}, \bibinfo {author} {\bibfnamefont {M.~R.}\ \bibnamefont {Norman}},
  \ and\ \bibinfo {author} {\bibfnamefont {P.}~\bibnamefont {Abbamonte}},\
  }\bibfield  {title} {\enquote {\bibinfo {title} {{Measurement of the dynamic
  charge response of materials using low-energy, momentum-resolved electron
  energy-loss spectroscopy (M-EELS)}},}\ }\href {\doibase
  10.21468/SciPostPhys.3.4.026} {\bibfield  {journal} {\bibinfo  {journal}
  {SciPost Phys.}\ }\textbf {\bibinfo {volume} {3}},\ \bibinfo {pages} {026}
  (\bibinfo {year} {2017})}\BibitemShut {NoStop}%
\bibitem [{\citenamefont {Mitrano}\ \emph {et~al.}(2018)\citenamefont
  {Mitrano}, \citenamefont {Husain}, \citenamefont {Vig}, \citenamefont
  {Kogar}, \citenamefont {Rak}, \citenamefont {Rubeck}, \citenamefont
  {Schmalian}, \citenamefont {Uchoa}, \citenamefont {Schneeloch}, \citenamefont
  {Zhong}, \citenamefont {Gu},\ and\ \citenamefont {Abbamonte}}]{Abbamonte1}%
  \BibitemOpen
  \bibfield  {author} {\bibinfo {author} {\bibfnamefont {M.}~\bibnamefont
  {Mitrano}}, \bibinfo {author} {\bibfnamefont {A.~A.}\ \bibnamefont {Husain}},
  \bibinfo {author} {\bibfnamefont {S.}~\bibnamefont {Vig}}, \bibinfo {author}
  {\bibfnamefont {A.}~\bibnamefont {Kogar}}, \bibinfo {author} {\bibfnamefont
  {M.~S.}\ \bibnamefont {Rak}}, \bibinfo {author} {\bibfnamefont {S.~I.}\
  \bibnamefont {Rubeck}}, \bibinfo {author} {\bibfnamefont {J.}~\bibnamefont
  {Schmalian}}, \bibinfo {author} {\bibfnamefont {B.}~\bibnamefont {Uchoa}},
  \bibinfo {author} {\bibfnamefont {J.}~\bibnamefont {Schneeloch}}, \bibinfo
  {author} {\bibfnamefont {R.}~\bibnamefont {Zhong}}, \bibinfo {author}
  {\bibfnamefont {G.~D.}\ \bibnamefont {Gu}}, \ and\ \bibinfo {author}
  {\bibfnamefont {P.}~\bibnamefont {Abbamonte}},\ }\bibfield  {title} {\enquote
  {\bibinfo {title} {Anomalous density fluctuations in a strange metal},}\
  }\href {\doibase 10.1073/pnas.1721495115} {\bibfield  {journal} {\bibinfo
  {journal} {Proceedings of the National Academy of Sciences}\ }\textbf
  {\bibinfo {volume} {115}},\ \bibinfo {pages} {5392} (\bibinfo {year}
  {2018})}\BibitemShut {NoStop}%
\bibitem [{\citenamefont {Husain}\ \emph {et~al.}(2019)\citenamefont {Husain},
  \citenamefont {Mitrano}, \citenamefont {Rak}, \citenamefont {Rubeck},
  \citenamefont {Uchoa}, \citenamefont {March}, \citenamefont {Dwyer},
  \citenamefont {Schneeloch}, \citenamefont {Zhong}, \citenamefont {Gu},\ and\
  \citenamefont {Abbamonte}}]{Abbamonte2}%
  \BibitemOpen
  \bibfield  {author} {\bibinfo {author} {\bibfnamefont {A.~A.}\ \bibnamefont
  {Husain}}, \bibinfo {author} {\bibfnamefont {M.}~\bibnamefont {Mitrano}},
  \bibinfo {author} {\bibfnamefont {M.~S.}\ \bibnamefont {Rak}}, \bibinfo
  {author} {\bibfnamefont {S.}~\bibnamefont {Rubeck}}, \bibinfo {author}
  {\bibfnamefont {B.}~\bibnamefont {Uchoa}}, \bibinfo {author} {\bibfnamefont
  {K.}~\bibnamefont {March}}, \bibinfo {author} {\bibfnamefont
  {C.}~\bibnamefont {Dwyer}}, \bibinfo {author} {\bibfnamefont
  {J.}~\bibnamefont {Schneeloch}}, \bibinfo {author} {\bibfnamefont
  {R.}~\bibnamefont {Zhong}}, \bibinfo {author} {\bibfnamefont {G.~D.}\
  \bibnamefont {Gu}}, \ and\ \bibinfo {author} {\bibfnamefont {P.}~\bibnamefont
  {Abbamonte}},\ }\bibfield  {title} {\enquote {\bibinfo {title} {Crossover of
  charge fluctuations across the strange metal phase diagram},}\ }\href
  {\doibase 10.1103/PhysRevX.9.041062} {\bibfield  {journal} {\bibinfo
  {journal} {Phys. Rev. X}\ }\textbf {\bibinfo {volume} {9}},\ \bibinfo {pages}
  {041062} (\bibinfo {year} {2019})}\BibitemShut {NoStop}%
\bibitem [{\citenamefont {Levallois}\ \emph {et~al.}(2016)\citenamefont
  {Levallois}, \citenamefont {Tran}, \citenamefont {Pouliot}, \citenamefont
  {Presura}, \citenamefont {Greene}, \citenamefont {Eckstein}, \citenamefont
  {Uccelli}, \citenamefont {Giannini}, \citenamefont {Gu}, \citenamefont
  {Leggett},\ and\ \citenamefont {van~der Marel}}]{Marel16}%
  \BibitemOpen
  \bibfield  {author} {\bibinfo {author} {\bibfnamefont {J.}~\bibnamefont
  {Levallois}}, \bibinfo {author} {\bibfnamefont {M.~K.}\ \bibnamefont {Tran}},
  \bibinfo {author} {\bibfnamefont {D.}~\bibnamefont {Pouliot}}, \bibinfo
  {author} {\bibfnamefont {C.~N.}\ \bibnamefont {Presura}}, \bibinfo {author}
  {\bibfnamefont {L.~H.}\ \bibnamefont {Greene}}, \bibinfo {author}
  {\bibfnamefont {J.~N.}\ \bibnamefont {Eckstein}}, \bibinfo {author}
  {\bibfnamefont {J.}~\bibnamefont {Uccelli}}, \bibinfo {author} {\bibfnamefont
  {E.}~\bibnamefont {Giannini}}, \bibinfo {author} {\bibfnamefont {G.~D.}\
  \bibnamefont {Gu}}, \bibinfo {author} {\bibfnamefont {A.~J.}\ \bibnamefont
  {Leggett}}, \ and\ \bibinfo {author} {\bibfnamefont {D.}~\bibnamefont
  {van~der Marel}},\ }\bibfield  {title} {\enquote {\bibinfo {title}
  {Temperature-dependent ellipsometry measurements of partial coulomb energy in
  superconducting cuprates},}\ }\href {\doibase 10.1103/PhysRevX.6.031027}
  {\bibfield  {journal} {\bibinfo  {journal} {Phys. Rev. X}\ }\textbf {\bibinfo
  {volume} {6}},\ \bibinfo {pages} {031027} (\bibinfo {year}
  {2016})}\BibitemShut {NoStop}%
\bibitem [{\citenamefont {Nag}\ \emph {et~al.}(2020)\citenamefont {Nag},
  \citenamefont {Zhu}, \citenamefont {Bejas}, \citenamefont {Li}, \citenamefont
  {Robarts}, \citenamefont {Yamase}, \citenamefont {Petsch}, \citenamefont
  {Song}, \citenamefont {Eisaki}, \citenamefont {Walters}, \citenamefont
  {Garc\'{\i}a-Fern\'andez}, \citenamefont {Greco}, \citenamefont {Hayden},\
  and\ \citenamefont {Zhou}}]{Hayden20}%
  \BibitemOpen
  \bibfield  {author} {\bibinfo {author} {\bibfnamefont {A.}~\bibnamefont
  {Nag}}, \bibinfo {author} {\bibfnamefont {M.}~\bibnamefont {Zhu}}, \bibinfo
  {author} {\bibfnamefont {M.}~\bibnamefont {Bejas}}, \bibinfo {author}
  {\bibfnamefont {J.}~\bibnamefont {Li}}, \bibinfo {author} {\bibfnamefont
  {H.~C.}\ \bibnamefont {Robarts}}, \bibinfo {author} {\bibfnamefont
  {H.}~\bibnamefont {Yamase}}, \bibinfo {author} {\bibfnamefont {A.~N.}\
  \bibnamefont {Petsch}}, \bibinfo {author} {\bibfnamefont {D.}~\bibnamefont
  {Song}}, \bibinfo {author} {\bibfnamefont {H.}~\bibnamefont {Eisaki}},
  \bibinfo {author} {\bibfnamefont {A.~C.}\ \bibnamefont {Walters}}, \bibinfo
  {author} {\bibfnamefont {M.}~\bibnamefont {Garc\'{\i}a-Fern\'andez}},
  \bibinfo {author} {\bibfnamefont {A.}~\bibnamefont {Greco}}, \bibinfo
  {author} {\bibfnamefont {S.~M.}\ \bibnamefont {Hayden}}, \ and\ \bibinfo
  {author} {\bibfnamefont {K.-J.}\ \bibnamefont {Zhou}},\ }\bibfield  {title}
  {\enquote {\bibinfo {title} {Detection of acoustic plasmons in hole-doped
  lanthanum and bismuth cuprate superconductors using resonant inelastic x-ray
  scattering},}\ }\href {\doibase 10.1103/PhysRevLett.125.257002} {\bibfield
  {journal} {\bibinfo  {journal} {Phys. Rev. Lett.}\ }\textbf {\bibinfo
  {volume} {125}},\ \bibinfo {pages} {257002} (\bibinfo {year}
  {2020})}\BibitemShut {NoStop}%
\bibitem [{\citenamefont {Bozovic}\ \emph {et~al.}(1987)\citenamefont
  {Bozovic}, \citenamefont {Kirillov}, \citenamefont {Kapitulnik},
  \citenamefont {Char}, \citenamefont {Hahn}, \citenamefont {Beasley},
  \citenamefont {Geballe}, \citenamefont {Kim},\ and\ \citenamefont
  {Heeger}}]{bozovic87}%
  \BibitemOpen
  \bibfield  {author} {\bibinfo {author} {\bibfnamefont {I.}~\bibnamefont
  {Bozovic}}, \bibinfo {author} {\bibfnamefont {D.}~\bibnamefont {Kirillov}},
  \bibinfo {author} {\bibfnamefont {A.}~\bibnamefont {Kapitulnik}}, \bibinfo
  {author} {\bibfnamefont {K.}~\bibnamefont {Char}}, \bibinfo {author}
  {\bibfnamefont {M.~R.}\ \bibnamefont {Hahn}}, \bibinfo {author}
  {\bibfnamefont {M.~R.}\ \bibnamefont {Beasley}}, \bibinfo {author}
  {\bibfnamefont {T.~H.}\ \bibnamefont {Geballe}}, \bibinfo {author}
  {\bibfnamefont {Y.~H.}\ \bibnamefont {Kim}}, \ and\ \bibinfo {author}
  {\bibfnamefont {A.~J.}\ \bibnamefont {Heeger}},\ }\bibfield  {title}
  {\enquote {\bibinfo {title} {Optical measurements on oriented thin
  ${\mathrm{yba}}_{2}$${\mathrm{cu}}_{3}$${\mathrm{o}}_{7\mathrm{\ensuremath{-}}\mathrm{\ensuremath{\delta}}}$
  films: Lack of evidence for excitonic superconductivity},}\ }\href {\doibase
  10.1103/PhysRevLett.59.2219} {\bibfield  {journal} {\bibinfo  {journal}
  {Phys. Rev. Lett.}\ }\textbf {\bibinfo {volume} {59}},\ \bibinfo {pages}
  {2219} (\bibinfo {year} {1987})}\BibitemShut {NoStop}%
\bibitem [{\citenamefont {Slakey}\ \emph {et~al.}(1991)\citenamefont {Slakey},
  \citenamefont {Klein}, \citenamefont {Rice},\ and\ \citenamefont
  {Ginsberg}}]{ginsberg91}%
  \BibitemOpen
  \bibfield  {author} {\bibinfo {author} {\bibfnamefont {F.}~\bibnamefont
  {Slakey}}, \bibinfo {author} {\bibfnamefont {M.~V.}\ \bibnamefont {Klein}},
  \bibinfo {author} {\bibfnamefont {J.~P.}\ \bibnamefont {Rice}}, \ and\
  \bibinfo {author} {\bibfnamefont {D.~M.}\ \bibnamefont {Ginsberg}},\
  }\bibfield  {title} {\enquote {\bibinfo {title} {Raman investigation of the
  ${\mathrm{yba}}_{2}$${\mathrm{cu}}_{3}$${\mathrm{o}}_{7}$ imaginary response
  function},}\ }\href {\doibase 10.1103/PhysRevB.43.3764} {\bibfield  {journal}
  {\bibinfo  {journal} {Phys. Rev. B}\ }\textbf {\bibinfo {volume} {43}},\
  \bibinfo {pages} {3764} (\bibinfo {year} {1991})}\BibitemShut {NoStop}%
\bibitem [{\citenamefont {{Husain}}\ \emph {et~al.}(2020)\citenamefont
  {{Husain}}, \citenamefont {{Huang}}, \citenamefont {{Mitrano}}, \citenamefont
  {{Rak}}, \citenamefont {{Rubeck}}, \citenamefont {{Guo}}, \citenamefont
  {{Yang}}, \citenamefont {{Sow}}, \citenamefont {{Maeno}}, \citenamefont
  {{Uchoa}}, \citenamefont {{Chiang}}, \citenamefont {{Batson}}, \citenamefont
  {{Phillips}},\ and\ \citenamefont {{Abbamonte}}}]{Abbamonte3}%
  \BibitemOpen
  \bibfield  {author} {\bibinfo {author} {\bibfnamefont {A.~A.}\ \bibnamefont
  {{Husain}}}, \bibinfo {author} {\bibfnamefont {E.~W.}\ \bibnamefont
  {{Huang}}}, \bibinfo {author} {\bibfnamefont {M.}~\bibnamefont {{Mitrano}}},
  \bibinfo {author} {\bibfnamefont {M.~S.}\ \bibnamefont {{Rak}}}, \bibinfo
  {author} {\bibfnamefont {S.~I.}\ \bibnamefont {{Rubeck}}}, \bibinfo {author}
  {\bibfnamefont {X.}~\bibnamefont {{Guo}}}, \bibinfo {author} {\bibfnamefont
  {H.}~\bibnamefont {{Yang}}}, \bibinfo {author} {\bibfnamefont
  {C.}~\bibnamefont {{Sow}}}, \bibinfo {author} {\bibfnamefont
  {Y.}~\bibnamefont {{Maeno}}}, \bibinfo {author} {\bibfnamefont
  {B.}~\bibnamefont {{Uchoa}}}, \bibinfo {author} {\bibfnamefont {T.~C.}\
  \bibnamefont {{Chiang}}}, \bibinfo {author} {\bibfnamefont {P.~E.}\
  \bibnamefont {{Batson}}}, \bibinfo {author} {\bibfnamefont {P.~W.}\
  \bibnamefont {{Phillips}}}, \ and\ \bibinfo {author} {\bibfnamefont
  {P.}~\bibnamefont {{Abbamonte}}},\ }\bibfield  {title} {\enquote {\bibinfo
  {title} {{Observation of Pines' Demon in Sr$_2$RuO$_4$}},}\ }\href@noop {}
  {\bibfield  {journal} {\bibinfo  {journal} {arXiv e-prints}\ ,\ \bibinfo
  {eid} {arXiv:2007.06670}} (\bibinfo {year} {2020})},\ \Eprint
  {http://arxiv.org/abs/2007.06670} {arXiv:2007.06670 [cond-mat.str-el]}
  \BibitemShut {NoStop}%
\bibitem [{\citenamefont {{Chowdhury}}\ \emph {et~al.}(2021)\citenamefont
  {{Chowdhury}}, \citenamefont {{Georges}}, \citenamefont {{Parcollet}},\ and\
  \citenamefont {{Sachdev}}}]{DCrmp}%
  \BibitemOpen
  \bibfield  {author} {\bibinfo {author} {\bibfnamefont {D.}~\bibnamefont
  {{Chowdhury}}}, \bibinfo {author} {\bibfnamefont {A.}~\bibnamefont
  {{Georges}}}, \bibinfo {author} {\bibfnamefont {O.}~\bibnamefont
  {{Parcollet}}}, \ and\ \bibinfo {author} {\bibfnamefont {S.}~\bibnamefont
  {{Sachdev}}},\ }\bibfield  {title} {\enquote {\bibinfo {title}
  {{Sachdev-Ye-Kitaev Models and Beyond: A Window into Non-Fermi Liquids}},}\
  }\href@noop {} {\bibfield  {journal} {\bibinfo  {journal} {arXiv e-prints}\
  ,\ \bibinfo {eid} {arXiv:2109.05037}} (\bibinfo {year} {2021})},\ \Eprint
  {http://arxiv.org/abs/2109.05037} {arXiv:2109.05037 [cond-mat.str-el]}
  \BibitemShut {NoStop}%
\bibitem [{\citenamefont {Sachdev}\ and\ \citenamefont {Ye}(1993)}]{SYK1}%
  \BibitemOpen
  \bibfield  {author} {\bibinfo {author} {\bibfnamefont {S.}~\bibnamefont
  {Sachdev}}\ and\ \bibinfo {author} {\bibfnamefont {J.}~\bibnamefont {Ye}},\
  }\bibfield  {title} {\enquote {\bibinfo {title} {Gapless spin-fluid ground
  state in a random quantum heisenberg magnet},}\ }\href {\doibase
  10.1103/PhysRevLett.70.3339} {\bibfield  {journal} {\bibinfo  {journal}
  {Phys. Rev. Lett.}\ }\textbf {\bibinfo {volume} {70}},\ \bibinfo {pages}
  {3339} (\bibinfo {year} {1993})}\BibitemShut {NoStop}%
\bibitem [{\citenamefont {Kitaev}(2015)}]{SYK2}%
  \BibitemOpen
  \bibfield  {author} {\bibinfo {author} {\bibfnamefont {A.}~\bibnamefont
  {Kitaev}},\ }\bibfield  {title} {\enquote {\bibinfo {title} {Proceedings of
  kitp: A simple model of quantum holography, entanglement in strongly
  correlated quantum matter},}\ }\href@noop {} {\  (\bibinfo {year}
  {2015})}\BibitemShut {NoStop}%
\bibitem [{\citenamefont {Fu}\ and\ \citenamefont {Sachdev}(2016)}]{SYK3}%
  \BibitemOpen
  \bibfield  {author} {\bibinfo {author} {\bibfnamefont {W.}~\bibnamefont
  {Fu}}\ and\ \bibinfo {author} {\bibfnamefont {S.}~\bibnamefont {Sachdev}},\
  }\bibfield  {title} {\enquote {\bibinfo {title} {Numerical study of fermion
  and boson models with infinite-range random interactions},}\ }\href {\doibase
  10.1103/PhysRevB.94.035135} {\bibfield  {journal} {\bibinfo  {journal} {Phys.
  Rev. B}\ }\textbf {\bibinfo {volume} {94}},\ \bibinfo {pages} {035135}
  (\bibinfo {year} {2016})}\BibitemShut {NoStop}%
\bibitem [{\citenamefont {Georges}\ \emph {et~al.}(2001)\citenamefont
  {Georges}, \citenamefont {Parcollet},\ and\ \citenamefont {Sachdev}}]{SYK4}%
  \BibitemOpen
  \bibfield  {author} {\bibinfo {author} {\bibfnamefont {A.}~\bibnamefont
  {Georges}}, \bibinfo {author} {\bibfnamefont {O.}~\bibnamefont {Parcollet}},
  \ and\ \bibinfo {author} {\bibfnamefont {S.}~\bibnamefont {Sachdev}},\
  }\bibfield  {title} {\enquote {\bibinfo {title} {Quantum fluctuations of a
  nearly critical heisenberg spin glass},}\ }\href {\doibase
  10.1103/PhysRevB.63.134406} {\bibfield  {journal} {\bibinfo  {journal} {Phys.
  Rev. B}\ }\textbf {\bibinfo {volume} {63}},\ \bibinfo {pages} {134406}
  (\bibinfo {year} {2001})}\BibitemShut {NoStop}%
\bibitem [{\citenamefont {Maldacena}\ and\ \citenamefont
  {Stanford}(2016)}]{SYK5}%
  \BibitemOpen
  \bibfield  {author} {\bibinfo {author} {\bibfnamefont {J.}~\bibnamefont
  {Maldacena}}\ and\ \bibinfo {author} {\bibfnamefont {D.}~\bibnamefont
  {Stanford}},\ }\bibfield  {title} {\enquote {\bibinfo {title} {Remarks on the
  sachdev-ye-kitaev model},}\ }\href {\doibase 10.1103/PhysRevD.94.106002}
  {\bibfield  {journal} {\bibinfo  {journal} {Phys. Rev. D}\ }\textbf {\bibinfo
  {volume} {94}},\ \bibinfo {pages} {106002} (\bibinfo {year}
  {2016})}\BibitemShut {NoStop}%
\bibitem [{\citenamefont {Sachdev}(2015)}]{SYK6}%
  \BibitemOpen
  \bibfield  {author} {\bibinfo {author} {\bibfnamefont {S.}~\bibnamefont
  {Sachdev}},\ }\bibfield  {title} {\enquote {\bibinfo {title}
  {Bekenstein-hawking entropy and strange metals},}\ }\href {\doibase
  10.1103/PhysRevX.5.041025} {\bibfield  {journal} {\bibinfo  {journal} {Phys.
  Rev. X}\ }\textbf {\bibinfo {volume} {5}},\ \bibinfo {pages} {041025}
  (\bibinfo {year} {2015})}\BibitemShut {NoStop}%
\bibitem [{\citenamefont {Parcollet}\ and\ \citenamefont
  {Georges}(1999)}]{Parcollet1}%
  \BibitemOpen
  \bibfield  {author} {\bibinfo {author} {\bibfnamefont {O.}~\bibnamefont
  {Parcollet}}\ and\ \bibinfo {author} {\bibfnamefont {A.}~\bibnamefont
  {Georges}},\ }\bibfield  {title} {\enquote {\bibinfo {title}
  {{Non-Fermi-liquid regime of a doped Mott insulator}},}\ }\href {\doibase
  10.1103/PhysRevB.59.5341} {\bibfield  {journal} {\bibinfo  {journal} {Phys.
  Rev. B}\ }\textbf {\bibinfo {volume} {59}},\ \bibinfo {pages} {5341}
  (\bibinfo {year} {1999})}\BibitemShut {NoStop}%
\bibitem [{\citenamefont {Song}\ \emph {et~al.}(2017)\citenamefont {Song},
  \citenamefont {Jian},\ and\ \citenamefont {Balents}}]{Balents}%
  \BibitemOpen
  \bibfield  {author} {\bibinfo {author} {\bibfnamefont {X.-Y.}\ \bibnamefont
  {Song}}, \bibinfo {author} {\bibfnamefont {C.-M.}\ \bibnamefont {Jian}}, \
  and\ \bibinfo {author} {\bibfnamefont {L.}~\bibnamefont {Balents}},\
  }\bibfield  {title} {\enquote {\bibinfo {title} {{Strongly Correlated Metal
  Built from Sachdev-Ye-Kitaev Models}},}\ }\href {\doibase
  10.1103/PhysRevLett.119.216601} {\bibfield  {journal} {\bibinfo  {journal}
  {Phys. Rev. Lett.}\ }\textbf {\bibinfo {volume} {119}},\ \bibinfo {pages}
  {216601} (\bibinfo {year} {2017})}\BibitemShut {NoStop}%
\bibitem [{\citenamefont {Chowdhury}\ \emph {et~al.}(2018)\citenamefont
  {Chowdhury}, \citenamefont {Werman}, \citenamefont {Berg},\ and\
  \citenamefont {Senthil}}]{DCsyk}%
  \BibitemOpen
  \bibfield  {author} {\bibinfo {author} {\bibfnamefont {D.}~\bibnamefont
  {Chowdhury}}, \bibinfo {author} {\bibfnamefont {Y.}~\bibnamefont {Werman}},
  \bibinfo {author} {\bibfnamefont {E.}~\bibnamefont {Berg}}, \ and\ \bibinfo
  {author} {\bibfnamefont {T.}~\bibnamefont {Senthil}},\ }\bibfield  {title}
  {\enquote {\bibinfo {title} {Translationally invariant non-fermi-liquid
  metals with critical fermi surfaces: Solvable models},}\ }\href {\doibase
  10.1103/PhysRevX.8.031024} {\bibfield  {journal} {\bibinfo  {journal} {Phys.
  Rev. X}\ }\textbf {\bibinfo {volume} {8}},\ \bibinfo {pages} {031024}
  (\bibinfo {year} {2018})}\BibitemShut {NoStop}%
\bibitem [{\citenamefont {Varma}\ \emph {et~al.}(1989)\citenamefont {Varma},
  \citenamefont {Littlewood}, \citenamefont {Schmitt-Rink}, \citenamefont
  {Abrahams},\ and\ \citenamefont {Ruckenstein}}]{Varma89}%
  \BibitemOpen
  \bibfield  {author} {\bibinfo {author} {\bibfnamefont {C.~M.}\ \bibnamefont
  {Varma}}, \bibinfo {author} {\bibfnamefont {P.~B.}\ \bibnamefont
  {Littlewood}}, \bibinfo {author} {\bibfnamefont {S.}~\bibnamefont
  {Schmitt-Rink}}, \bibinfo {author} {\bibfnamefont {E.}~\bibnamefont
  {Abrahams}}, \ and\ \bibinfo {author} {\bibfnamefont {A.~E.}\ \bibnamefont
  {Ruckenstein}},\ }\bibfield  {title} {\enquote {\bibinfo {title}
  {{Phenomenology of the normal state of Cu-O high-temperature
  superconductors}},}\ }\href {\doibase 10.1103/PhysRevLett.63.1996} {\bibfield
   {journal} {\bibinfo  {journal} {Phys. Rev. Lett.}\ }\textbf {\bibinfo
  {volume} {63}},\ \bibinfo {pages} {1996} (\bibinfo {year}
  {1989})}\BibitemShut {NoStop}%
\bibitem [{\citenamefont {Zhang}(2017)}]{Zhang17}%
  \BibitemOpen
  \bibfield  {author} {\bibinfo {author} {\bibfnamefont {P.}~\bibnamefont
  {Zhang}},\ }\bibfield  {title} {\enquote {\bibinfo {title} {{Dispersive
  Sachdev-Ye-Kitaev model: Band structure and quantum chaos}},}\ }\href
  {\doibase 10.1103/PhysRevB.96.205138} {\bibfield  {journal} {\bibinfo
  {journal} {Phys. Rev. B}\ }\textbf {\bibinfo {volume} {96}},\ \bibinfo
  {pages} {205138} (\bibinfo {year} {2017})}\BibitemShut {NoStop}%
\bibitem [{\citenamefont {Haldar}\ \emph {et~al.}(2018)\citenamefont {Haldar},
  \citenamefont {Banerjee},\ and\ \citenamefont {Shenoy}}]{shenoy}%
  \BibitemOpen
  \bibfield  {author} {\bibinfo {author} {\bibfnamefont {A.}~\bibnamefont
  {Haldar}}, \bibinfo {author} {\bibfnamefont {S.}~\bibnamefont {Banerjee}}, \
  and\ \bibinfo {author} {\bibfnamefont {V.~B.}\ \bibnamefont {Shenoy}},\
  }\bibfield  {title} {\enquote {\bibinfo {title} {{Higher-dimensional
  Sachdev-Ye-Kitaev non-Fermi liquids at Lifshitz transitions}},}\ }\href
  {\doibase 10.1103/PhysRevB.97.241106} {\bibfield  {journal} {\bibinfo
  {journal} {Phys. Rev. B}\ }\textbf {\bibinfo {volume} {97}},\ \bibinfo
  {pages} {241106} (\bibinfo {year} {2018})}\BibitemShut {NoStop}%
\bibitem [{\citenamefont {Jose}\ \emph {et~al.}(2022)\citenamefont {Jose},
  \citenamefont {Seo},\ and\ \citenamefont {Uchoa}}]{Jose}%
  \BibitemOpen
  \bibfield  {author} {\bibinfo {author} {\bibfnamefont {G.}~\bibnamefont
  {Jose}}, \bibinfo {author} {\bibfnamefont {K.}~\bibnamefont {Seo}}, \ and\
  \bibinfo {author} {\bibfnamefont {B.}~\bibnamefont {Uchoa}},\ }\bibfield
  {title} {\enquote {\bibinfo {title} {Non-fermi liquid behavior in the
  sachdev-ye-kitaev model for a one-dimensional incoherent semimetal},}\ }\href
  {\doibase 10.1103/PhysRevResearch.4.013145} {\bibfield  {journal} {\bibinfo
  {journal} {Phys. Rev. Res.}\ }\textbf {\bibinfo {volume} {4}},\ \bibinfo
  {pages} {013145} (\bibinfo {year} {2022})}\BibitemShut {NoStop}%
\bibitem [{\citenamefont {Patel}\ \emph {et~al.}(2018)\citenamefont {Patel},
  \citenamefont {McGreevy}, \citenamefont {Arovas},\ and\ \citenamefont
  {Sachdev}}]{SSmagneto}%
  \BibitemOpen
  \bibfield  {author} {\bibinfo {author} {\bibfnamefont {A.~A.}\ \bibnamefont
  {Patel}}, \bibinfo {author} {\bibfnamefont {J.}~\bibnamefont {McGreevy}},
  \bibinfo {author} {\bibfnamefont {D.~P.}\ \bibnamefont {Arovas}}, \ and\
  \bibinfo {author} {\bibfnamefont {S.}~\bibnamefont {Sachdev}},\ }\bibfield
  {title} {\enquote {\bibinfo {title} {Magnetotransport in a model of a
  disordered strange metal},}\ }\href {\doibase 10.1103/PhysRevX.8.021049}
  {\bibfield  {journal} {\bibinfo  {journal} {Phys. Rev. X}\ }\textbf {\bibinfo
  {volume} {8}},\ \bibinfo {pages} {021049} (\bibinfo {year}
  {2018})}\BibitemShut {NoStop}%
\bibitem [{\citenamefont {Esterlis}\ \emph {et~al.}(2021)\citenamefont
  {Esterlis}, \citenamefont {Guo}, \citenamefont {Patel},\ and\ \citenamefont
  {Sachdev}}]{largeN}%
  \BibitemOpen
  \bibfield  {author} {\bibinfo {author} {\bibfnamefont {I.}~\bibnamefont
  {Esterlis}}, \bibinfo {author} {\bibfnamefont {H.}~\bibnamefont {Guo}},
  \bibinfo {author} {\bibfnamefont {A.~A.}\ \bibnamefont {Patel}}, \ and\
  \bibinfo {author} {\bibfnamefont {S.}~\bibnamefont {Sachdev}},\ }\bibfield
  {title} {\enquote {\bibinfo {title} {Large-$n$ theory of critical fermi
  surfaces},}\ }\href {\doibase 10.1103/PhysRevB.103.235129} {\bibfield
  {journal} {\bibinfo  {journal} {Phys. Rev. B}\ }\textbf {\bibinfo {volume}
  {103}},\ \bibinfo {pages} {235129} (\bibinfo {year} {2021})}\BibitemShut
  {NoStop}%
\bibitem [{\citenamefont {Fu}\ \emph {et~al.}(2017)\citenamefont {Fu},
  \citenamefont {Gaiotto}, \citenamefont {Maldacena},\ and\ \citenamefont
  {Sachdev}}]{Fu:2016vas}%
  \BibitemOpen
  \bibfield  {author} {\bibinfo {author} {\bibfnamefont {W.}~\bibnamefont
  {Fu}}, \bibinfo {author} {\bibfnamefont {D.}~\bibnamefont {Gaiotto}},
  \bibinfo {author} {\bibfnamefont {J.}~\bibnamefont {Maldacena}}, \ and\
  \bibinfo {author} {\bibfnamefont {S.}~\bibnamefont {Sachdev}},\ }\bibfield
  {title} {\enquote {\bibinfo {title} {Supersymmetric sachdev-ye-kitaev
  models},}\ }\href {\doibase 10.1103/PhysRevD.95.026009} {\bibfield  {journal}
  {\bibinfo  {journal} {Phys. Rev. D}\ }\textbf {\bibinfo {volume} {95}},\
  \bibinfo {pages} {026009} (\bibinfo {year} {2017})}\BibitemShut {NoStop}%
\bibitem [{\citenamefont {Murugan}\ \emph {et~al.}(2017)\citenamefont
  {Murugan}, \citenamefont {Stanford},\ and\ \citenamefont
  {Witten}}]{Murugan:2017eto}%
  \BibitemOpen
  \bibfield  {author} {\bibinfo {author} {\bibfnamefont {J.}~\bibnamefont
  {Murugan}}, \bibinfo {author} {\bibfnamefont {D.}~\bibnamefont {Stanford}}, \
  and\ \bibinfo {author} {\bibfnamefont {E.}~\bibnamefont {Witten}},\
  }\bibfield  {title} {\enquote {\bibinfo {title} {{More on Supersymmetric and
  2d Analogs of the SYK Model}},}\ }\href {\doibase 10.1007/JHEP08(2017)146}
  {\bibfield  {journal} {\bibinfo  {journal} {JHEP}\ }\textbf {\bibinfo
  {volume} {08}},\ \bibinfo {pages} {146} (\bibinfo {year} {2017})},\ \Eprint
  {http://arxiv.org/abs/1706.05362} {arXiv:1706.05362 [hep-th]} \BibitemShut
  {NoStop}%
\bibitem [{\citenamefont {Patel}\ and\ \citenamefont
  {Sachdev}(2018)}]{Patel:2018zpy}%
  \BibitemOpen
  \bibfield  {author} {\bibinfo {author} {\bibfnamefont {A.~A.}\ \bibnamefont
  {Patel}}\ and\ \bibinfo {author} {\bibfnamefont {S.}~\bibnamefont
  {Sachdev}},\ }\bibfield  {title} {\enquote {\bibinfo {title} {{Critical
  strange metal from fluctuating gauge fields in a solvable random model}},}\
  }\href {\doibase 10.1103/PhysRevB.98.125134} {\bibfield  {journal} {\bibinfo
  {journal} {Phys. Rev. B}\ }\textbf {\bibinfo {volume} {98}},\ \bibinfo
  {pages} {125134} (\bibinfo {year} {2018})},\ \Eprint
  {http://arxiv.org/abs/1807.04754} {arXiv:1807.04754 [cond-mat.str-el]}
  \BibitemShut {NoStop}%
\bibitem [{\citenamefont {Marcus}\ and\ \citenamefont
  {Vandoren}(2019)}]{Marcus:2018tsr}%
  \BibitemOpen
  \bibfield  {author} {\bibinfo {author} {\bibfnamefont {E.}~\bibnamefont
  {Marcus}}\ and\ \bibinfo {author} {\bibfnamefont {S.}~\bibnamefont
  {Vandoren}},\ }\bibfield  {title} {\enquote {\bibinfo {title} {{A new class
  of SYK-like models with maximal chaos}},}\ }\href {\doibase
  10.1007/JHEP01(2019)166} {\bibfield  {journal} {\bibinfo  {journal} {JHEP}\
  }\textbf {\bibinfo {volume} {01}},\ \bibinfo {pages} {166} (\bibinfo {year}
  {2019})},\ \Eprint {http://arxiv.org/abs/1808.01190} {arXiv:1808.01190
  [hep-th]} \BibitemShut {NoStop}%
\bibitem [{\citenamefont {Wang}(2020)}]{Wang:2019bpd}%
  \BibitemOpen
  \bibfield  {author} {\bibinfo {author} {\bibfnamefont {Y.}~\bibnamefont
  {Wang}},\ }\bibfield  {title} {\enquote {\bibinfo {title} {{Solvable
  Strong-coupling Quantum Dot Model with a Non-Fermi-liquid Pairing
  Transition}},}\ }\href {\doibase 10.1103/PhysRevLett.124.017002} {\bibfield
  {journal} {\bibinfo  {journal} {Phys. Rev. Lett.}\ }\textbf {\bibinfo
  {volume} {124}},\ \bibinfo {pages} {017002} (\bibinfo {year} {2020})},\
  \Eprint {http://arxiv.org/abs/1904.07240} {arXiv:1904.07240
  [cond-mat.str-el]} \BibitemShut {NoStop}%
\bibitem [{\citenamefont {Esterlis}\ and\ \citenamefont
  {Schmalian}(2019)}]{JS}%
  \BibitemOpen
  \bibfield  {author} {\bibinfo {author} {\bibfnamefont {I.}~\bibnamefont
  {Esterlis}}\ and\ \bibinfo {author} {\bibfnamefont {J.}~\bibnamefont
  {Schmalian}},\ }\bibfield  {title} {\enquote {\bibinfo {title} {{Cooper
  pairing of incoherent electrons: An electron-phonon version of the
  Sachdev-Ye-Kitaev model}},}\ }\href {\doibase 10.1103/PhysRevB.100.115132}
  {\bibfield  {journal} {\bibinfo  {journal} {Phys. Rev. B}\ }\textbf {\bibinfo
  {volume} {100}},\ \bibinfo {pages} {115132} (\bibinfo {year}
  {2019})}\BibitemShut {NoStop}%
\bibitem [{\citenamefont {Wang}\ and\ \citenamefont
  {Chubukov}(2020)}]{Wang:2020dtj}%
  \BibitemOpen
  \bibfield  {author} {\bibinfo {author} {\bibfnamefont {Y.}~\bibnamefont
  {Wang}}\ and\ \bibinfo {author} {\bibfnamefont {A.~V.}\ \bibnamefont
  {Chubukov}},\ }\bibfield  {title} {\enquote {\bibinfo {title} {{Quantum Phase
  Transition in the Yukawa-SYK Model}},}\ }\href {\doibase
  10.1103/PhysRevResearch.2.033084} {\bibfield  {journal} {\bibinfo  {journal}
  {Phys. Rev. Res.}\ }\textbf {\bibinfo {volume} {2}},\ \bibinfo {pages}
  {033084} (\bibinfo {year} {2020})},\ \Eprint
  {http://arxiv.org/abs/2005.07205} {arXiv:2005.07205 [cond-mat.str-el]}
  \BibitemShut {NoStop}%
\bibitem [{\citenamefont {Kim}\ \emph {et~al.}(2021)\citenamefont {Kim},
  \citenamefont {Altman},\ and\ \citenamefont {Cao}}]{Kim:2020jpz}%
  \BibitemOpen
  \bibfield  {author} {\bibinfo {author} {\bibfnamefont {J.}~\bibnamefont
  {Kim}}, \bibinfo {author} {\bibfnamefont {E.}~\bibnamefont {Altman}}, \ and\
  \bibinfo {author} {\bibfnamefont {X.}~\bibnamefont {Cao}},\ }\bibfield
  {title} {\enquote {\bibinfo {title} {{Dirac Fast Scramblers}},}\ }\href
  {\doibase 10.1103/PhysRevB.103.L081113} {\bibfield  {journal} {\bibinfo
  {journal} {Phys. Rev. B}\ }\textbf {\bibinfo {volume} {103}},\ \bibinfo
  {pages} {081113} (\bibinfo {year} {2021})},\ \Eprint
  {http://arxiv.org/abs/2010.10545} {arXiv:2010.10545 [cond-mat.str-el]}
  \BibitemShut {NoStop}%
\bibitem [{\citenamefont {{Aldape}}\ \emph {et~al.}(2020)\citenamefont
  {{Aldape}}, \citenamefont {{Cookmeyer}}, \citenamefont {{Patel}},\ and\
  \citenamefont {{Altman}}}]{Adalpe20}%
  \BibitemOpen
  \bibfield  {author} {\bibinfo {author} {\bibfnamefont {E.~E.}\ \bibnamefont
  {{Aldape}}}, \bibinfo {author} {\bibfnamefont {T.}~\bibnamefont
  {{Cookmeyer}}}, \bibinfo {author} {\bibfnamefont {A.~A.}\ \bibnamefont
  {{Patel}}}, \ and\ \bibinfo {author} {\bibfnamefont {E.}~\bibnamefont
  {{Altman}}},\ }\bibfield  {title} {\enquote {\bibinfo {title} {{Solvable
  Theory of a Strange Metal at the Breakdown of a Heavy Fermi Liquid}},}\
  }\href@noop {} {\  (\bibinfo {year} {2020})},\ \Eprint
  {http://arxiv.org/abs/2012.00763} {arXiv:2012.00763 [cond-mat.str-el]}
  \BibitemShut {NoStop}%
\bibitem [{\citenamefont {{Wang}}\ \emph {et~al.}(2021)\citenamefont {{Wang}},
  \citenamefont {{Davis}}, \citenamefont {{Pan}}, \citenamefont {{Wang}},\ and\
  \citenamefont {{Meng}}}]{WangMeng21}%
  \BibitemOpen
  \bibfield  {author} {\bibinfo {author} {\bibfnamefont {W.}~\bibnamefont
  {{Wang}}}, \bibinfo {author} {\bibfnamefont {A.}~\bibnamefont {{Davis}}},
  \bibinfo {author} {\bibfnamefont {G.}~\bibnamefont {{Pan}}}, \bibinfo
  {author} {\bibfnamefont {Y.}~\bibnamefont {{Wang}}}, \ and\ \bibinfo {author}
  {\bibfnamefont {Z.~Y.}\ \bibnamefont {{Meng}}},\ }\bibfield  {title}
  {\enquote {\bibinfo {title} {{Phase diagram of the spin-1/2
  Yukawa-Sachdev-Ye-Kitaev model: Non-Fermi liquid, insulator, and
  superconductor}},}\ }\href {\doibase 10.1103/PhysRevB.103.195108} {\bibfield
  {journal} {\bibinfo  {journal} {Phys. Rev. B}\ }\textbf {\bibinfo {volume}
  {103}},\ \bibinfo {eid} {195108} (\bibinfo {year} {2021})},\ \Eprint
  {http://arxiv.org/abs/2102.10755} {arXiv:2102.10755 [cond-mat.str-el]}
  \BibitemShut {NoStop}%
\bibitem [{\citenamefont {Lee}(1989)}]{PALee89}%
  \BibitemOpen
  \bibfield  {author} {\bibinfo {author} {\bibfnamefont {P.~A.}\ \bibnamefont
  {Lee}},\ }\bibfield  {title} {\enquote {\bibinfo {title} {{Gauge field,
  Aharonov-Bohm flux, and high-${T}_{c}$ superconductivity}},}\ }\href
  {\doibase 10.1103/PhysRevLett.63.680} {\bibfield  {journal} {\bibinfo
  {journal} {Phys. Rev. Lett.}\ }\textbf {\bibinfo {volume} {63}},\ \bibinfo
  {pages} {680} (\bibinfo {year} {1989})}\BibitemShut {NoStop}%
\bibitem [{\citenamefont {Altshuler}\ \emph {et~al.}(1994)\citenamefont
  {Altshuler}, \citenamefont {Ioffe},\ and\ \citenamefont {Millis}}]{AIM}%
  \BibitemOpen
  \bibfield  {author} {\bibinfo {author} {\bibfnamefont {B.~L.}\ \bibnamefont
  {Altshuler}}, \bibinfo {author} {\bibfnamefont {L.~B.}\ \bibnamefont
  {Ioffe}}, \ and\ \bibinfo {author} {\bibfnamefont {A.~J.}\ \bibnamefont
  {Millis}},\ }\bibfield  {title} {\enquote {\bibinfo {title} {Low-energy
  properties of fermions with singular interactions},}\ }\href {\doibase
  10.1103/PhysRevB.50.14048} {\bibfield  {journal} {\bibinfo  {journal} {Phys.
  Rev. B}\ }\textbf {\bibinfo {volume} {50}},\ \bibinfo {pages} {14048}
  (\bibinfo {year} {1994})}\BibitemShut {NoStop}%
\bibitem [{\citenamefont {Polchinski}(1994)}]{Polchinski:1993ii}%
  \BibitemOpen
  \bibfield  {author} {\bibinfo {author} {\bibfnamefont {J.}~\bibnamefont
  {Polchinski}},\ }\bibfield  {title} {\enquote {\bibinfo {title} {{Low-energy
  dynamics of the spinon gauge system}},}\ }\href {\doibase
  10.1016/0550-3213(94)90449-9} {\bibfield  {journal} {\bibinfo  {journal}
  {Nucl. Phys. B}\ }\textbf {\bibinfo {volume} {422}},\ \bibinfo {pages} {617}
  (\bibinfo {year} {1994})},\ \Eprint {http://arxiv.org/abs/cond-mat/9303037}
  {arXiv:cond-mat/9303037} \BibitemShut {NoStop}%
\bibitem [{\citenamefont {{Metlitski}}\ and\ \citenamefont
  {{Sachdev}}(2010)}]{metlitski1}%
  \BibitemOpen
  \bibfield  {author} {\bibinfo {author} {\bibfnamefont {M.~A.}\ \bibnamefont
  {{Metlitski}}}\ and\ \bibinfo {author} {\bibfnamefont {S.}~\bibnamefont
  {{Sachdev}}},\ }\bibfield  {title} {\enquote {\bibinfo {title} {{Quantum
  phase transitions of metals in two spatial dimensions. I. Ising-nematic
  order}},}\ }\href {\doibase 10.1103/PhysRevB.82.075127} {\bibfield  {journal}
  {\bibinfo  {journal} {Phys. Rev. B}\ }\textbf {\bibinfo {volume} {82}},\
  \bibinfo {eid} {075127} (\bibinfo {year} {2010})},\ \Eprint
  {http://arxiv.org/abs/1001.1153} {arXiv:1001.1153 [cond-mat.str-el]}
  \BibitemShut {NoStop}%
\bibitem [{\citenamefont {{Lee}}(2009)}]{sungsik1}%
  \BibitemOpen
  \bibfield  {author} {\bibinfo {author} {\bibfnamefont {S.-S.}\ \bibnamefont
  {{Lee}}},\ }\bibfield  {title} {\enquote {\bibinfo {title} {{Low-energy
  effective theory of Fermi surface coupled with U(1) gauge field in 2+1
  dimensions}},}\ }\href {\doibase 10.1103/PhysRevB.80.165102} {\bibfield
  {journal} {\bibinfo  {journal} {Phys. Rev. B}\ }\textbf {\bibinfo {volume}
  {80}},\ \bibinfo {eid} {165102} (\bibinfo {year} {2009})},\ \Eprint
  {http://arxiv.org/abs/0905.4532} {arXiv:0905.4532 [cond-mat.str-el]}
  \BibitemShut {NoStop}%
\bibitem [{\citenamefont {Mross}\ \emph {et~al.}(2010)\citenamefont {Mross},
  \citenamefont {McGreevy}, \citenamefont {Liu},\ and\ \citenamefont
  {Senthil}}]{mross}%
  \BibitemOpen
  \bibfield  {author} {\bibinfo {author} {\bibfnamefont {D.~F.}\ \bibnamefont
  {Mross}}, \bibinfo {author} {\bibfnamefont {J.}~\bibnamefont {McGreevy}},
  \bibinfo {author} {\bibfnamefont {H.}~\bibnamefont {Liu}}, \ and\ \bibinfo
  {author} {\bibfnamefont {T.}~\bibnamefont {Senthil}},\ }\bibfield  {title}
  {\enquote {\bibinfo {title} {{Controlled expansion for certain
  non-Fermi-liquid metals}},}\ }\href {\doibase 10.1103/PhysRevB.82.045121}
  {\bibfield  {journal} {\bibinfo  {journal} {Phys. Rev. B}\ }\textbf {\bibinfo
  {volume} {82}},\ \bibinfo {pages} {045121} (\bibinfo {year}
  {2010})}\BibitemShut {NoStop}%
\bibitem [{\citenamefont {{Guo}}\ \emph {et~al.}(2022)\citenamefont {{Guo}},
  \citenamefont {{Patel}}, \citenamefont {{Esterlis}},\ and\ \citenamefont
  {{Sachdev}}}]{SS22}%
  \BibitemOpen
  \bibfield  {author} {\bibinfo {author} {\bibfnamefont {H.}~\bibnamefont
  {{Guo}}}, \bibinfo {author} {\bibfnamefont {A.~A.}\ \bibnamefont {{Patel}}},
  \bibinfo {author} {\bibfnamefont {I.}~\bibnamefont {{Esterlis}}}, \ and\
  \bibinfo {author} {\bibfnamefont {S.}~\bibnamefont {{Sachdev}}},\ }\bibfield
  {title} {\enquote {\bibinfo {title} {{Large $N$ theory of critical Fermi
  surfaces II: conductivity}},}\ }\href@noop {} {\bibfield  {journal} {\bibinfo
   {journal} {arXiv e-prints}\ ,\ \bibinfo {eid} {arXiv:2207.08841}} (\bibinfo
  {year} {2022})},\ \Eprint {http://arxiv.org/abs/2207.08841} {arXiv:2207.08841
  [cond-mat.str-el]} \BibitemShut {NoStop}%
\bibitem [{\citenamefont {Prange}\ and\ \citenamefont {Kadanoff}(1964)}]{PK}%
  \BibitemOpen
  \bibfield  {author} {\bibinfo {author} {\bibfnamefont {R.~E.}\ \bibnamefont
  {Prange}}\ and\ \bibinfo {author} {\bibfnamefont {L.~P.}\ \bibnamefont
  {Kadanoff}},\ }\bibfield  {title} {\enquote {\bibinfo {title} {Transport
  theory for electron-phonon interactions in metals},}\ }\href {\doibase
  10.1103/PhysRev.134.A566} {\bibfield  {journal} {\bibinfo  {journal} {Phys.
  Rev.}\ }\textbf {\bibinfo {volume} {134}},\ \bibinfo {pages} {A566} (\bibinfo
  {year} {1964})}\BibitemShut {NoStop}%
\bibitem [{\citenamefont {Kim}\ \emph {et~al.}(1995)\citenamefont {Kim},
  \citenamefont {Lee},\ and\ \citenamefont {Wen}}]{qbe}%
  \BibitemOpen
  \bibfield  {author} {\bibinfo {author} {\bibfnamefont {Y.~B.}\ \bibnamefont
  {Kim}}, \bibinfo {author} {\bibfnamefont {P.~A.}\ \bibnamefont {Lee}}, \ and\
  \bibinfo {author} {\bibfnamefont {X.-G.}\ \bibnamefont {Wen}},\ }\bibfield
  {title} {\enquote {\bibinfo {title} {Quantum boltzmann equation of composite
  fermions interacting with a gauge field},}\ }\href {\doibase
  10.1103/PhysRevB.52.17275} {\bibfield  {journal} {\bibinfo  {journal} {Phys.
  Rev. B}\ }\textbf {\bibinfo {volume} {52}},\ \bibinfo {pages} {17275}
  (\bibinfo {year} {1995})}\BibitemShut {NoStop}%
\bibitem [{\citenamefont {Nave}\ and\ \citenamefont {Lee}(2007)}]{nave}%
  \BibitemOpen
  \bibfield  {author} {\bibinfo {author} {\bibfnamefont {C.~P.}\ \bibnamefont
  {Nave}}\ and\ \bibinfo {author} {\bibfnamefont {P.~A.}\ \bibnamefont {Lee}},\
  }\bibfield  {title} {\enquote {\bibinfo {title} {Transport properties of a
  spinon fermi surface coupled to a u(1) gauge field},}\ }\href {\doibase
  10.1103/PhysRevB.76.235124} {\bibfield  {journal} {\bibinfo  {journal} {Phys.
  Rev. B}\ }\textbf {\bibinfo {volume} {76}},\ \bibinfo {pages} {235124}
  (\bibinfo {year} {2007})}\BibitemShut {NoStop}%
\bibitem [{\citenamefont {Keldysh}(1964)}]{Keldysh}%
  \BibitemOpen
  \bibfield  {author} {\bibinfo {author} {\bibfnamefont {L.~V.}\ \bibnamefont
  {Keldysh}},\ }\bibfield  {title} {\enquote {\bibinfo {title} {Diagram
  technique for nonequilibrium processes},}\ }\href@noop {} {\bibfield
  {journal} {\bibinfo  {journal} {J. Exptl. Theoret. Phys. (U.S.S.R.)}\
  }\textbf {\bibinfo {volume} {47}},\ \bibinfo {pages} {1515} (\bibinfo {year}
  {1964})}\BibitemShut {NoStop}%
\bibitem [{\citenamefont {Chubukov}(2005)}]{Andrey3}%
  \BibitemOpen
  \bibfield  {author} {\bibinfo {author} {\bibfnamefont {A.~V.}\ \bibnamefont
  {Chubukov}},\ }\bibfield  {title} {\enquote {\bibinfo {title} {Ward
  identities for strongly coupled eliashberg theories},}\ }\href {\doibase
  10.1103/PhysRevB.72.085113} {\bibfield  {journal} {\bibinfo  {journal} {Phys.
  Rev. B}\ }\textbf {\bibinfo {volume} {72}},\ \bibinfo {pages} {085113}
  (\bibinfo {year} {2005})}\BibitemShut {NoStop}%
\bibitem [{\citenamefont {Klein}\ \emph {et~al.}(2018)\citenamefont {Klein},
  \citenamefont {Lederer}, \citenamefont {Chowdhury}, \citenamefont {Berg},\
  and\ \citenamefont {Chubukov}}]{Andrey1}%
  \BibitemOpen
  \bibfield  {author} {\bibinfo {author} {\bibfnamefont {A.}~\bibnamefont
  {Klein}}, \bibinfo {author} {\bibfnamefont {S.}~\bibnamefont {Lederer}},
  \bibinfo {author} {\bibfnamefont {D.}~\bibnamefont {Chowdhury}}, \bibinfo
  {author} {\bibfnamefont {E.}~\bibnamefont {Berg}}, \ and\ \bibinfo {author}
  {\bibfnamefont {A.}~\bibnamefont {Chubukov}},\ }\bibfield  {title} {\enquote
  {\bibinfo {title} {Dynamical susceptibility near a long-wavelength critical
  point with a nonconserved order parameter},}\ }\href {\doibase
  10.1103/PhysRevB.97.155115} {\bibfield  {journal} {\bibinfo  {journal} {Phys.
  Rev. B}\ }\textbf {\bibinfo {volume} {97}},\ \bibinfo {pages} {155115}
  (\bibinfo {year} {2018})}\BibitemShut {NoStop}%
\bibitem [{\citenamefont {Chubukov}\ and\ \citenamefont
  {Maslov}(2017)}]{Andrey2}%
  \BibitemOpen
  \bibfield  {author} {\bibinfo {author} {\bibfnamefont {A.~V.}\ \bibnamefont
  {Chubukov}}\ and\ \bibinfo {author} {\bibfnamefont {D.~L.}\ \bibnamefont
  {Maslov}},\ }\bibfield  {title} {\enquote {\bibinfo {title} {Optical
  conductivity of a two-dimensional metal near a quantum critical point: The
  status of the extended drude formula},}\ }\href {\doibase
  10.1103/PhysRevB.96.205136} {\bibfield  {journal} {\bibinfo  {journal} {Phys.
  Rev. B}\ }\textbf {\bibinfo {volume} {96}},\ \bibinfo {pages} {205136}
  (\bibinfo {year} {2017})}\BibitemShut {NoStop}%
\bibitem [{\citenamefont {Beach}\ \emph {et~al.}(2000)\citenamefont {Beach},
  \citenamefont {Gooding},\ and\ \citenamefont {Marsiglio}}]{pade1}%
  \BibitemOpen
  \bibfield  {author} {\bibinfo {author} {\bibfnamefont {K.~S.~D.}\
  \bibnamefont {Beach}}, \bibinfo {author} {\bibfnamefont {R.~J.}\ \bibnamefont
  {Gooding}}, \ and\ \bibinfo {author} {\bibfnamefont {F.}~\bibnamefont
  {Marsiglio}},\ }\bibfield  {title} {\enquote {\bibinfo {title} {Reliable
  pad\'e analytical continuation method based on a high-accuracy symbolic
  computation algorithm},}\ }\href {\doibase 10.1103/PhysRevB.61.5147}
  {\bibfield  {journal} {\bibinfo  {journal} {Phys. Rev. B}\ }\textbf {\bibinfo
  {volume} {61}},\ \bibinfo {pages} {5147} (\bibinfo {year}
  {2000})}\BibitemShut {NoStop}%
\bibitem [{\citenamefont {Sch\"ott}\ \emph {et~al.}(2016)\citenamefont
  {Sch\"ott}, \citenamefont {Locht}, \citenamefont {Lundin}, \citenamefont
  {Gr$\aa{}$n\"as}, \citenamefont {Eriksson},\ and\ \citenamefont
  {Di~Marco}}]{pade2}%
  \BibitemOpen
  \bibfield  {author} {\bibinfo {author} {\bibfnamefont {J.}~\bibnamefont
  {Sch\"ott}}, \bibinfo {author} {\bibfnamefont {I.~L.~M.}\ \bibnamefont
  {Locht}}, \bibinfo {author} {\bibfnamefont {E.}~\bibnamefont {Lundin}},
  \bibinfo {author} {\bibfnamefont {O.}~\bibnamefont {Gr$\aa{}$n\"as}},
  \bibinfo {author} {\bibfnamefont {O.}~\bibnamefont {Eriksson}}, \ and\
  \bibinfo {author} {\bibfnamefont {I.}~\bibnamefont {Di~Marco}},\ }\bibfield
  {title} {\enquote {\bibinfo {title} {Analytic continuation by averaging
  pad\'e approximants},}\ }\href {\doibase 10.1103/PhysRevB.93.075104}
  {\bibfield  {journal} {\bibinfo  {journal} {Phys. Rev. B}\ }\textbf {\bibinfo
  {volume} {93}},\ \bibinfo {pages} {075104} (\bibinfo {year}
  {2016})}\BibitemShut {NoStop}%
\bibitem [{\citenamefont {Lucas}\ and\ \citenamefont
  {Das~Sarma}(2018)}]{PhysRevB.97.115449}%
  \BibitemOpen
  \bibfield  {author} {\bibinfo {author} {\bibfnamefont {A.}~\bibnamefont
  {Lucas}}\ and\ \bibinfo {author} {\bibfnamefont {S.}~\bibnamefont
  {Das~Sarma}},\ }\bibfield  {title} {\enquote {\bibinfo {title} {Electronic
  sound modes and plasmons in hydrodynamic two-dimensional metals},}\ }\href
  {\doibase 10.1103/PhysRevB.97.115449} {\bibfield  {journal} {\bibinfo
  {journal} {Phys. Rev. B}\ }\textbf {\bibinfo {volume} {97}},\ \bibinfo
  {pages} {115449} (\bibinfo {year} {2018})}\BibitemShut {NoStop}%
\bibitem [{\citenamefont {Mandal}(2022)}]{mandal}%
  \BibitemOpen
  \bibfield  {author} {\bibinfo {author} {\bibfnamefont {I.}~\bibnamefont
  {Mandal}},\ }\bibfield  {title} {\enquote {\bibinfo {title} {Zero sound and
  plasmon modes for non-fermi liquids},}\ }\href {\doibase
  https://doi.org/10.1016/j.physleta.2022.128292} {\bibfield  {journal}
  {\bibinfo  {journal} {Physics Letters A}\ }\textbf {\bibinfo {volume}
  {447}},\ \bibinfo {pages} {128292} (\bibinfo {year} {2022})}\BibitemShut
  {NoStop}%
\bibitem [{\citenamefont {Gu}\ \emph {et~al.}(2020)\citenamefont {Gu},
  \citenamefont {Kitaev}, \citenamefont {Sachdev},\ and\ \citenamefont
  {Tarnopolsky}}]{Gu2020NotesOT}%
  \BibitemOpen
  \bibfield  {author} {\bibinfo {author} {\bibfnamefont {Y.}~\bibnamefont
  {Gu}}, \bibinfo {author} {\bibfnamefont {A.~Y.}\ \bibnamefont {Kitaev}},
  \bibinfo {author} {\bibfnamefont {S.}~\bibnamefont {Sachdev}}, \ and\
  \bibinfo {author} {\bibfnamefont {G.~M.}\ \bibnamefont {Tarnopolsky}},\
  }\bibfield  {title} {\enquote {\bibinfo {title} {Notes on the complex
  sachdev-ye-kitaev model},}\ }\href
  {https://link.springer.com/article/10.1007/JHEP02(2020)157#citeas} {\bibfield
   {journal} {\bibinfo  {journal} {Journal of High Energy Physics}\ }\textbf
  {\bibinfo {volume} {2020}},\ \bibinfo {pages} {1} (\bibinfo {year}
  {2020})}\BibitemShut {NoStop}%
\bibitem [{\citenamefont {Guo}\ \emph {et~al.}(2020)\citenamefont {Guo},
  \citenamefont {Gu},\ and\ \citenamefont {Sachdev}}]{linear}%
  \BibitemOpen
  \bibfield  {author} {\bibinfo {author} {\bibfnamefont {H.}~\bibnamefont
  {Guo}}, \bibinfo {author} {\bibfnamefont {Y.}~\bibnamefont {Gu}}, \ and\
  \bibinfo {author} {\bibfnamefont {S.}~\bibnamefont {Sachdev}},\ }\bibfield
  {title} {\enquote {\bibinfo {title} {Linear in temperature resistivity in the
  limit of zero temperature from the time reparameterization soft mode},}\
  }\href {\doibase 10.1016/j.aop.2020.168202} {\bibfield  {journal} {\bibinfo
  {journal} {Annals of Physics}\ }\textbf {\bibinfo {volume} {418}} (\bibinfo
  {year} {2020}),\ 10.1016/j.aop.2020.168202}\BibitemShut {NoStop}%
\bibitem [{\citenamefont {Shankar}(1994)}]{RevModPhys.66.129}%
  \BibitemOpen
  \bibfield  {author} {\bibinfo {author} {\bibfnamefont {R.}~\bibnamefont
  {Shankar}},\ }\bibfield  {title} {\enquote {\bibinfo {title}
  {Renormalization-group approach to interacting fermions},}\ }\href {\doibase
  10.1103/RevModPhys.66.129} {\bibfield  {journal} {\bibinfo  {journal} {Rev.
  Mod. Phys.}\ }\textbf {\bibinfo {volume} {66}},\ \bibinfo {pages} {129}
  (\bibinfo {year} {1994})}\BibitemShut {NoStop}%
\bibitem [{\citenamefont {Abrikosov}\ and\ \citenamefont
  {Khalatnikov}(1959)}]{Abrikosov_1959}%
  \BibitemOpen
  \bibfield  {author} {\bibinfo {author} {\bibfnamefont {A.~A.}\ \bibnamefont
  {Abrikosov}}\ and\ \bibinfo {author} {\bibfnamefont {I.~M.}\ \bibnamefont
  {Khalatnikov}},\ }\bibfield  {title} {\enquote {\bibinfo {title} {The theory
  of a fermi liquid (the properties of liquid 3he at low temperatures)},}\
  }\href {\doibase 10.1088/0034-4885/22/1/310} {\bibfield  {journal} {\bibinfo
  {journal} {Reports on Progress in Physics}\ }\textbf {\bibinfo {volume}
  {22}},\ \bibinfo {pages} {329} (\bibinfo {year} {1959})}\BibitemShut
  {NoStop}%
\end{thebibliography}%
\end{document}